\documentclass[manuscript]{acmart}

\AtBeginDocument{%
  \providecommand\BibTeX{{%
    \normalfont B\kern-0.5em{\scshape i\kern-0.25em b}\kern-0.8em\TeX}}}

\setcopyright{acmlicensed}

\acmJournal{IMWUT}
\acmYear{2024} 
\acmVolume{8} 
\acmNumber{1} 
\acmArticle{XX} 
\acmMonth{3}
\acmDOI{10.1145/XXXXXXX}

\usepackage{_macros}

\begin{document}

\title{Exploring Uni-manual Around Ear Off-Device Gestures for Earables}

\author{Shaikh Shawon Arefin Shimon}
\email{ssarefin@uwaterloo.ca}
\orcid{0000-0001-5007-2828}
\affiliation{%
  \institution{School of Computer Science, University of Waterloo}
  \streetaddress{200 University Ave W}
  \city{Waterloo}
  \state{Ontario}
  \country{Canada}
  \postcode{N2L 3G1}
}

\author{Ali Neshati}
\orcid{0000-0002-0405-1169}
\email{ali.neshati@ontariotechu.ca}
\affiliation{%
  \institution{Ontario Tech University}
  \streetaddress{2000 Simcoe St N}
  \city{Oshawa}
  \state{Ontario}
  \country{Canada}
  \postcode{L1G 0C5}
}

\author{Junwei Sun}
\email{junwei.sun@huawei.com}
\orcid{0009-0001-5933-0122}
\author{Qiang Xu}
\email{qiang.xu1@huawei.com}
\orcid{0000-0002-0077-1003}
\affiliation{%
  \institution{Huawei Human-Machine Interaction Lab}
  \streetaddress{19 Allstate Pkwy}
  \city{Markham}
  \state{Ontario}
  \country{Canada}
}

\author{Jian Zhao}
\orcid{0000-0001-5008-4319}
\email{jianzhao@uwaterloo.ca}
\affiliation{%
  \institution{School of Computer Science, University of Waterloo}
  \streetaddress{200 University Ave W}
  \city{Waterloo}
  \state{Ontario}
  \country{Canada}
  \postcode{N2L 3G1}
}

\renewcommand{\shortauthors}{Shimon et al.}

\begin{abstract}
Small form factor limits physical input space in earable (i.e., ear-mounted wearable) devices. 
Off-device earable inputs in alternate mid-air and on-skin around-ear interaction spaces using uni-manual gestures can address this input space limitation. 
Segmenting these alternate interaction spaces to create multiple gesture regions for reusing off-device gestures can expand earable input vocabulary by a large margin.
Although prior earable interaction research has explored off-device gesture preferences and recognition techniques in such interaction spaces, supporting gesture reuse over multiple gesture regions needs \rev{further exploration}. 
We collected and analyzed 7560 uni-manual gesture motion data from 18 participants to explore earable gesture reuse by segmentation of on-skin and mid-air spaces around the ear.
\rev{Our results show that gesture performance degrades significantly beyond 3 mid-air and 5 on-skin around-ear gesture regions for different uni-manual gesture classes (\eg,~\textit{swipe}, \textit{pinch}, \textit{tap}).}
We also present qualitative findings on most and least preferred regions (and associated boundaries) by end-users for different uni-manual gesture shapes across both interaction spaces for earable devices. 
Our results complement earlier elicitation studies and interaction technologies for earables to help expand the gestural input vocabulary and potentially drive future commercialization of such devices.   
\end{abstract}




\maketitle

\section{Introduction}

Wireless earbuds are one of the most popular wearables~\cite{9195351} among consumers.
It is now commonplace to see people wearing such wearables in public settings (\eg, office, gym, public transportation) to discretely listen to music, take calls, or filter out external noise using noise suppression features for creating personal quiet zones.  
However, the small form factor and input area constraint in wireless earbuds restrict physical touch interaction to simple on-device taps and swipes, thus limiting their input space.
Off-the-shelve wireless earbuds increasingly integrate different sensors~\cite{song2021designing} to extend their functionality beyond personal audio and hearing aid support~\cite{Earable-Computing-Romit-UIUC, plazak2018survey, GrandViewResearch_EarableReport}.
Wireless earbuds such as \textit{Apple AirPods}~\cite{apple_airpods_pro}, or \textit{Samsung Galaxy Buds Pro}~\cite{samsung_galaxy_buds_pro} are pre-equipped with IMU, motion\rev{,} and health sensors for supporting health sensing and fitness tracking.
These kinds of sensor-infused, ear-mounted computing devices supporting additional functionalities (\eg,~health sensing\rev{~\cite{ne2021hearables}}, activity recognition\rev{~\cite{eSense-HAR-Kyushu-Sozolab}}, gesture sensing\rev{~\cite{min2018exploring}}, wearable public display\rev{~\cite{Stanke_Headphone_Public_Displays}}) beyond traditional personal audio or hearing aids are defined as \textit{earable} devices~\cite{Earable-Taxonomy} in the ubiquitous computing community.

Earable platforms have gained significant traction among ubiquitous computing researchers~\cite{Earable-Computing-Romit-UIUC, eSense-EarlyPrototype, plazak2018survey}, especially in supporting gestural sensing to counter the on-device interaction space limitation problem.
The research community has proposed several off-device approaches~\cite{Earable-Taxonomy} (\eg,~manual on-skin/mid-air motion gestures, head movements, facial expressions, voiced and silent commands) leveraging gesture sensing using off-the-shelve or custom hardware to expand the earable input vocabulary.
Among the proposed approaches, mid-air and \rev{on-skin uni-manual off-device around-ear gestures} are popular strategies~\cite{cao2023earace, Earable-Computing-Romit-UIUC}.
Such hand-to-face gestures tend to occur naturally and unconsciously with significant daily frequency among end users~\cite{nicas2008study, kwok2015face}.
Besides gesture recognition techniques for such off-device gestures~\cite{Earable-Taxonomy} for earables, prior gesture elicitation studies~\cite{Smartbuds-Elicitation-Hanae-UWaterloo, chen2020exploring} looked at eliciting gesture vocabulary with individual qualitative metrics such as \textit{task suitability}, \textit{gesture usability}\rev{,} and \textit{social comfort}. 
These studies revealed that end-users have a high agreement regarding uni-manual horizontal~(front/back) and vertical~(up/down) \textit{swipe}, \textit{pinch (in/out)}, and \textit{tap} gestures for off-device earable interactions.

\begin{figure}[tb!]
    \centering
     \begin{subfigure}[b]{0.73\linewidth}
         \centering
         \includegraphics[width=0.9\linewidth]{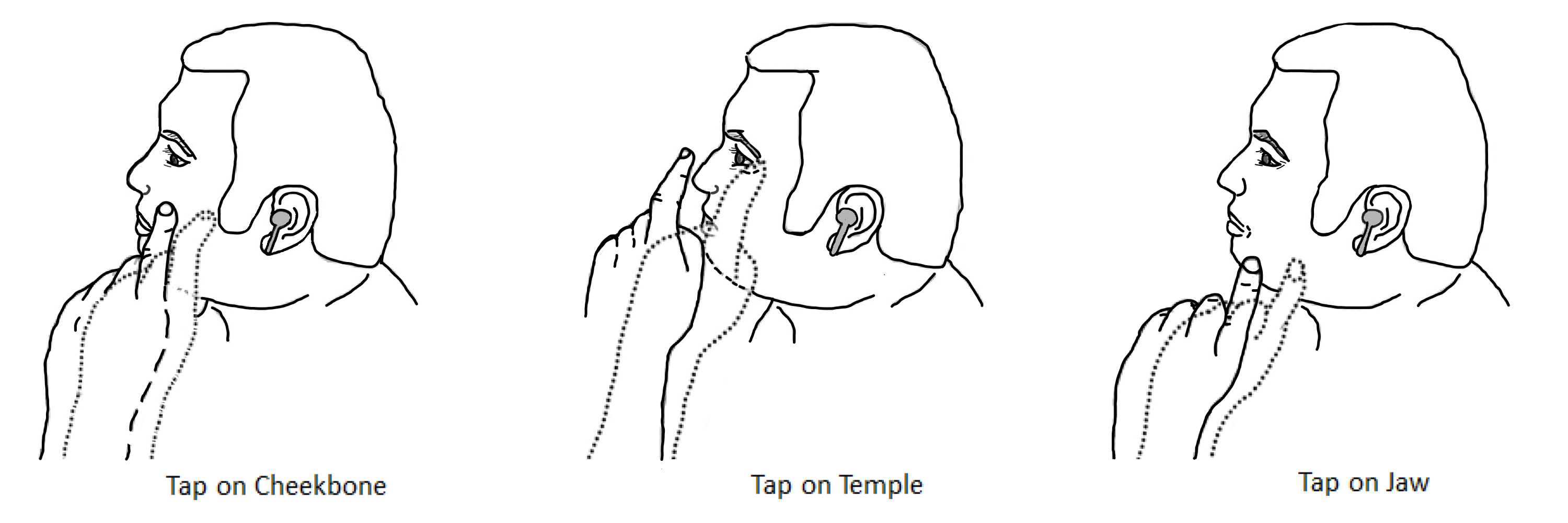}
         \vspace*{-0.2cm} 
         \captionsetup{justification=centering}
         \caption{}
         \label{fig:Taps at different locations}
    \end{subfigure}
    \hfill
    \begin{subfigure}[b]{0.24\linewidth}
         \centering
         \includegraphics[width=0.9\linewidth]{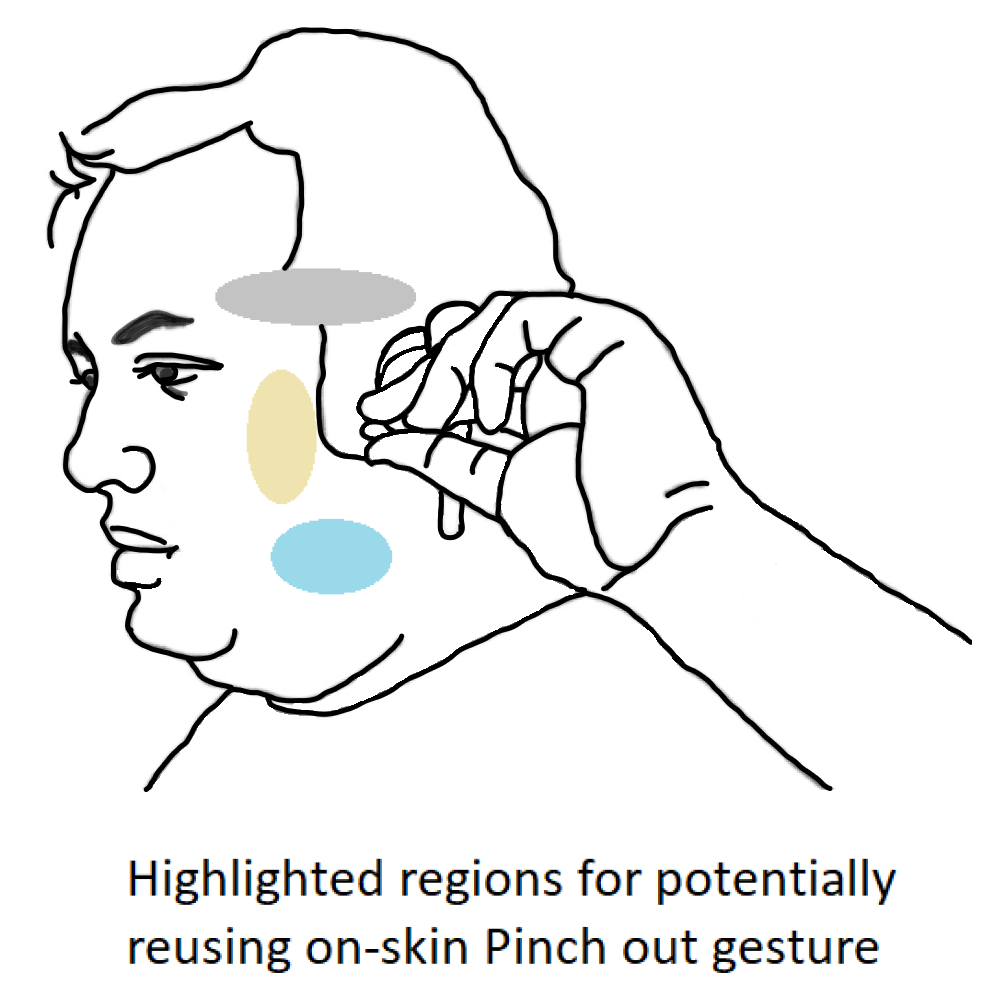}
         \vspace*{-0.2cm} 
         \captionsetup{justification=centering}
         \caption{}
         \label{fig:-PotentialApplications-Sideview}
    \end{subfigure}
    \label{fig: face sketches for gesture representation}
    \captionsetup{justification=centering}
    \vspace*{-0.3cm}    
    \caption{Reusing gestures at different regions for off-device Earable interaction: (a) Off-device, around-ear and on-skin taps at 3 distinct gesture regions. (b) Potential gesture regions with highlighted boundaries.}
\end{figure}  

Such on-skin and mid-air gestures could be fine-grained by reusing them across different gesture regions in an interaction space (\rev{Figure }\ref{fig:Taps at different locations}). 
For example, a \rev{\textit{tap}} on the temple could toggle the \textit{noise cancellation} feature on an earable device. 
In contrast, a \rev{\textit{tap}} on the cheek could toggle the \rev{\textit{microphone muting}} feature on the earable device. 
Understanding the threshold for the number of unique gesture regions in an interaction space for different gestures and their associated region boundaries is necessary to support gesture reuse in off-device and around-ear earable interaction.
Segmenting an interaction space into multiple gesture regions could greatly expand the gestural input vocabulary while reusing a few gesture classes (\textit{swipe}, \textit{tap}, \textit{pinch}), improving gesture memorability~\cite{nacenta2013memorability}.
Although prior literature explored off-device uni-manual gesture reuse across different segmented regions for outer-ear touches to support gesture reusability, similar exploration is lacking for around-ear on-skin space (i.e., face, head, and neck) and above-ear mid-air spaces for supporting similar gestures.

We address this research gap by studying the reuse of different gesture classes at different mid-air and on-skin gesture regions and the effect of increasing the number of gesture regions on quantitative and qualitative gesture performance metrics for mid-air and on-skin interaction spaces. 
By leveraging some common uni-manual off-device and around-ear gestures (\ie, {swipe}, {tap}, and {pinch}) proposed in earlier earable gesture elicitation studies~\cite{Smartbuds-Elicitation-Hanae-UWaterloo, FacialExpressionSmartGlass}, we collected 7,560 gesture motion data from 18 participants using a \textit{Vicon} motion camera for analyzing their performance against an increasing number of gesture regions in mid-air and on-skin interaction spaces.
We also examined how participants defined gesture region boundaries in the off-device interaction spaces under consideration.

The following outlines the contributions of our work:
\begin{itemize}
    \item Comparison of mid-air and around-ear on-skin space for uni-manual, off-device earable interaction across different gesture classes;
    \item Understanding of end-user preference shift from one interaction space to another for increasing gesture reuse;
    \item Analysis of the effects of different factors (\ie, the number of gesture regions and associated boundaries) on interaction space segmentation for uni-manual, off-device earable gestures.
\end{itemize}

Collectively, the findings from our gesture motion experiment on end-users for off-device, around ear earable gestures provide a guideline for implementing gesture reuse across different interaction spaces by complimenting earlier gesture recognition technology research for earables and qualitative exploration of gesture preferences for off-device earable interaction.

\section{Related Work}
\label{section: Related Work}

\subsection{Gesture Recognition Support and Device Format for Earables}
\label{subsection: Related Work - Technologies and Devices}

Prior research on interaction technologies for alternate earable interaction spaces looked at face touch~\cite{kakaraparthi2021facesense, EarBuddy-UWashington}, ear touch~\cite{Earput2013Lissermann, kikuchi2017eartouch, zhang2022toward}, mid-air manual motion gesture~\cite{EarGest, imagimob, metzger2004freedigiter, jin2021sonicasl, tamaki2009brainy, 6710137, 5650512, 6200559}, head movements~\cite{eSense-HeadTracking, vanderdonckt2019head}, facial expressions~\cite{PPGface, amesaka2019facial, FaceListener, li2022eario}, voice~\cite{berdasco2019user}, and silent~\cite{EarCommand_Jin, zeng2023msilent, srivastava2022muteit, shariff2022cw} command recognition to solve earable input space limitation by leveraging various sensors (IMU, proximity, infrared, PPG, ECG) and microphones present in off-the-shelf, or custom earable hardware.
Besides these sensor-based approaches, computer vision has also been utilized in various instances~\cite{tamaki2009brainy, 6710137, 5650512, 6200559} to detect various mid-air or hand-to-face gestures using cameras mounted on earable devices.
However, such camera-based, body-mounted gesture recognition platforms \rev{have practical} usability and privacy concerns~\cite{EarCommand_Jin}.
Although research into end-user acceptability for gestures shows that facial expression-based gestures are acceptable in private settings, they are less socially acceptable in public settings~\cite{koelle2020social} due to the possibility of attracting unwarranted, negative attention~\cite{eastwood2003negative}.
Voice or silent command-based interactions require memorization of a large set of unique \rev{commands, and} are not well suited for user interface navigational tasks~\cite{farringdon1999co} in other wearable or interactive devices. 
Physical uni-manual interactions on ear helix (\eg,~ear bends, pinch, \rev{slides)} proposed in Lissermann \etal~\cite{Earput2013Lissermann} can potentially displace earable devices, making on-ear physical touch-based interaction difficult for expanding earable input space.
Head/neck-based movement can put significant physical strain after repetitive use. 
Large manual gestures away from body suffer from \textit{``gorilla arm''} fatigue \rev{syndrome~\cite{hincapie2014consumed}} and can attract unwarranted, negative attention~\cite{eastwood2003negative} in public spaces, similar to facial and head/neck motion gestures.

In natural, unconstrained settings when one or both hands are free for gestural interaction, studies have shown that end-users prefer making uni-manual hand-to-face gestures on skin or mid-air closer to \rev{the} head, compared to other large body gestures~\cite{hossain2021exploring}.
Such hand-to-face gestures are unconscious and spontaneous motions~\cite{barroso1980self, ekman1972hand, harrigan1987self, krout1954experimental, mueller2019self} with observed frequency from 15.7 to 23 contacts/hr~\cite{nicas2008study}.
Such subtle hand-to-face gestures are more socially acceptable ~\cite{koelle2020social} \rev{than} large manual motion gestures.
Previous research on earables~\cite{Earput2013Lissermann, Smartbuds-Elicitation-Hanae-UWaterloo} points out some real-world scenarios of uni-manual interaction with ear-based devices in real-world settings.
Such interaction also closely matches end-user expectations of wearable earbud devices, where user behavior is to \rev{primarily} interact with one earbud with one hand at a time in a public space, freeing up the alternate hand for other possible tasks. 

These reasons make uni-manual touch gestures on the skin around the ear (face, head, and neck\rev{) and} mid-air gestures around the head viable choices for expanding earable input vocabulary.
Most of the previously discussed research on on-skin and mid-air earable interaction spaces do show that it is possible to recognize both gesture classes and gesture location over different areas around the face. 
However, the prior literature lacks a comprehensive analysis of earable gesture spaces to support gesture reuse by leveraging gesture location and gesture class recognition techniques. 
This inspired our research focus on \rev{off-device} gesture space exploration \rev{to support} gesture reuse in \rev{earables and identify} gesture regions of interest. 
Research into emotional user experience elements of earphone form factors~\cite {yoo2019analysis} \rev{shows} in-ear earables (\ie,~wireless earbuds) have higher wearing style \rev{satisfaction and} preference for discrete interaction with device in public spaces ~\cite{grandviewresearchSmartHeadphones} compared to behind-the-ear or on-ear (\eg, headphones) earable formats. 
As such, we limited our gesture reuse study to in-ear earable devices such as wireless earbuds.

\subsection{Interaction Space Segmentation for Gestural Input Devices}
\label{subsection: related work - Interaction Space Segmentation for Gestural Input Devices}

Wong \etal~\cite{Wong2020BezelInitiatedSwipeSmartwatch} explored \textit{bezel-initiated swipe} (BIS---a form of uni-manual touch interaction) reuse over multiple gesture regions ~(6/ 8/ 12/ 16/ 24/ 36) on smartwatch bezels.
\textit{Accuracy}, \textit{task completion time}, \rev{and other} round-bezel-specific metrics (\ie, \textit{absolute}/\textit{relative offset}) were used to measure BIS performance against increasing gesture regions for input reuse. 
Rey \etal~\cite{Rey2022BezelToBezel} proposed a similar uni-manual smartwatch gesture called \textit{bezel-to-bezel (B2B)} \rev{interaction and} explored the reuse of the B2B gesture over 4, 6\rev{, and} 8 bezel regions.
Dezfuli \etal~\cite{Dezfuli_Palm_RC} explored segmentation of an imaginary gesture space over \rev{hand/palm} surface for using on-skin \textit{swipe} and \rev{\textit{tap}-like} gestures to control a smart \rev{TV}. 
Their observations revealed that up to 5 landmarks on the palm could be correctly targeted using uni-manual touch gestures with alternate hands. 
Gil \etal~\cite{gil2023thumbair} proposed an \rev{in-air thumb} typing system (\textit{ThumbAir}) on a commercial head-mounted display \rev{(HMD), where} they explored viable gesture region positions and an optimal number of regions for dividing the input space for different letter input groups. 
Their analysis leveraged similar general metrics to~\cite{Wong2020BezelInitiatedSwipeSmartwatch, Rey2022BezelToBezel} (\eg,~ \textit{accuracy}, \textit{gesture time}\rev{) and} gesture motion-specific metrics (\rev{\eg,~\textit{wrist angle change}}) to compare the different numbers of regions on both hands for typing input. 
Although the above works focus on different types of wearable devices compared to earables, the analysis metrics and study design still provide essential guidelines for performing analysis of wearable gesture reuses over multiple gesture regions in alternate interaction spaces.

Besides proposing custom earable hardware to detect physical touch gestures on the outer ear, 
Lissermann \etal~\cite{Earput2013Lissermann} explored the ear helix design space for touch gesture reuse across multiple regions, examined the relationship between gesture accuracy and increasing gesture regions for gesture reuse, and suggested potential applications leveraging this design space for earable interaction. 
However, their exploration into region segmentation for earable interaction is limited only to the outer ear and explored in the context of a single touch or tap gesture.
The effect of other touch primitives \rev{(\eg,~swipes and pinches)} on the outer ear \rev{has} not been explored in prior works. 
Prior earable research also lacks exploration of around-ear on-skin/mid-air gesture reuse.

To address this research gap, we focus on exploring the segmentation of on-skin (\ie,~face/neck or other parts of the head) and mid-air (\ie,~above-ear and around-head), around-ear interaction space to support gesture reuse across different gesture regions.
Our work is closely related to and inspired by gesture region segmentation analysis on ear helix for uni-manual physical interaction in \textit{Earput}~\cite{Earput2013Lissermann}. 
Based on findings from previous earable elicitation studies~\cite{chen2020exploring, Smartbuds-Elicitation-Hanae-UWaterloo}, we limit our gesture classes to primitive gestures such as \textit{tap}, \textit{swipe}, and \textit{pinch} for exploring gesture reusability across multiple gesture regions.
\section{Methodology}
\label{section: methodology}

\subsection{Research Questions}
\label{subsection: Research Questions}

In this work, we experimented with 7 gestures~(\textit{swipe up}, \textit{swipe down}, \textit{swipe front}, \textit{swipe back}, \textit{pinch in}, \textit{pinch out}, and \textit{tap}).
These \rev{were} the most common gestures proposed in the previous earable elicitation studies~\cite{chen2020exploring, Smartbuds-Elicitation-Hanae-UWaterloo} and had the most ease of performance, social acceptability, and memorability in various settings ~\cite{ahlstrom2014you}.
We did not leverage the ``\textit{cover ear}" gesture elicited from ~\cite{Smartbuds-Elicitation-Hanae-UWaterloo} study as our study focused only on around-ear gestures.
We explored how users perform the gestures mentioned above within the following two interaction spaces: 
\begin{itemize}
    \item \textbf{On-skin, around-ear space}: 
    This space includes all areas on the skin above the neck, excluding both ears.  
    \item \textbf{Mid-air space}:
    This space includes mid-air space above and around the ear without any limitation of distance from the ear. 
\end{itemize}

We aimed to analyze the gesture motion data from the \rev{7 gestures as mentioned earlier} and identify region boundaries for gesture reuse in \rev{on-skin and mid-air} interaction spaces to answer the following research questions:

\begin{itemize}
    \item \textbf{RQ1}: Whether and how does gesture performance vary between in mid-air and on-skin space?
    
    \item \textbf{RQ2}: Whether and how does gesture performance vary when the number of segments increases in a chosen interaction space?
    
    \item \textbf{RQ3}: Whether and how does end-user consensus exist on most and least preferred regions across a fixed number of gesture regions in a particular interaction space? 
    
\end{itemize}

\subsection{Experimental Variables and Hypotheses}
\label{subsection: Experimental Variables and Hypotheses}
To answer our research questions, we designed an experiment using the following 3 independent variables:
\begin{itemize}
    \item \textbf{IV1} - \textit{Interaction space}: \rev{On-skin and} Mid-air.
    
    \item \textbf{IV2} - \textit{Gesture shape}: Tap, Swipe Up, Swipe Down, Swipe Front, Swipe Back, Pinch In, and Pinch Out.
    
    \item \textbf{IV3} - \textit{Number of gesture regions}: 3, 5, and 7.
    
\end{itemize}

The following dependent variables were leveraged to measure the performance of gestures across different interaction spaces and among different \rev{numbers} of gesture regions in a particular interaction space.
\begin{itemize}
    \item \textbf{DV1} - \textit{Gesture time}:
    Time window measured from the end of a \textit{gesture delimiter} to the end of the gesture recording, during which the user performs a given gesture $G$, measured in milliseconds (ms). 
    This metric was chosen based on prior related interaction space segmentation research ~\cite{Wong2020BezelInitiatedSwipeSmartwatch, Rey2022BezelToBezel}. 
    Details on calculating DV1 with the help of delimiter gestures are discussed in Section \ref{subsection: Apparatus}.
    
    \item \textbf{DV2} - \textit{Gesture path length}:
    Length of gesture $G$ traversed during the above (DV1) time window, measured in millimeters (mm). 
    This metric was developed specifically for this study, as we wanted to analyze the starting and ending of a gesture across different regions in an interaction space.
    
    \item \textbf{DV3} - \textit{Gesture accuracy}:
    The percentage of the actual number of times ($n$) a gesture $G$ was successfully done on~(on-skin)~/~above~(mid-air) target gesture regions in an interaction space ($S$) with respect to the required number of trials. 
    This metric was chosen based on prior research ~\cite{Wong2020BezelInitiatedSwipeSmartwatch, Rey2022BezelToBezel}. 
    
\end{itemize}

These quantitative criteria allow us to compare the performance of gestures in each interaction space and observe the effect of the number of segments in the interaction space on gesture performance, thus helping us investigate \textbf{RQ1} and \textbf{RQ2}. 
We formulated the following two null hypotheses based on our research questions.

\begin{itemize}
    \item $\mathbf{H^{RQ1}_{0}}$: User performance ({DV1}, {DV2}, {DV3}) does not vary when gesture $G$ ({IV2}) is performed in different interaction spaces ({IV1}).    
    
    \item $\mathbf{H^{RQ2}_{0}}$: User performance ({DV1}, {DV2}, {DV3}) for gesture $G$ ({IV2}) does not vary when the number of segments ({IV3}) increase for a specific interaction space ({IV1}).
    
\end{itemize}

In addition, to investigate \textbf{RQ3}, we aimed to employ qualitative observations based on NASA TLX surveys, collected gesture region definitions, questionnaire results, \rev{and participant discussions}. 

\section{Experiment}
\label{section: Experiment}

\subsection{Participants}
\label{subsection: Participants}
We recruited 18 participants (11 males and 7 females) aged between 22 and 40 ($M = 26.37$, $SD = 5.520$). 
All the participants were right-handed and wore a watch in their left hand. 
Participants reported varying levels of earbud usage in their regular life. 
Overall, 17 out of 18 participants reported using wireless earbuds to make calls, listen to music, and participate in online meetings using phones and workstations (period: 0--25 hours/week, $M = 9.1$, $SD = 7.12$). 
The study was divided into two halves corresponding to two interaction spaces with a break in between.
Each trial lasted 15-25 seconds, and each half of the study took 75 to 90 minutes on average. 
The total study took around 2.5 to 3 hours per participant\rev{, with} pre-study and post-study times.
Each participant was paid at a rate of \$15 per hour for their time and effort. 

\subsection{Apparatus}
\label{subsection: Apparatus}

\begin{figure}[tb!]
    \centering
    \includegraphics[width=\linewidth]{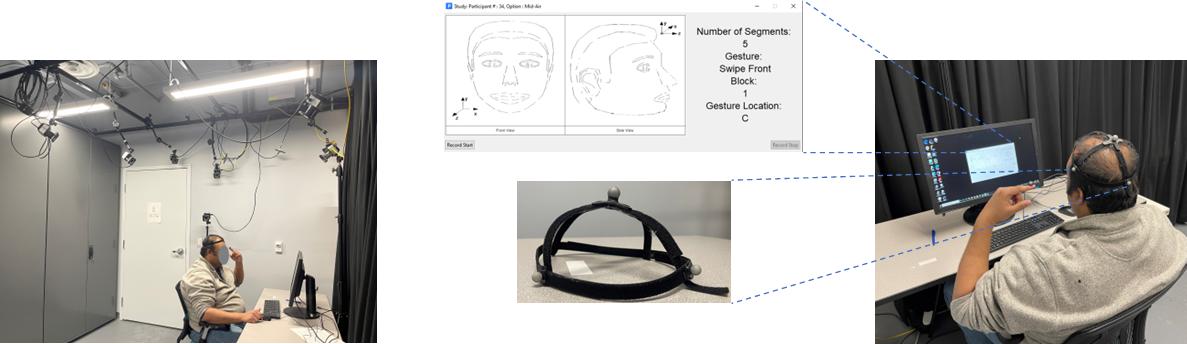}
    \captionsetup{justification=centering}
    \vspace*{-0.75cm}    
    \caption{Experimental setup with Vicon: the participant is performing gestural tasks shown on the screen}
    \label{fig:-01-ViconSetupWithParticipant}
\end{figure}

We designed a desktop application to record, filter, and analyze the gesture data, henceforth called \textit{Gesture Recording Data Collection Application}~({GRDA}).
GRDA was configured on a desktop PC connected to 8 \textit{Vicon} cameras and a 24-inch screen (Figure \ref{fig:-01-ViconSetupWithParticipant}).
This PC served as the \textit{Vicon} camera server and was configured with an \textit{Intel Xeon} processor with 32 GB RAM and 16 GB graphics \rev{card} attached with a standard keyboard and mouse for running \textit{Vicon Tracker 3.9} application.

The gesture recording apparatus was set up to support both right and left-handed participants. 
As all participants in our study were right-handed, they were asked to create uni-manual gestures using their non-dominant (left) hand so that the dominant (right) hand could control GRDA using a keyboard and mouse for data recording.
The \textit{Vicon} cameras recorded the gestural inputs using markers attached to the participant's head and left hand~\footnote{\label{footnote:non-dominant-hand} ~Non-dominant hand used for creating unimanual gestures. For left-handed participants, markers would be worn on the right hand.}.
For gesture tracking and reference purposes, 3 markers were attached to different parts of the head using velcro straps (Figure \ref{fig:-01-ViconSetupWithParticipant}).
The markers were positioned in the following places: 
1) center top on the crown of the head;
2) back of the head; and
3) side of the head behind the temple and above the helix of the left ear~\footnote{\label{footnote:non-dominant-ear} ~Closest ear to the hand used for creating uni-manual gesture. For left-handed participants, markers would be worn on the right hand and the head marker above the ear helix would be placed above the right ear.}. 
Finger markers were attached to the left index (and thumb) finger~{\footref{footnote:non-dominant-hand}} using double-sided tape.
One marker was attached to the index finger for all tasks, as shown in Figure \ref{fig:-01-ViconSetupWithParticipant}.
For specific gestural inputs (pinch in/pinch out) involving two fingers, another marker was also placed on the left~\footref{footnote:non-dominant-hand} thumb finger.
Before running GRDA for our experiment, \rev{the} \textit{Vicon Tracker 3.9} application was turned on to track the gesture motion data using the Vicon markers and continuously send the data stream back to GRDA for recording purposes.
The participant controlled gesture data recording using this GRDA application (Figure \ref{fig:-01-ViconSetupWithParticipant}), which filtered and categorized the marker data stream collected from \rev{the} \textit{Vicon Tracker 3.9} application running on the backend.
All uni-manual gestures started from a fixed position (marked by a white tape in the table containing the Vicon server PC in the rightmost image of Figure \ref{fig:-01-ViconSetupWithParticipant}).
Before any actual gesture, participants performed one of the following two \textit{Delimiter Gestures} above the table's fixed (taped) position.
\begin{itemize}
    \item \textit{Delimiter Gesture A}: 
        Wiggle the index finger (3 times) with the index and thumb fingers of the marker-wearing left\footref{footnote:non-dominant-hand} hand. 
        Used only before two-finger uni-manual gestures (pinch in/out) using index and thumb fingers.
    
    \item \textit{Delimiter Gesture B}:
        Wiggle with the \rev{marker-wearing hand's index finger (3 times)}.
        Used before all single finger-based uni-manual gestures (tap, horizontal/vertical swipe) and \rev{during \textit{head model generation}} phase \rev{(Section \ref{subsubsection-Head Model Generation})}.
\end{itemize}
The purpose of these delimiter gestures was to help: 
1) accurately distinguish between the thumb and index finger of the marker-wearing hand for two-finger gestures and
2) mark the starting of the gesture for both the participant and the gesture apparatus.
After the delimiter gesture was made and the motion for the targeted uni-manual gesture began, gesture metric measurement automatically started once the marker-wearing fingers reached within 30 cm from the center~\footnote{For the purpose of our calculation, the center of the head is located on the vertical plane bisecting the body, located between the eyes and the top of the tragus in the outer ear on the horizontal field-of-view~(FOV) plane.} of the head (Figure \ref{fig:-head-center} in Appendix \ref{appendix:grda_application}).
The metric measurement stopped once the gesture was completed, and the hand moved out of the above range to filter out gesture motion noise and evaluate only the relevant gesture motion. 
Gesture recording automatically stopped when the participant completed the task and rested his hand on the starting position.

\subsection{Design}
\label{subsection: Experimental Design}

We employed a within-subjects design for our experiment. 
According to the independent variables ({IV1-3}), each participant was provided with 2 interaction spaces (on-skin and mid-air) $\times$ 3 segment numbers (3, 5, and 7) $\times$ 7 gestures (\textit{Tap}, \textit{Swipe Up}, \textit{Swipe Down}, \textit{Swipe Front}, \textit{Swipe Back}, \textit{Pinch In}, and \textit{Pinch Out}) = 42 experimental conditions.
For each condition, the participant was requested to define the region boundaries on a paper containing a side view and a front view of a face, and the regions were then labeled with identifiers\rev. 
Using a \textit{Latin square} design, we counter-balanced the interaction space (IV1) and the number of segments (IV3) variables across the whole experiment.
The participant was first given a particular interaction space, within which the participant was provided with a specific number of segments. 
For each combination of these two variables, the participant was asked to perform each of the seven gestures ({IV2}) in a randomized order, and there were 2 repetitions for each gesture task. 
The participant could perform different numbers of trials based on the number of segments within the IV1 and IV3 combination. 
For example, with mid-air and 5 segments, the participant would be asked to perform each of the seven gestures for 5 segments $\times$ 2 repetitions = 10 times. 
The 5 segments labeled using region identifiers (A--E) were shown randomly across the tasks.
Therefore, each participant performed: 2 interaction spaces $\times$ (3~+~5~+~7) segments $\times$ 7 gestures $\times$ 2 repetitions = 420 trials.
We collected 7,560 gestural data points for 18 participants. 

\subsection{Procedure}

The experiment was conducted in an office environment with adequate soundproofing to provide a calm environment.
During the experiment, the participants sat in front of a table containing the \textit{Vicon} server PC and designated starting position (marked by white tape). 
They operated GRDA via keyboard and mouse to record the gesture, as in Figure \ref{fig:-01-ViconSetupWithParticipant}. 
Before the study started, the purpose of the research was explained to the participants.
\rev{They were encouraged to ask the researcher questions for clarification} regarding any steps of the study.
Participants were informed that they could take breaks or abort the experiment \rev{at any time}.
After explaining the study purpose and break structures, participant consent was obtained using a consent form.
Then, participants began the two-phased \rev{study} (described in \rev{Sections} \ref{subsubsection-Head Model Generation} and \ref{subsubsection-Gesture Data Collection}). 
\rev{During both phases, all gesture motions started from the table's designated position (marked with white tape)}. 
At the end of the study, demographic data was collected from the participants using a \textit{Qualtrics} form.

\subsubsection{Phase 1: Head Model Generation}
\label{subsubsection-Head Model Generation}

\begin{figure}[tb!]
    \centering
    \begin{subfigure}{0.32\linewidth}
    \centering
        \includegraphics[width=\linewidth]{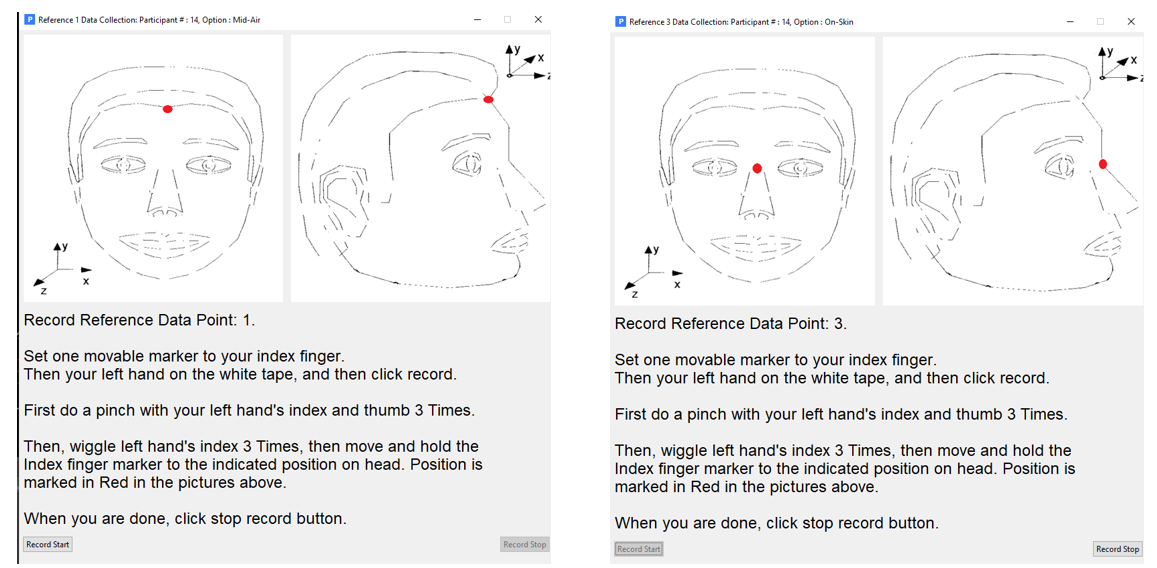}
        \captionsetup{justification=centering}
        \vspace*{-5mm}
        \caption{}
        \label{fig:-GRDA_App_FacePoints}
    \end{subfigure}
    \hfill
    \begin{subfigure}{0.32\linewidth}
    \centering
        \includegraphics[width=\linewidth]{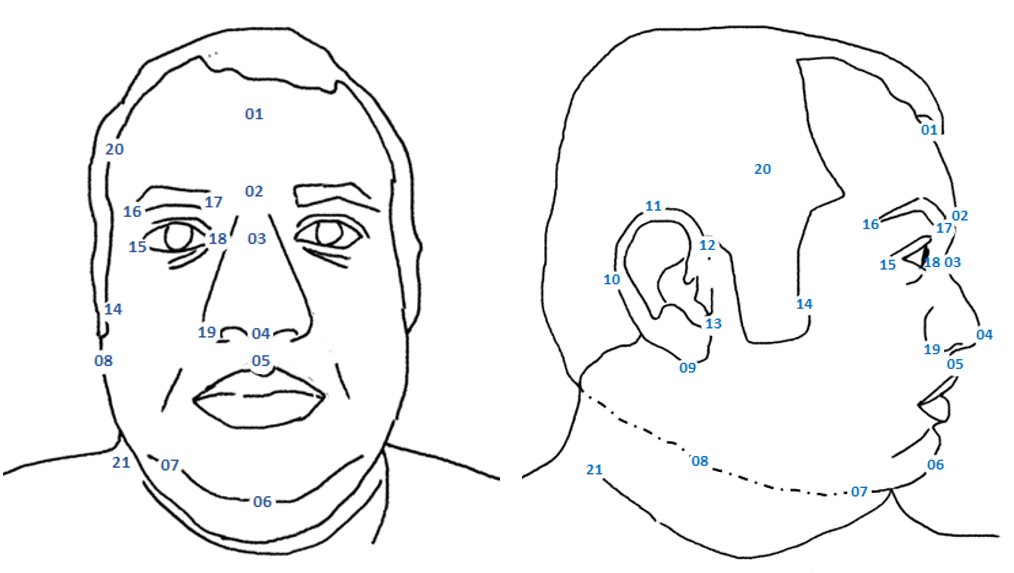}
        \captionsetup{justification=centering}
        \vspace*{-0.5cm}
        \caption{}
        \label{fig:-Actual21_Facepoints}
    \end{subfigure}
    \hfill
    \begin{subfigure}{0.32\linewidth}
    \centering
        \includegraphics[width=\linewidth]{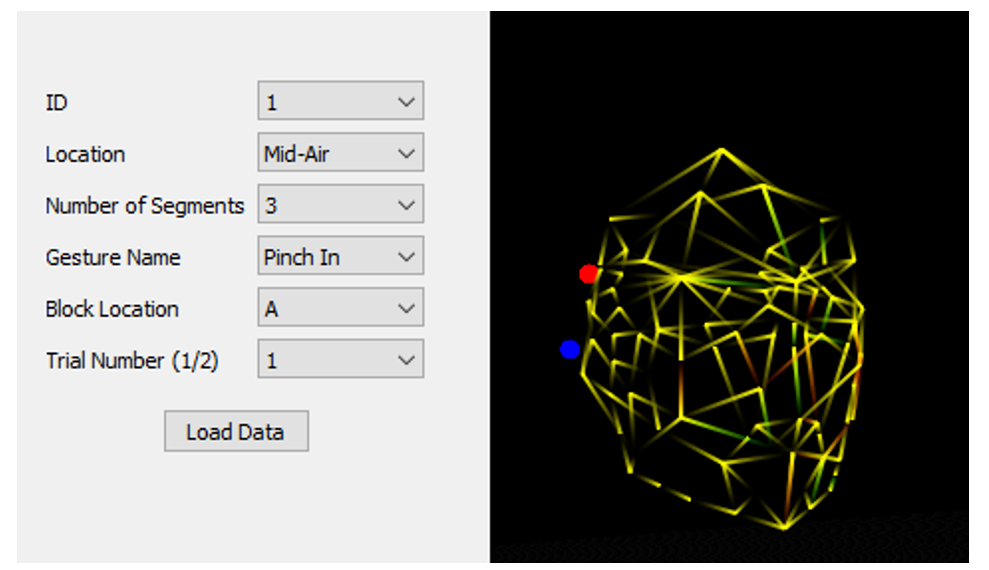}
        \captionsetup{justification=centering}
        \vspace*{-0.5cm}
        \caption{}
        \label{fig:-GRDA_Headmodels}
    \end{subfigure}
    \captionsetup{justification=centering}
    \vspace*{-0.3cm}    
    \caption{(a)~Collection of original reference points on the face using GRDA for head reconstruction. (b)~\rev{The location} of 21 original reference points on the face. (c) Snapshot of GRDA with a reconstructed head model.}
    \label{fig:-12-ReconstructedHeadModel}
\end{figure}

As each participant has a unique facial structure, we first generated a 3D head model for each participant using the \textit{Vicon} motion tracking camera and markers to collect meaningful gesture data in Phase 2.
Figure \ref{fig:-GRDA_Headmodels} shows a sample of the reconstructed head model with GRDA. 
The head model was created by collecting 21 reference points on different head \rev{parts} (\eg,~\textit{chin}, \textit{earlobe}, \textit{cheek}, \textit{top of nose}, \textit{temple}) and performing data augmentation.

For collecting each reference point data, GRDA showed the position of the reference point for the face, head, and neck.
Once a referent point was shown (Figure {\ref{fig:-12-ReconstructedHeadModel}A}), the gesture recording task was explained in the application, with the researcher providing additional clarifications.
Once participants were ready for reference data collection, they were asked to reposition their left~\footref{footnote:non-dominant-hand} hand at the designated starting point on the table. 
Once the hand is positioned, participants were instructed to start the recording via GRDA using the available dominant (right) hand, then perform the delimiter gesture using the marker wearing left {\footref{footnote:non-dominant-hand}} hand, followed by holding the left index finger ~\footnote{\rev{For a right-handed participant, the left index finger contains a \textit{Vicon} marker, which switches to right index finger for a left-handed participant.}} on the designated reference point on the head for 3 seconds before stopping the recording.
After the recording was completed, a pop-up recording confirmation window allowed the participants to proceed to the following reference point recording or redo the current recording if necessary.
After all reference points were collected, the generated head model was shown on GRDA, allowing participants to adjust the head model using a custom toolkit integrated into GRDA.
Once the head model \rev{was} finalized, participants proceeded to Phase 2 to perform the actual gesture data collection task.

\subsubsection{Phase 2: Gesture Data Collection}
\label{subsubsection-Gesture Data Collection}

After confirmation of a head model, participants were shown 42 gesture
conditions in an approach described in the study design (Section \ref{subsection: Experimental Design}). 
For each condition, they were asked to define preferred region boundaries for the targeted interaction space and the number of around-ear gesture regions.
They were also informed that region boundaries for different gesture conditions could vary. 
The participants were allowed to explore targeted interaction space in a practice session for each condition to build a mental model of gesture space segmentation and finalize the gesture region boundaries.
For on-skin gestures, participants were reminded that touching the outer ear was prohibited, and touching any other region above the neck to the head crown was allowed. 
During mid-air gesture sessions, they were informed that physical touch interaction with the body part was prohibited. 
However, participants were clarified that mid-air regions could be directly above the outer ear and located near or away from the body, based on their preferences.

After completing a practice session, a side and front view image of the head and neck region on paper was presented to the participants to define the gesture region boundaries. 
Each region was marked alphabetically, starting with the letter \textbf{A}.
The researcher asked specific questions about the positioning of each region to different body parts (i.e., eyebrow, top of nose, cheekbone) and took notes while the participant defined the region boundaries. 
Once boundaries were defined, a particular gesture region identifier (\textbf{A}-\textbf{G}) was presented to the participants on the desktop application (Figure \ref{fig:-01-ViconSetupWithParticipant}) for recording uni-manual gestures for that region. 
After starting the gesture recording, participants began uni-manual motion from the starting position with the appropriate delimiter gesture, followed by the target gesture over the intended boundary before turning off the recording. 
Controlling the recording via GRDA was done using the right (dominant) hand, and simultaneously, the other hand was used for uni-manual gestures. 
After performing all the gesture trials for a specific gesture condition (interaction space and associated number of regions), a \textit{Qualtrics} survey was presented in GRDA to rate the condition using a NASA TLX survey and identify preferred gesture region locations for that specific condition.
After completing all conditions, an \rev{end-of-study} \textit{Qualtrics} survey was used to collect additional information regarding gesture preferences for both interaction spaces from participants.
\section{Quantitative Results}
\label{section - quantitative results}

Using normality tests, we tested the data distribution of gesture metrics DV1, DV2, and DV3. When applicable, we followed up with testing homogeneity of variance for the distributions using Bartlett’s test for choosing appropriate parametric or non-parametric tests. 
Appendix \ref{appendix:stat_analysis} provides the supporting statistical analysis accompanying quantitive results outlined in Figures \ref{fig:MidAirVsOnSkin_PerformanceMetrics} and \ref{fig:PerformanceMetrics-EveryGesture}.
Appendix \ref{appendix:mean_and_media} outlines the mean and median gesture metric values across interaction spaces and the number of regions. 
Appendix \ref{appendix:DV1}, \ref{appendix:DV2}, and \ref{appendix:DV3} outline normality and significance test results to test the effect of interaction spaces (RQ1) and increasing around-ear regions (RQ2). 

As the layouts and size for mid-air and on-skin regions associated with different gestures could vary between participants, \textit{gesture time}~(DV1) and \textit{path length}~(DV2) for different gestures could not be paired across interaction spaces for addressing RQ1 or matched across different numbers of around-ear gesture regions within each interaction space for addressing RQ2. 
However, as \textit{gesture accuracy} (DV3) was not directly related to gesture region layout/size at different around-ear interaction spaces, this metric could be paired/matched for both research questions. These paired (on-skin and mid-air factors) or matched (3/5/7 regions) criteria guided our choice between paired/matched parametric and non-parametric tests for RQ1 and RQ2.

\subsection{Effect of Interaction Space Choice on Performance Metrics (\textbf{RQ1})}
\label{subsection - Effect of Interaction Space Choice on Gesture Shapes}

\begin{figure}[tb!]
    \centering
    \includegraphics[width=1.0\textwidth,keepaspectratio]{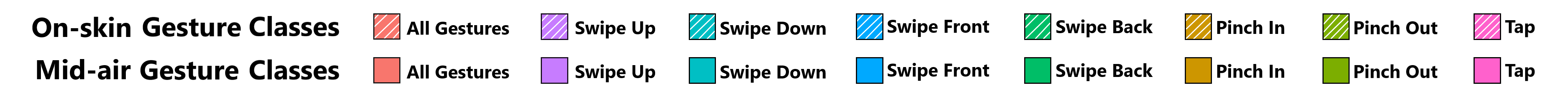}
    \includegraphics[width=1.0\textwidth,keepaspectratio]{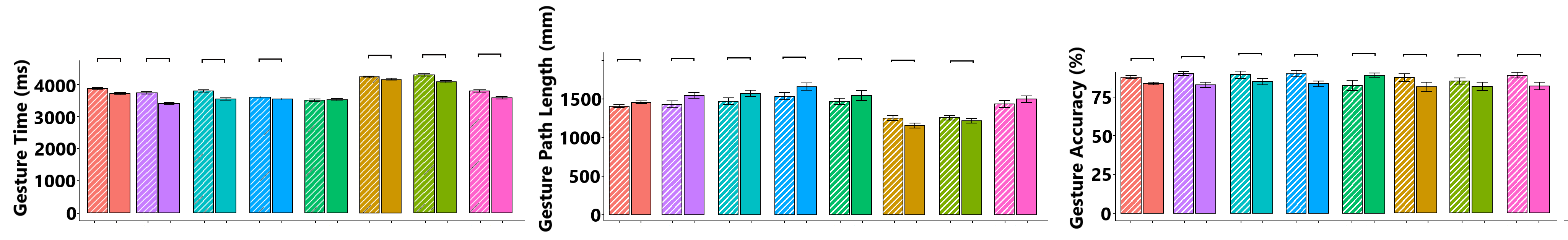}
    \vspace*{-0.8cm}
    \captionsetup{justification=centering}  
    \caption{Performance metrics for mid-air and on-skin gestures with 95\% confidence intervals. Number of regions around the ears have been ignored.} 
    \label{fig:MidAirVsOnSkin_PerformanceMetrics}   
\end{figure}

\begin{figure}[tb!]
    \centering
    \includegraphics[width=1.0\textwidth,keepaspectratio]{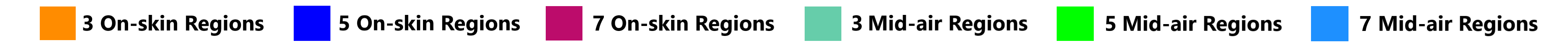}   
    \includegraphics[width=1.0\textwidth,keepaspectratio]{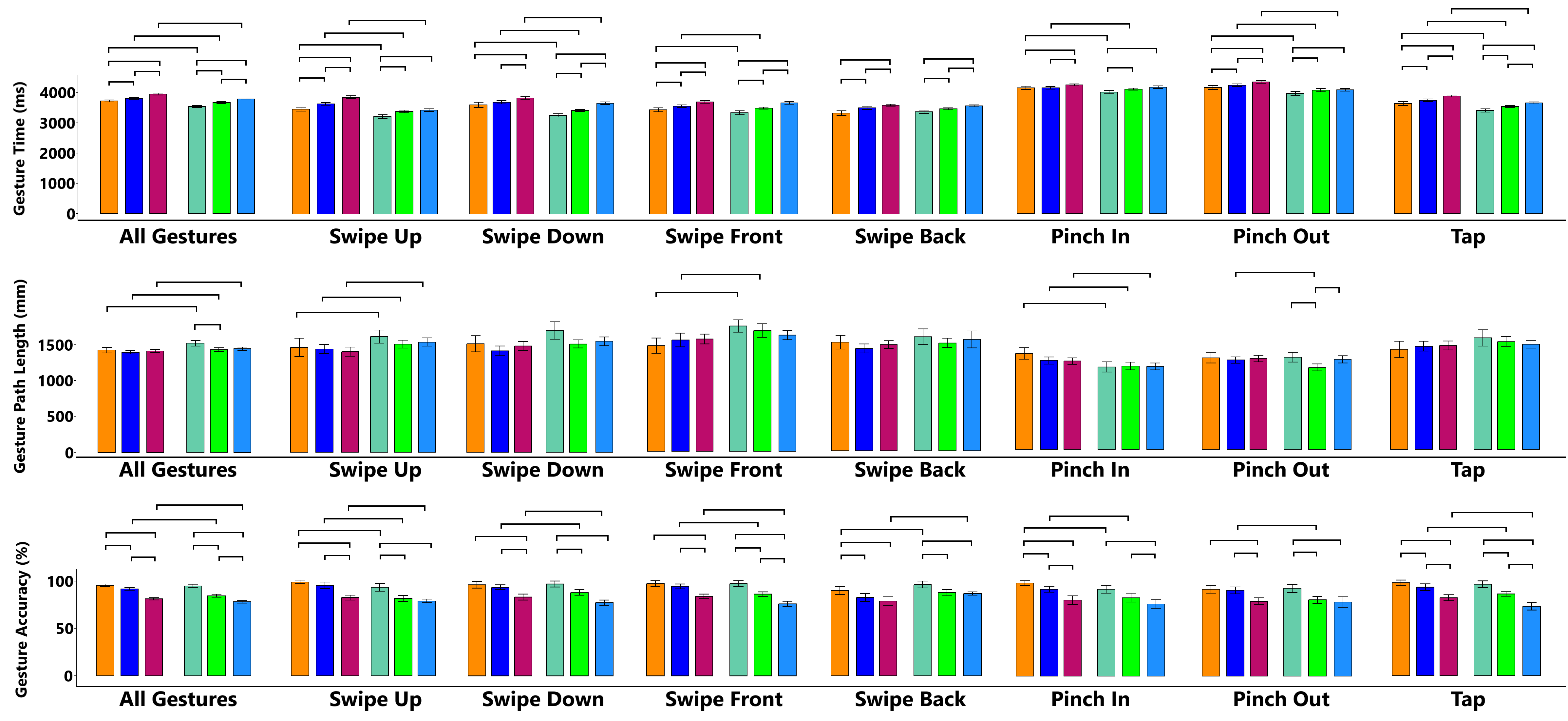}
    \vspace*{-0.8cm}
    \captionsetup{justification=centering}
    \caption{Performance metrics for mid-air and on-skin gestures across all numbers of gesture regions around the ear with 95\% confidence intervals.}  
    \label{fig:PerformanceMetrics-EveryGesture}   
\end{figure}

When analyzing \textit{gesture time}~(DV1) for individual gesture classes, we observed that overall (Figure \ref{fig:MidAirVsOnSkin_PerformanceMetrics})\rev{,} all gestures except \textit{swipe back} took significantly more time to perform on-skin than in mid-air in general (Table \ref{tab:DV1_vs_RQ1} in Appendix \ref{appendix:DV1}). 
Diving deeper (Figure \ref{fig:PerformanceMetrics-EveryGesture}), we observed that on-skin \textit{swipe back} gestures took more time than mid-air counterparts across all (3/5/7) numbers of regions around the ear. 
However, the difference was statistically insignificant (Table \ref{tab:DV1_vs_RQ1}).
For all other gestures, we observed \rev{that} on-skin gesture time \rev{was} significantly higher than mid-air gestures for 3 and 5 around-ear regions. 
For 7 around-ear regions, only vertical (up/down) \textit{swipe}, \textit{pinch} (in/out), and \textit{tap} gestures took significantly longer to perform on \rev{the} skin. 
Although on-skin gesture time for horizontal (front/back) \textit{swipe} gestures in 7 regions was higher, it was not statistically significant.

Our \textit{path length}~(DV2) analysis also showed a significant effect of interaction space choice on gestures (Figure \ref{fig:MidAirVsOnSkin_PerformanceMetrics} - all gestures).
In general, mid-air swipe and tap gesture paths were longer, and the difference was statistically significant for \textit{swipe} gestures overall (Table \ref{tab:DV2_vs_RQ1} in Appendix \ref{appendix:DV2}). 
However, comparing the number of regions, we only found mid-air \textit{swipe up} and \textit{swipe front} gestures to be significantly longer for both 3 and 5 around-ear regions (Table \ref{tab:DV2_vs_RQ1}).
Mid-air \textit{swipe up} gesture was also significantly longer for 7 around-ear regions. 
However, the gesture path length increase was not statistically significant for 3, 5, and 7 mid-air regions for \textit{swipe back} and \textit{swipe down}. 
A noticeable deviation for DV2 was observed for pinch gestures. 
Mid-air pinch gesture paths were significantly shorter than on-skin (Figure \ref{fig:MidAirVsOnSkin_PerformanceMetrics}).
Analyzing the number of regions, we observed significantly shorter mid-air path \rev{lengths} for all numbers (3/5/7) of around-ear regions for \textit{pinch in} \rev{gestures}.
Although the \textit{pinch out} gesture paths were shorter for different numbers of mid-air around-ear regions, the differences were significant only for 7 region segmentations.

\rev{On-skin gesture accuracy (DV3) was significantly higher (Figure \ref{fig:MidAirVsOnSkin_PerformanceMetrics}) than mid-air for For \textit{swipe up}, \textit{swipe down}, \textit{swipe front}, \textit{pinch in}, \textit{pinch out}, and \textit{tap} gestures.}
Analyzing further individual gesture shapes and the number of around-ear regions, we observed that gesture accuracy for mid-air swipe back was better for all numbers of around-ear regions. 
The differences were statistically significant for both 3 and 7 region segmentation. 
\rev{For all other gestures and the number of regions, mid-air gestures were less accurate than on-skin.}
Overall (Figure \ref{fig:PerformanceMetrics-EveryGesture} - all gestures), gesture accuracy \rev{remained} relatively similar for 3 around-ear regions and starts diverging as the number of regions increases in mid-air and on-skin spaces. 
Figure \ref{fig:PerformanceMetrics-EveryGesture} highlights the statistically significant differences in DV3 metric, and Table \ref{tab:DV3_vs_RQ1} in Appendix \ref{appendix:DV3} outlines associated statistical test results.

To summarize, The effect of interaction space choice is less prominent for fewer ($< 5$) around-ear regions - the performance starts to significantly degrade for mid-air gestures compared to on-skin gestures as the number of gesture regions increases ($\geq 5$). 
For a lower number of regions ($=3$), mid-air and on-skin gestures have overall similar performances, with mid-air gestures being faster than (and as accurate as) on-skin gestures while trading off subtility in terms of larger gesture motion (higher DV2).

\subsection{Effect of Increasing Gesture Regions on Performance Metrics (\textbf{RQ2})}
\label{subsection-SegmentSize-Gestures}

We found that increasing both on-skin and mid-air gesture regions for gesture reuse had a significant effect on both \textit{Gesture Time} (DV1) (Table \ref{tab:DV1_vs_RQ2}) and \textit{Gesture Accuracy}~(DV3) ~(Table \ref{tab:DV3_vs_RQ2}) across all gesture classes (Figure \ref{fig:PerformanceMetrics-EveryGesture} - ``\textit{All Gestures}'').
However, we did not find significant changes in \textit{Gesture Path Length}~(DV2) for overall gestures in \rev{the} on-skin interaction space. 
We did observe a significant drop in path length (Table \ref{tab:DV2_vs_RQ2}) when mid-air, around-ear regions increased from 3 to 5.
\rev{However,} the changes from 5 to 7 regions, or 3 to 7 regions, were insignificant.

Analyzing RQ2 for gesture time (DV1) by individual on-skin gesture class, we observed that gesture time increase was insignificant for on-skin \textit{swipe down} and \textit{pinch in} gestures when gesture regions increased from 3 to 5.
However, the increase in gesture time was significant for all other gesture region increases (3 to 5, 5 to 7, or 3 to 7) for all other gesture classes. 
In comparison, gesture time significantly increases for all mid-air gesture classes \rev{when mid-air} gesture regions increase from 3 to 5 and 3 to 7. 
Except for mid-air \textit{swipe up}, \textit{pinch in}, and \textit{pinch out} gestures, increasing the number of gesture regions from 5 to 7 was also significant for all other gestures. 
This indicated a potential upper bound in terms of \textit{gesture time} (DV2) for mid-air (3 regions) and on-skin (5-region) spaces for gesture reuse.
For gesture path length (DV2), we found a significant effect only for mid-air \textit{pinch out} gestures when considering the effect of increasing around-ear regions on individual gesture classes (Table \ref{tab:DV2_vs_RQ2}). 
There was no significant effect on DV2 for all other mid-air and on-skin gestures. 
For \textit{pinch out}, DV2 significantly increased from 5 to 7 segmentation.

Analyzing gesture accuracy (DV3) for individual mid-air and on-skin gesture classes, we observed a significant decrease in accuracy for all mid-air and on-skin gesture classes when the number of around-ear regions was increased from 3 to 7.
Although gesture accuracy decreased for all gesture classes on increasing on-skin regions from 3 to 5, only \textit{swipe back} and \textit{pinch in} gestures showed a significant decrease.
Gesture accuracy further decreased when the number of on-skin regions was increased from 5 to 7, and except for \textit{swipe back}, all decreases were significant.
When mid-air gesture regions increased from 3 to 5, gesture accuracy dropped for all gesture classes, and except for \textit{pinch in}, all accuracy drops were significant. 
However, upon increasing mid-air regions further to 7, the drop in gesture accuracy was insignificant for most gesture classes except \textit{swipe down} and \textit{tap}.
Table \ref{tab:DV3_vs_RQ2} in Appendix \ref{appendix:DV3} shows the statistical test summary for testing individual gesture classes for DV3 metric. 

The above metrics indicate that gesture performance does not vary significantly when the number of around-ear, on-skin regions \rev{increase} from 3 to 5. 
However, further on-skin region segmentation for gesture reuse can significantly degrade on-skin gesture performance. 
In contrast, overall mid-air gesture performance degrades significantly when mid-air regions increase from 3 to 5. However, further segmentation of mid-air gesture space does not significantly affect overall gesture performance. 
Overall, this indicates a threshold of 3 mid-air regions and 5 on-skin regions for supporting gesture reuse.
\section{Qualitative Observations}
\label{section, qualitative Observations}

\subsection{Best and Worst Regions}
\label{subsection, best and worst regions}

\begin{figure}[tb!]
    \centering
   {\includegraphics[width=0.98\textwidth,keepaspectratio]{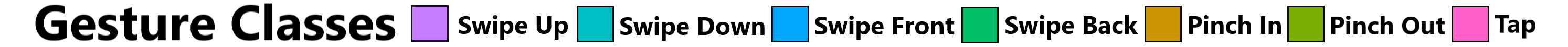}}
   \frame{\includegraphics[width=0.897\textwidth,keepaspectratio]{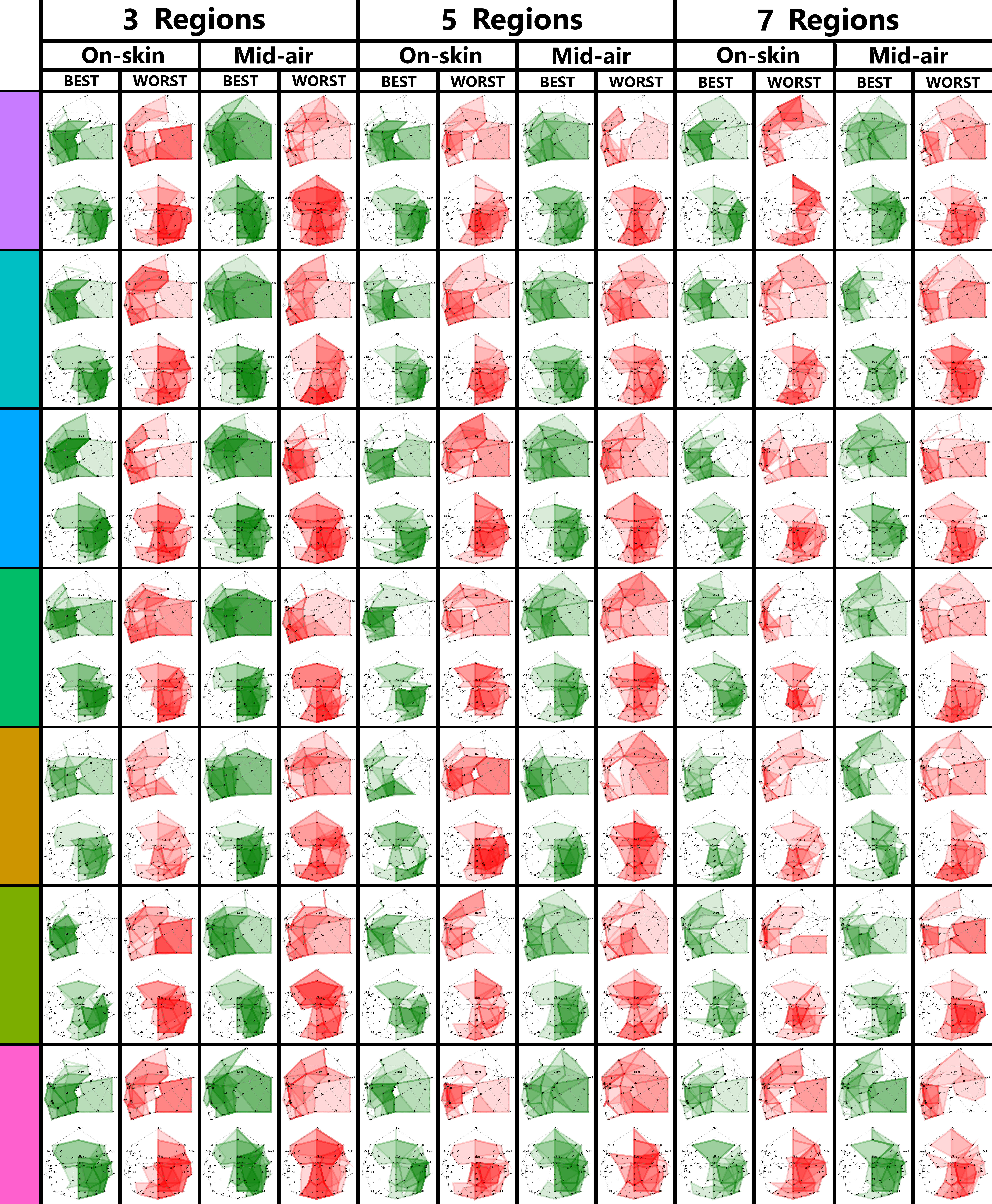}}
   \captionsetup{justification=centering}
   \vspace*{-0.2cm}
   \caption{Most \& least preferred off-device, around-ear on-skin and mid-air gesture regions ~\footref{side-view-footnote} selected by all 18 participants.\\
   Green = Most Preferred and Red = Least Preferred. Darker areas indicate higher overlaps. }
   \label{fig: Most and Least Preferred Regions for On-Skin and MidAir space}
\end{figure}

To answer \textbf{RQ3} outlined in Section~\ref{subsection: Research Questions}, we analyzed the gesture region boundaries collected in Phase 2 of our study (Section \ref{subsubsection-Gesture Data Collection}).
Figure \ref{fig: Most and Least Preferred Regions for On-Skin and MidAir space} outlines the most (\textit{green}) and least (\textit{red}) preferred regions across 3, 5, and 7 gesture regions in on-skin and mid-air interaction spaces, categorized by individual gesture classes (swipe/pinch/tap). 
The darker shades indicate where participants' preferences for the most and least preferred regions overlap {\footnote{\label{side-view-footnote}~Figures represent a participant creating \rev{a} uni-manual gesture with \rev{the} left (non-dominant) hand. Side views represent left-hand side views for participants. Participants use the left hand to create uni-manual gestures on the side closest to hand used to create gestures. Section \ref{subsection: Apparatus} provides details on non-dominant hand usage.}}. 

\begin{figure}[tb!]
    \centering 
    \captionsetup{justification=centering}
    \begin{subfigure}[b]{0.25\textwidth}
        \centering
        \includegraphics[width=0.9\textwidth]{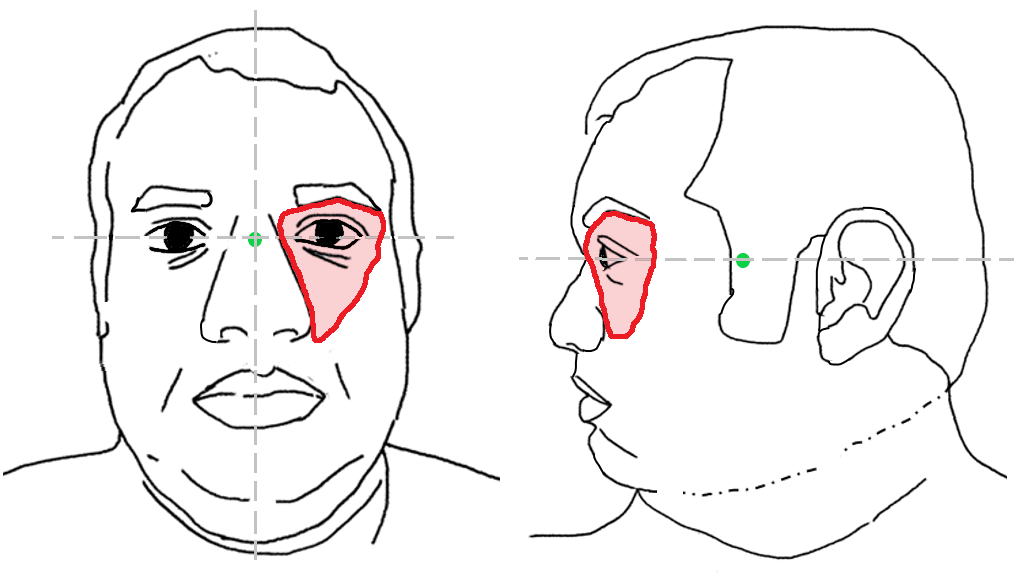}
        \vspace*{-0.25cm}
        \caption{}
        \label{fig:-avoidance-around-eyes}
    \end{subfigure}%
    \begin{subfigure}[b]{0.25\textwidth}
        \centering
        \includegraphics[width=0.9\textwidth]{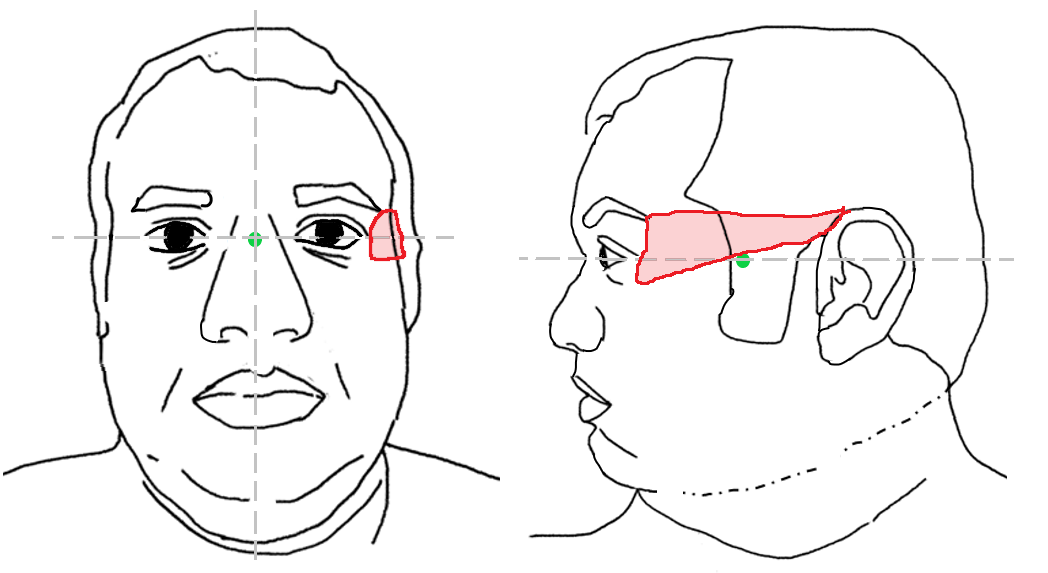}   
        \vspace*{-0.25cm}
        \caption{}
        \label{fig:-avoidance-around-temple}
    \end{subfigure}%
    \begin{subfigure}[b]{0.25\textwidth}
        \centering
        \includegraphics[width=0.9\textwidth]{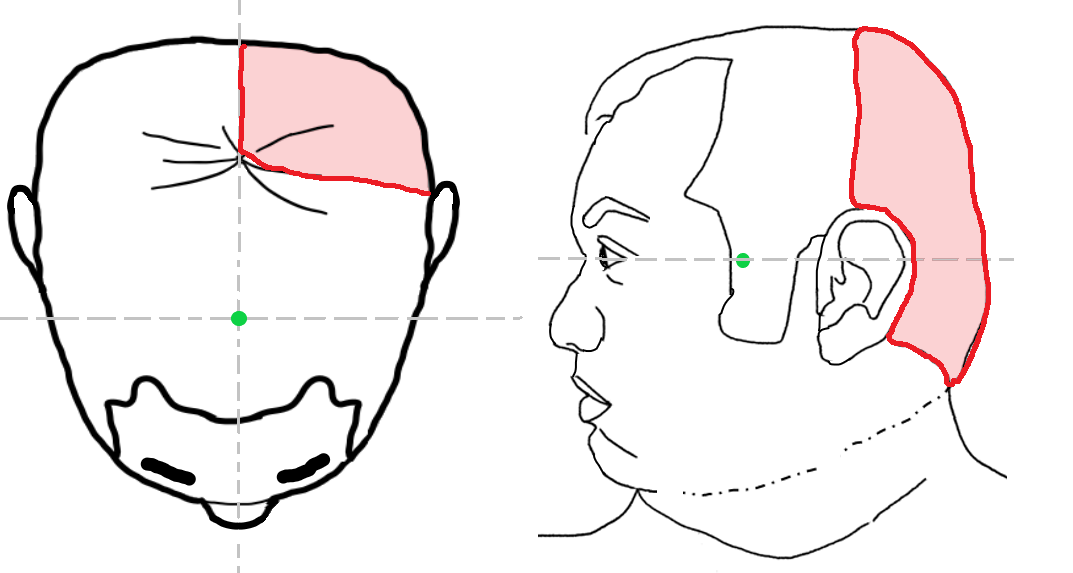} 
        \vspace*{-0.25cm}          
        \caption{}
        \label{fig:-avoidance-back-of-hair}
    \end{subfigure}%
    \vspace*{-0.3cm}
    \caption{Highlighted areas~\footref{side-view-footnote} around different parts of the face for avoiding on-skin gestures.}
    \label{fig:-highlighted face areas}
\end{figure}

Participants preferred to limit the gesture region within the frontal or peripheral view as much as possible for both on-skin and mid-air interaction spaces. 
Overlap of best and worst regions, alongside participant discussions, revealed some noticeable differences between mid-air and on-skin gesture region positioning. 
For on-skin gestures, user-defined gesture region boundaries usually follow along the contour lines of the face.
Participants showed considerable dislike for interacting with regions around the eyes. 
The area bounded by the \textit{eyebrow}, \textit{inner canthi} (corner of the eye on the side of the nose), \textit{nasal ala}, and \textit{outer canthi} (other corner of the eye near the temple) shown in Figure \ref{fig:-avoidance-around-eyes} are either not considered within different gesture regions or are usually considered among the worst-ranked regions for different gestures by almost all of the participants.
Participants tended to avoid this area to protect the eye from accidental finger poking, keeping the frontal vision unblocked, and reducing Potential manual interaction conflict with eyewear. 
Another region of interest for on-skin gestures spans around the \textit{temple} from the \textit{outer canthi} to the top of the outer ear (Figure \ref{fig:-avoidance-around-temple}). 
Analysis of on-skin region boundaries shows that some participants avoided this area or rated it less desirable than other on-skin regions.
These participants were regular or occasional eyewear users; and were concerned about potential frame displacement from manual interaction with the temple area.
However, some non-eyewear user participants identified this region as suitable for on-skin \textit{tap} and horizontal \textit{swipe} gestures.
The area above and behind the \textit{outer ear} (Figure \ref{fig:-avoidance-back-of-hair}) is generally avoided for all on-skin \rev{interactions} across 3/5/7 region segmentation.
Participants report increased hand and arm joint bending compared to gestures within the frontal view, peripheral view, and under the \textit{outer ear}. 
However, participants preferred these areas in Figure \ref{fig:-avoidance-back-of-hair} for on-skin interaction over the area around the \textit{eye} and \textit{temple} (Figures \ref{fig:-avoidance-around-eyes} and \ref{fig:-avoidance-around-temple}) as the inconvenience of resistance presented by hair region is less than an obstacle to vision or eyewear.

\begin{figure}[tb!]
    \centering 
    \captionsetup{justification=centering}
    \begin{subfigure}[b]{0.25\textwidth}
        \centering
        \includegraphics[width=0.9\textwidth]{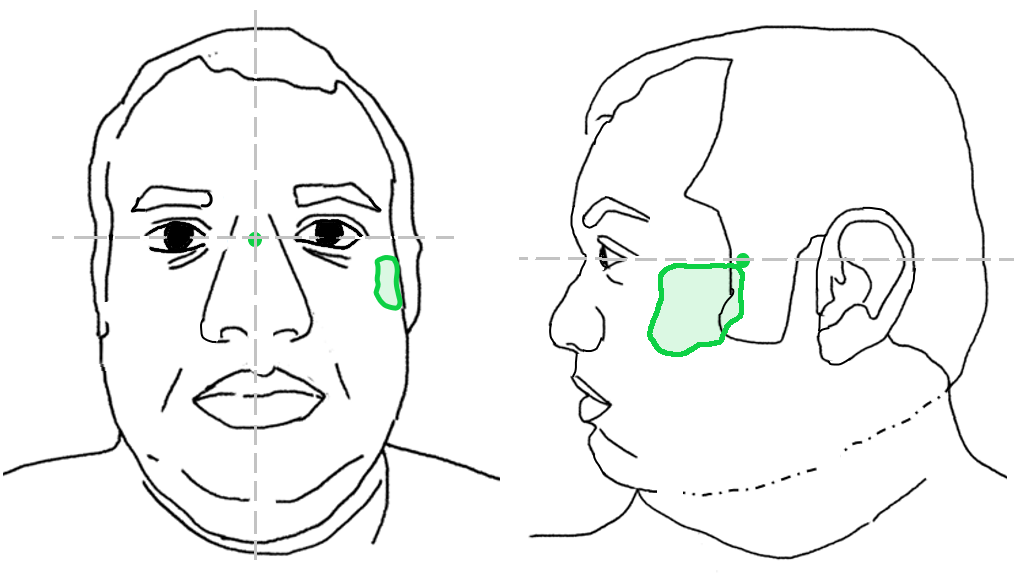}   
        \vspace*{-0.25cm}           
        \caption{}
        \label{fig:-preference-around-cheekbone}
    \end{subfigure}%
    \begin{subfigure}[b]{0.25\textwidth}
        \centering
        \includegraphics[width=0.9\textwidth]{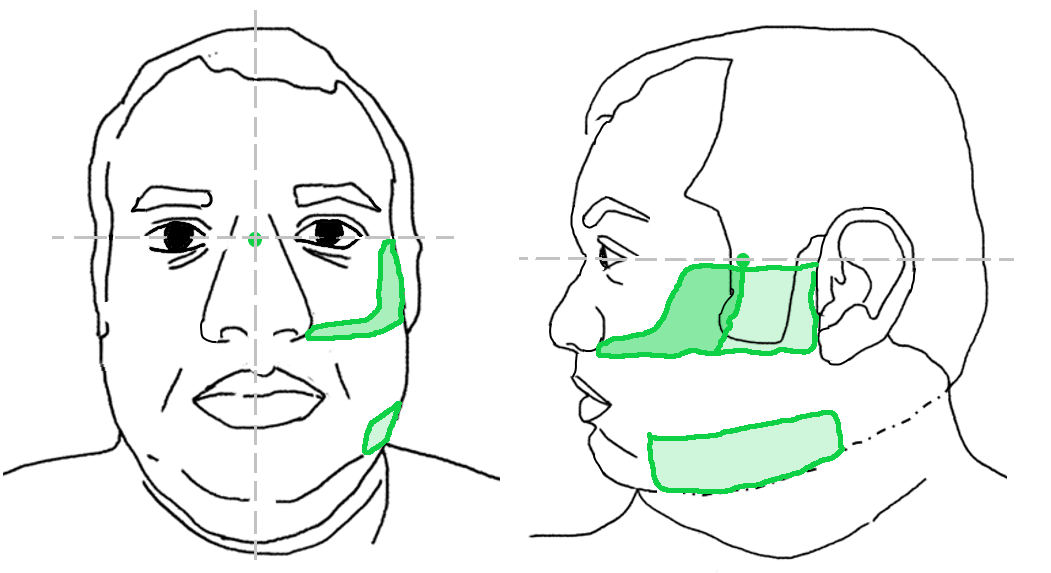}   
        \vspace*{-0.25cm}           
        \caption{}
        \label{fig:-preference-around-cheekbone-horizontal}
    \end{subfigure}%
    \begin{subfigure}[b]{0.25\textwidth}
        \centering
        \includegraphics[width=0.9\textwidth]{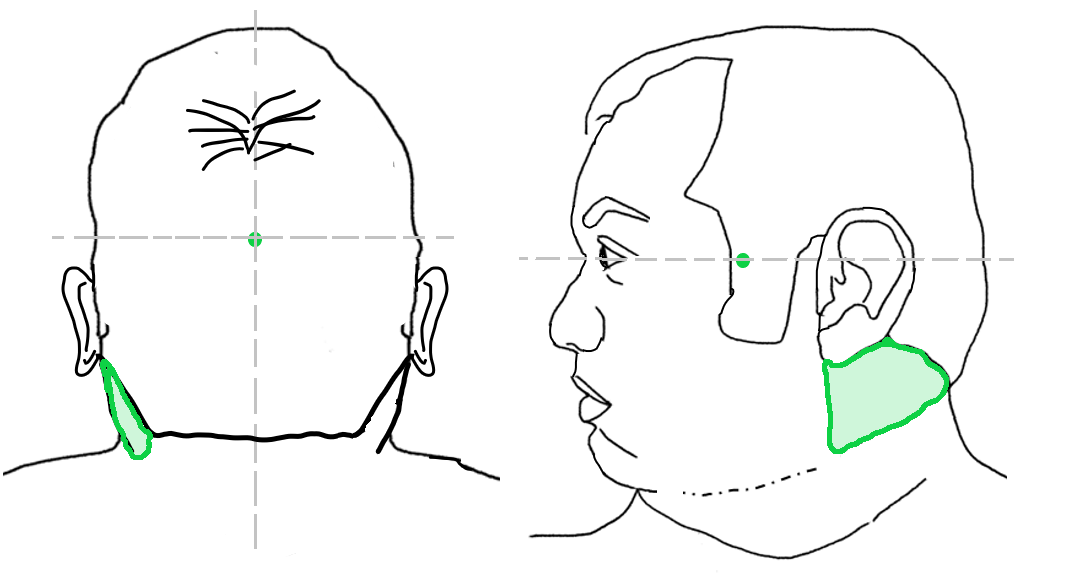}  
        \vspace*{-0.25cm}           
        \caption{}
        \label{fig:-preference-under-ears}
    \end{subfigure}%
    \begin{subfigure}[b]{0.25\textwidth}
        \centering
        \includegraphics[width=0.9\textwidth]{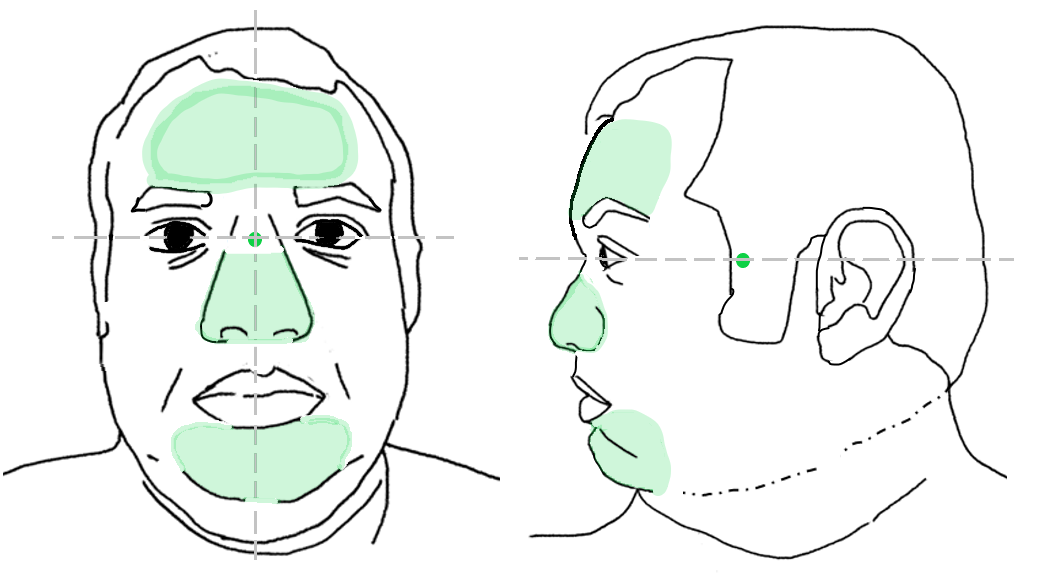}
        \vspace*{-0.25cm}        
        \caption{}
        \label{fig:-preference-front-of-face}
    \end{subfigure}%
    \vspace*{-0.3cm}
    \caption[]{Highlighted preferred areas~\footref{side-view-footnote} around different parts of the face for on-skin gestures.}
    \label{fig:-highlighted face areas2}
\end{figure}

The highlighted area around the \textit{cheekbone} (Figure \ref{fig:-preference-around-cheekbone}) was overwhelmingly preferred for tap and vertical swipe starting and ending points. 
Participants preferred the rigidity of the \textit{cheekbone} for tactile feedback, and the gestures mentioned above in this highlighted region were highly accurate, regardless of the number of around-ear regions. 
Also, compared to other gesture regions with facial hair, this area offers less resistance to on-skin interaction, making it a viable candidate for mapping frequent tasks supported by the aforementioned on-skin gestures. 
The highlighted \rev{areas} around the \textit{cheekbone} and \textit{jawline} in Figure \ref{fig:-preference-around-cheekbone-horizontal} are preferred across all number of regions for on-skin horizontal pinch and swipe gestures, with the area around the \textit{cheekbone} having higher preference than the \textit{jawline}.

Participants do not generally prefer the area under the ear (Figure \ref{fig:-preference-under-ears}) for reusing gestures in 3 on-skin regions.
Nevertheless, for 5 and 7 regions, this area provides the participants with another convenient spot for vertical swipe, pinch, and tap gestures. 
Participants reported uncomfortable hand and arm joint bending in this area for horizontal on-skin swipe and pinch gestures. 
This indicates that on-skin interaction with this area should \rev{be limited} to vertical motion gestures and taps. 
Like the area under the ear, highlighted areas in Figure \ref{fig:-preference-front-of-face} within the frontal view regions provided convenient, segmented regions for on-skin interaction. 
\rev{All highlighted} areas (\ie,~ \textit{chin}, \textit{nose}, \textit{forehead}) provided sufficient spaces for supporting on-skin swipes and taps. 
Participants reported horizontal pinches as more natural, subtle gestures suited for public spaces than vertical pinches for the above areas.
Facial landmarks and contours in these highlighted regions make it easier to define region boundaries and have better gesture accuracy.

For mid-air gestures, gesture regions are larger with minimal obstacles than on-skin space. 
However, the lack of tactile feedback results in relying solely on frontal and peripheral views for feedback. 
For gestures outside of the field of view, participants solely rely on muscle/joint-based perception of the location of the hand and finger relative to the outer ear. 
Overall, participants are \rev{comfortable making} uni-manual gesture motions in most of the mid-air space around the head except for the top and back. 
Due to uncomfortable hand and arm joint \rev{bending and} site positioning outside the field of view, most participants do not include these two areas within the consideration of gesture region boundary definition or mark these as the least preferred areas.
Another reported area of interest in mid-air gesture space is in front of the eyes within a direct field of view.
Most participants felt more comfortable performing gestures outside this region but within peripheral view (over areas like the \textit{cheek}, \textit{jawline}, or \textit{temple}) to avoid blocking the vision. 
Unlike on-skin gestures, participants were not wholly averted to make mid-air gestures directly above the area around the eyes highlighted in Figure \ref{fig:-avoidance-around-eyes} within frontal view. 
Participants reported eyewear not being a significant roadblock to performing mid-air uni-manual gestures, as gestures tended not to displace the eyewear frames. 
Although outside of the peripheral view, the area around the ear (directly over/ above/ below/ behind/ in-front) was preferred. 
Participants reported a better spatial sense of the relative location of the hand to the \textit{ear}, especially when wearing an earable device such as a wireless earbud. 
They also explained that such uni-manual motions around the ear mimic interacting with an earable device. They are considered subtle and natural motion in public spaces due to the popularity of earbud usage in public and private spaces.

\subsection{NASA-TLX Survey Results}
\label{subsection:-NASA-TLX Survey Results}

\begin{figure}[tb!]
    \centering
    \includegraphics[width=1.00\textwidth,keepaspectratio]{figures/NasaTLX/NasaBlockTLX-Labels.png}
    \includegraphics[width=1.00\textwidth,keepaspectratio]{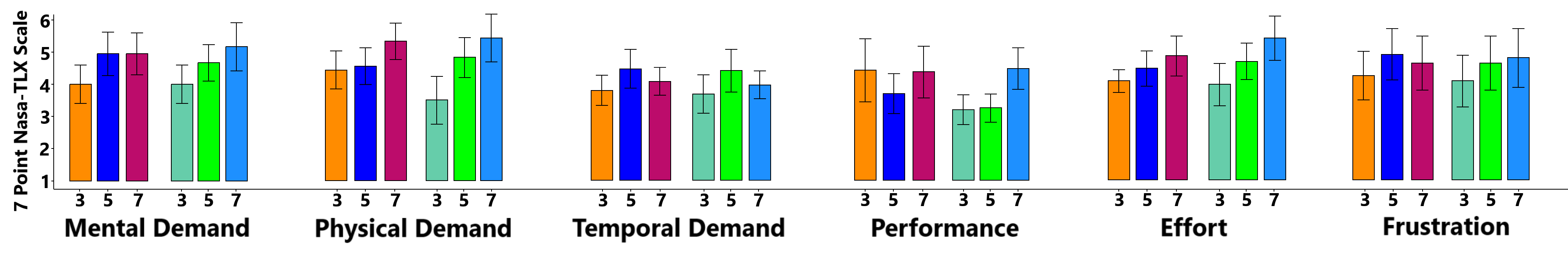}
    \vspace*{-0.8cm}
    \captionsetup{justification=centering}
    \caption{NASA TLX survey results for comparing increasing gesture region segments across mid-air and on-skin around-ear interaction space, ignoring gesture types (with 95\% confidence interval error bars)}
   \label{fig:-NasaTLX}
\end{figure}

During the study, participants completed a 7-point NASA TLX survey (Figure \ref{fig:-NasaTLX}) for each specific interaction space and region segmentation to measure their perceived workload.
This provided insight into changes in the perceived workload for increasing around-ear regions for gesture repetition. 

\textit{Mental demand} increased steadily when the number of around-ear regions increased from 3 to 5 for mid-air and on-skin gestures, plateauing for further on-skin region increase (to 7).
In contrast, it steadily increased from 5 to 7 mid-air gesture region increase.
Similar to \textit{mental demand}, \textit{physical demand} for mid-air gestures increased when the number of mid-air gestures was increased from 3 to 5 and from 5 to 7.
Reported \textit{physical demand} for reusing on-skin gestures across 3 and 5 segments was relatively similar.
However, it considerably increased when the number of regions increased from 5 to 7.
On-skin and mid-air gestures reported similar \textit{temporal demand} patterns.
Increased time pressure for reusing gestures across 5 gesture regions for both interaction spaces by the participants compared to 3/7 gesture regions. 

Participants reported that their perceived \textit{performance} was similar for 3 and 5 mid-air regions around the ear and considerably better than their on-skin counterparts. 
Surprisingly, for on-skin space, self-reported success in performing gestures over the intended region was better for 5 regions than 3 regions. 
The perceived \textit{performance} was considerably worse for 7 regions and relatively similar for mid-air and on-skin. 
The participants reported a steady increase \rev{in} both on-skin and mid-air \rev{gesture} \textit{effort} when the number of around-ear gesture regions increased. 
\rev{Similarly, \textit{frustration} increased steadily with the number of mid-air around-ear gesture regions.}
However, \textit{frustration} increased from 3 to 5 on-skin regions and slightly reduced when on-skin regions increased to 7, \rev{remaining} higher than frustration for gestures in 3 on-skin regions. 
Moreover, the self-reported effort showed a steady increase for on-skin interaction space, but interestingly, frustration slightly reduced from 5 to 7 segments.
Participants attributed this reduction in on-skin frustration to tactile feedback in finding target regions for more regions. 
Overall, these workload results support the threshold findings for mid-air and on-skin segmentations in Section \ref{subsection-SegmentSize-Gestures}.

\subsection{Reported Gesture Ranking}
\label{subsection:-Gesture Ranking}

\begin{figure}[tb!]
    \centering
    \includegraphics[width=0.9\textwidth]{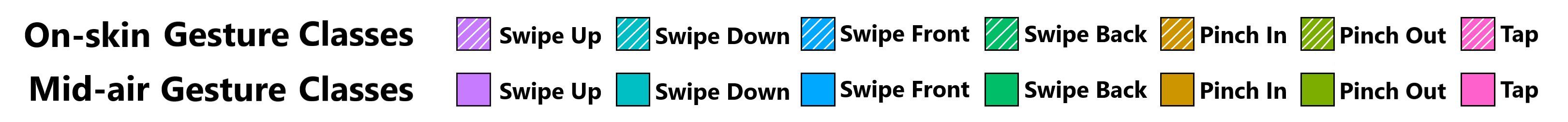}     
    \includegraphics[width=1.0\textwidth]{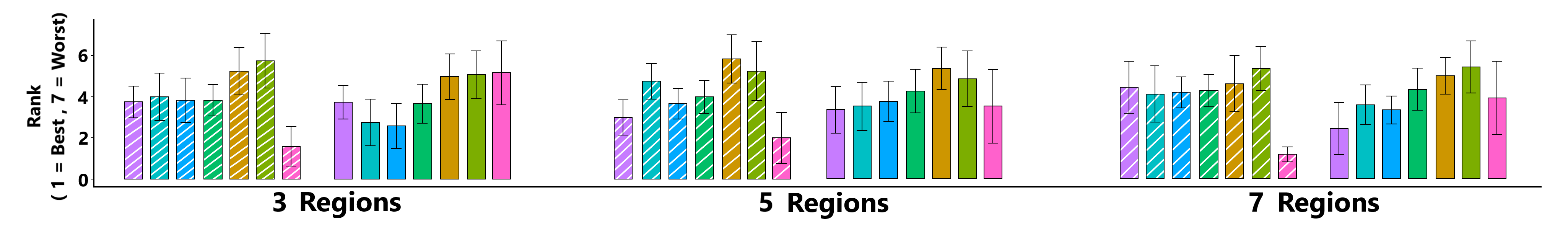}  
    \vspace*{-0.8cm}
    \captionsetup{justification=centering}
    \caption{Gesture rank for each block (with 95\% confidence interval error bars).}
    \label{fig:-Gesture Preference For Each Block}
\end{figure}

Participants ranked each of the 7 gestures against each other for different numbers of mid-air, and on-skin around-ear regions. 
Figure \ref{fig:-Gesture Preference For Each Block} shows each block's self-reported mean ranking of different gestures. 
In general, participants reported that touch sensation helped identify the position of on-skin gestures for the target gesture regions, even when the regions were outside of the direct or peripheral view of the participants. 

We found that \textit{tap} remained the most preferred gesture for on-skin interaction space as the number of around-ear gesture regions increased.
For mid-air gestures, \textit{swipe down} and \textit{swipe front} gestures received the best ranking for 3 regions.
For 5 mid-air regions, although \textit{tap} received the best ranking, it was closely followed by \textit{swipe up}, \textit{swipe down}, and \textit{swipe front} gestures.
When the number of around-ear mid-air gesture regions increased to 7, the ranking for \textit{tap} fell behind \textit{swipe up}, \textit{swipe down}, and \textit{swipe front} gestures. 

Analysis of \textit{Vicon} motion data and observation of participants' gesture motion show that all on-skin and mid-air \textit{tap} gestures started 35~-~40 cm away from the head in the mid-air position within the visual space.
As \textit{tap} gestures approached the head, they took up smaller visual space in central or peripheral vision than other gesture classes. 
Skin touch sensory feedback at the end of on-skin \textit{tap} gestures allowed the participant to quickly recognize if the gesture was over the correct gesture region. 
Also, motion data analysis showed that \textit{tap} gestures involved fewer steps than \textit{swipe} or \textit{pinch} gestures. 
The former involved pulling the index finger towards a target region on/over the face and rapidly pulling away the hand. Compared to this motion, swipe/pinch gestures involved moving towards a targeted region on/around the face, followed by moving the index finger(s) in a horizontal/vertical direction and then rapidly pulling away from the face. 
The skin touch feedback, alongside reduced movement complexity, contributed to the \textit{tap} gesture’s ranking as the primary choice for all numbers of around-ear gesture regions on the skin.

The lack of sensory feedback from skin touch in mid-air \textit{tap} gestures makes the participant solely reliant on frontal or peripheral vision to perceive \textit{tap} accuracy.
This, coupled with the more prominent horizontal/vertical \textit{swipe} motions over large mid-air regions, contributes to a better ranking of \textit{swipe} over \textit{tap} in 3 mid-air regions.
With the increase in mid-air gesture regions, the maximum available area for each region starts to shrink. 
This results in smaller mid-air \textit{swipe} motions being limited within the gesture boundary, providing it less visual prominence than similar \textit{swipe} gestures for a smaller number of mid-air regions. 
The decrease in \textit{swipe} prominence contributes to a slightly worsening ranking of \textit{swipe} gestures in 5 mid-air regions compared to 3 regions, making all \textit{swipe} and \textit{tap} gestures receive similar rankings from participants. 
Vicon motion analysis also shows that for mid-air and on-skin \textit{tap} gestures, gesture motion is usually bound within the same gesture region and does not mostly cross region boundaries from start to finish, making it difficult for participants to recognize mid-air \textit{tap} gestures over incorrect regions until the motion is complete. 
In 7 mid-air regions, the reduction in maximum available area for each region, as well as the abovementioned tap motion pattern, contributes to increased gesture attempts (compared to 3 and 5 mid-air regions and other swipe gestures) over different mid-air regions to make corrections between the starting and the ending of the gesture recording window -resulting in the shift in self-reported rankings in figure \ref{fig:-Gesture Preference For Each Block}, as well as increasing average path length for mid-air tap gestures in figure \ref{fig:MidAirVsOnSkin_PerformanceMetrics}).

For mid-air horizontal/vertical \textit{swipe} gestures, \textit{Vicon} motion data analysis revealed larger motions compared to their on-skin counterparts (\rev{Figures} \ref{fig:MidAirVsOnSkin_PerformanceMetrics} and \ref{fig:PerformanceMetrics-EveryGesture}). 
Mid-air \textit{swipe up}, \textit{swipe down}, and \textit{swipe front} received comparable self-reported ranking than \textit{tap} in the same interaction space (considerably better in 3 mid-air regions, and slightly better in 5 and 7 regions) as they started and ended within the direct or peripheral vision, usually crossing from one segmented region to another to reach the destination region. 
As such, participants were more confident that the swipe was done across the correct gesture region.

Participants reported discomfort in the backward bending of wrist and elbow joints for both on-skin and mid-air \textit{swipe back} gestures. 
For example, “I feel very uncomfortable swiping my hand back compared to the other gestures.” -P09.
The participants did not report similar discomfort for other gesture types across both interaction spaces. 
Also, for mid-air \textit{swipe back} gestures, the starting position was \textit{mainly} within the direct/peripheral view of the participants. 
However, for gesture regions that went beyond the peripheral view of the participants, the ending of such gestures could not be visually inspected by the participants. 
For such regions, the lack of tactile feedback reduced confidence in the perceived performance of mid-air swipe back over the intended gesture region, leading to worse ranking compared to other mid-air gestures across all numbers of around-ear regions.
However, this effect was not observed across different numbers of on-skin regions due to tactile feedback, and the gesture ranking of on-skin swipes was relatively similar.

Because of increased complexity, \textit{pinch} gestures took considerably more time than other gestures across all numbers of around-ear regions.
The relative complexity of gesture motions compared to \textit{swipe} and \textit{tap} gestures resulted in users preferring later gestures to pinch gestures in mid-air and on-skin regions. 
These reported rankings, alongside the gesture performance metrics discussed in Section \ref{section - quantitative results}, allow potential gesture designers to compare different gestures to choose them to map with appropriate tasks for different usage scenarios.

\section{Discussion}
\label{section: discussion}

\subsection{Interaction Space and Number of Gesture Regions}
\label{subsection:- discussion: DV1_and_DV2}

Our analysis shows that mid-air gestures were faster than on-skin gestures across 3, 5, and 7 around-ear regions (Figure \rev{\ref{fig:PerformanceMetrics-EveryGesture}}).
However, on-skin gestures demonstrated significantly better accuracy than mid-air gestures for 5 and 7 around-ear regions.
For 3 regions around the ear, there was no discernible difference in performance between these two interaction spaces. 
When considering accuracy, response time, and self-perceived performance from the NASA-TLX rating, it becomes evident that the most optimal condition for conducting interactions is 3 mid-air regions.
Several potential factors contribute to this efficiency gap. 
\begin{itemize}
    \item Mid-air gestures do not require physical contact with a surface, allowing for quicker initiation and execution.
    \item The absence of physical constraints associated with on-skin gestures provides more fluid and intuitive movement, enabling users to perform actions swiftly.
    \item The presence of larger region boundaries and more region location possibilities within the peripheral view compared to the same number of on-skin regions.
\end{itemize}

The following comments collected during the study revealed more insights into the preference for mid-air gestures compared to on-skin gestures for 3 gesture regions:

\pqt{For 3 Segmentation, I can create larger mid-air gesture regions than on-skin and avoid skin scraping.}{P08}

However, as the number of around-ear gesture regions increased from 3 regions, participants started reporting a preference shift towards on-skin interaction space for most gestures because of tactile sensation from skin touches. 
Perceived workload metrics except self-reported performance become similar for 5 regions (Figure \ref{fig:-NasaTLX}) for both interaction spaces. 
Although gestures over 5 mid-air regions are faster and have a higher self-reported preference, statistical analysis shows accuracy to be significantly better for 5 on-skin regions \rev{than} mid-air counterparts (Figure \ref{fig:PerformanceMetrics-EveryGesture}). 
Figure \ref{fig:-NasaTLX} also shows a noticeable improvement in self-reported performance when the number of on-skin regions increases from 3 to 5, supporting the preference shift observation towards on-skin for 5 regions.
When the number of gesture regions increases to 7, we observe a more significant drop in gesture accuracy and an increase in time for mid-air regions compared to on-skin regions. 
Overall results and observations from participant comments indicate \rev{that on-skin gestures are preferred over mid-air} for 7 regions.
The following comment by participants explains tactile feedback as a possible reason for the preference shift in interaction space for a higher number of gesture regions:
\pqt{It is difficult to gesture over the correct place when the number of regions increases to 5 and beyond. 
Touching the skin helps me to quickly identify if I am making a gesture over the correct region.
In mid-air gestures, it is challenging to make gestures over the correct boundary with only visual feedback; even more so when gestures are in the air behind the ears.
}{P10}

Overall, participants favored a 3-region configuration in both mid-air and on-skin interaction spaces and indicated that the configuration could be increased up to 5 regions for on-skin spaces because of tactile feedback.
Participants did not prefer 7 regions in both interaction spaces. 
These findings are similar to the B2B \cite{Rey2022BezelToBezel} study analyzing gesture reuse over multiple regions for smartwatches, indicating that fewer regions can enhance interaction performance in terms of accuracy and time efficiency.

\subsection{Observations on Gesture Region Size, Location and Boundaries}
\label{subsection:- Gesture Region Area}

\begin{figure}[tb!]
    \centering 
    \captionsetup{justification=centering}
    \begin{subfigure}[b]{0.3\textwidth}
        \centering
        {\includegraphics[width=0.9\textwidth]{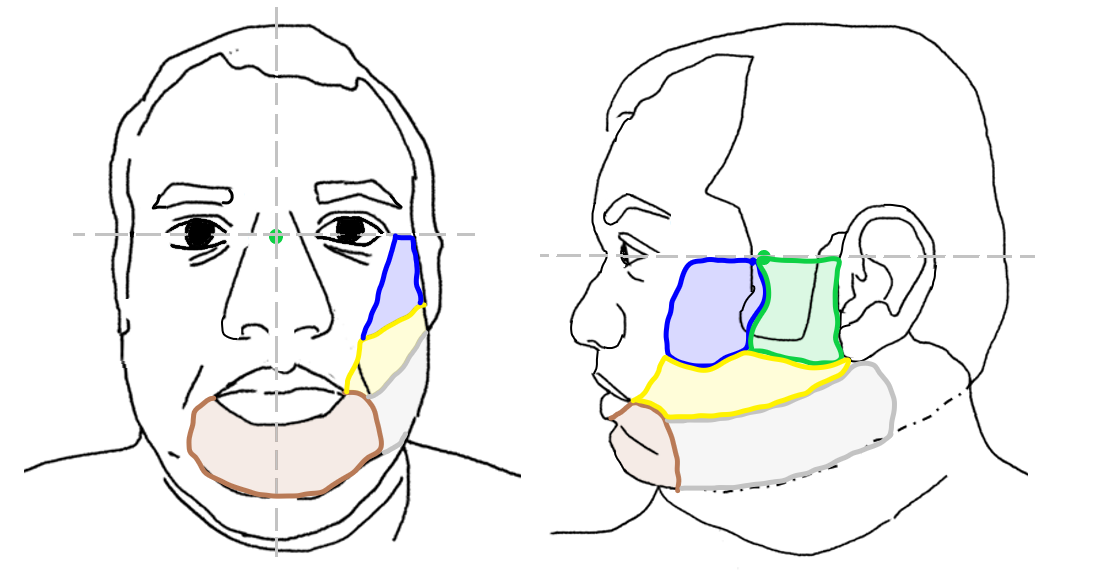}}   
        \vspace*{-0.2cm}
        \caption{}
        \label{fig:-region-size-and-location-ex1}
    \end{subfigure}%
    \begin{subfigure}[b]{0.3\textwidth}
        \centering
        {\includegraphics[width=0.2\textwidth]{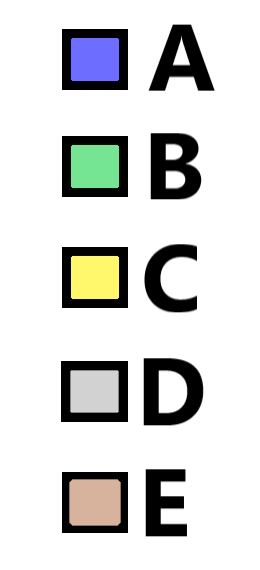}}   
    \end{subfigure}%
    \begin{subfigure}[b]{0.3\textwidth}
        \centering
        {\includegraphics[width=0.9\textwidth]{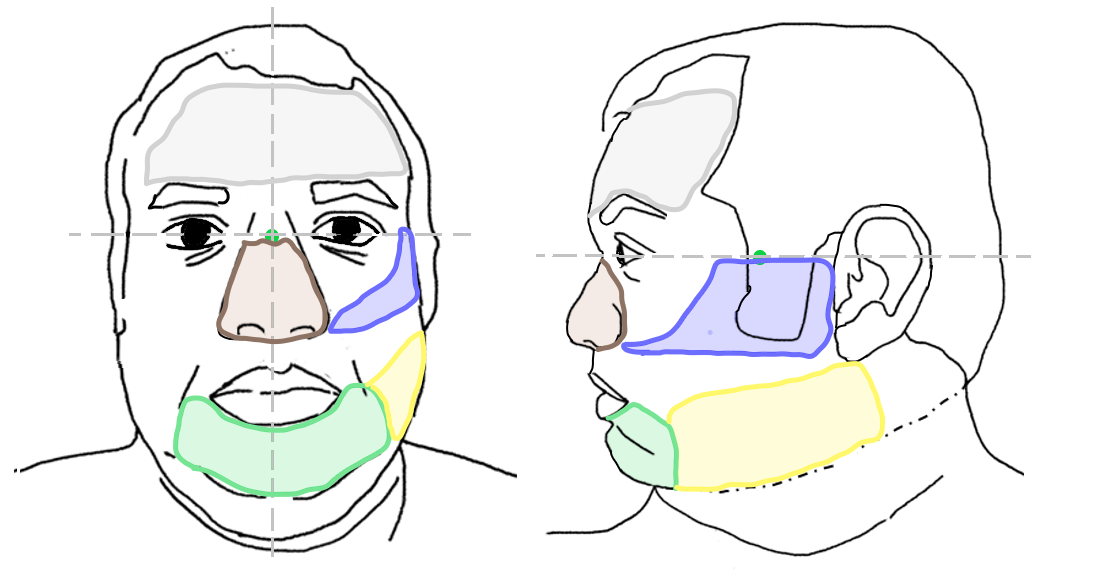}}   
        \vspace*{-0.2cm}
        \caption{}
        \label{fig:-region-size-and-location-ex2}
    \end{subfigure}%
    \vspace*{-0.3cm}
    \caption{Sample 5 on-skin gesture region boundary from two participants showing gesture region packing density.}
    \label{fig:-highlighted face areas5}
\end{figure}

We observed participants using facial contour lines or buffer spaces for defining on-skin gesture space boundaries and tended to create loosely packed gesture regions whenever possible. 
In minor instances where participants did not create region boundaries along facial contour lines (\eg,~dividing the jawline spanning from the end of the chin to the jaw angle into two regions), we observed incorrect swipe and pinch gestures as they could not correctly start or stop over the destined region. 
As on-skin gesture regions increased to 5 and 7, participants still found techniques to create loosely packed gesture regions and large gesture region boundaries with facial contours or buffer space separations by choosing areas outside the field of view. 
However, for on-skin gestures, gesture accuracy was more affected by gesture region packing structure and facial contour lines defining the gesture region boundaries than the size of the gesture region. 
We observed more incorrect gestures for densely packed gesture regions (\eg, multiple regions on the cheek) and region boundaries that did not follow facial contour lines. 
Figure \ref{fig:-region-size-and-location-ex1} shows a densely packed on-skin gesture region structure. 
The regions in this structure were reused for swipe back and swipe down gestures by the participant. 
Although the participant could perform most \textit{swipe back} gestures over all regions correctly, \rev{the} \textit{swipe down} gesture over region \textbf{C} incorrectly ended over region \textbf{D} over multiple attempts due to the region packing structure and shape of region \textbf{C}. 
The participant later remarked that during swipe down over the cheekbone (region \textbf{A}), gestures did not cross over to region \textbf{C} on the cheek as the region boundary (end of \textit{Zygomatic bone}) was evident from tactile feedback, which was not the case for region \textbf{C}. 
Figure \ref{fig:-region-size-and-location-ex2} shows another region packing structure created by another participant where buffer spaces exist between most gesture regions. 
Like Figure \ref{fig:-region-size-and-location-ex1}, regions on the cheek and jawbone border each other, and the region boundary (\textit{end of chin}) is prominent. 
These loose region packing structures reported higher overall accuracy than those in Figure \ref{fig:-region-size-and-location-ex1}. 
We also observed a similar relationship between gesture performance and region packing structure relationship for 7 regions, where more regions were packed without facial contour lines as gesture region boundaries, resulting in increased incorrect gestures. 
These observations match the results obtained by Dezfuli \etal~\cite{Dezfuli_Palm_RC} for imaginary user interfaces, where on-skin gesture effectiveness decreases with increasing gesture region packing density and expand their gesture reuse findings from palm-based interfaces to earable devices.

We observed a similar compounding effect of region size, location, and facial landmarks on mid-air gesture performances. 
Smaller gesture regions within the field of view were overall more accurate than larger gesture regions outside the field of view (\eg,~ behind /above the ear). 
Within the field of view, facial areas such as the \textit{nose}, \textit{corner of the eye}, \textit{corner of the mouth}, and \textit{starting of the cheekbone} served as visuospatial referent points, and mid-air gestures closer to these points were more accurately performed over the correct region. 
As on-skin gestures started from mid-air, hand proprioception for the above-mentioned visuospatial referent points also positively impacted gesture performance for on-skin regions around these landmarks. 
These findings are similar to the finger-pointing study using visuospatial reference points by Gustafson \etal~\cite{Gustafson_Imaginary_Interfaces} for imaginary mid-air interfaces.

\subsection{Implications for Design}
\label{subsection:- Implications for Design}

\begin{figure}[tb!]
    \centering 
    \captionsetup{justification=centering}
    \begin{subfigure}[b]{0.5\textwidth}
        \centering
        {\includegraphics[width=0.8\textwidth]{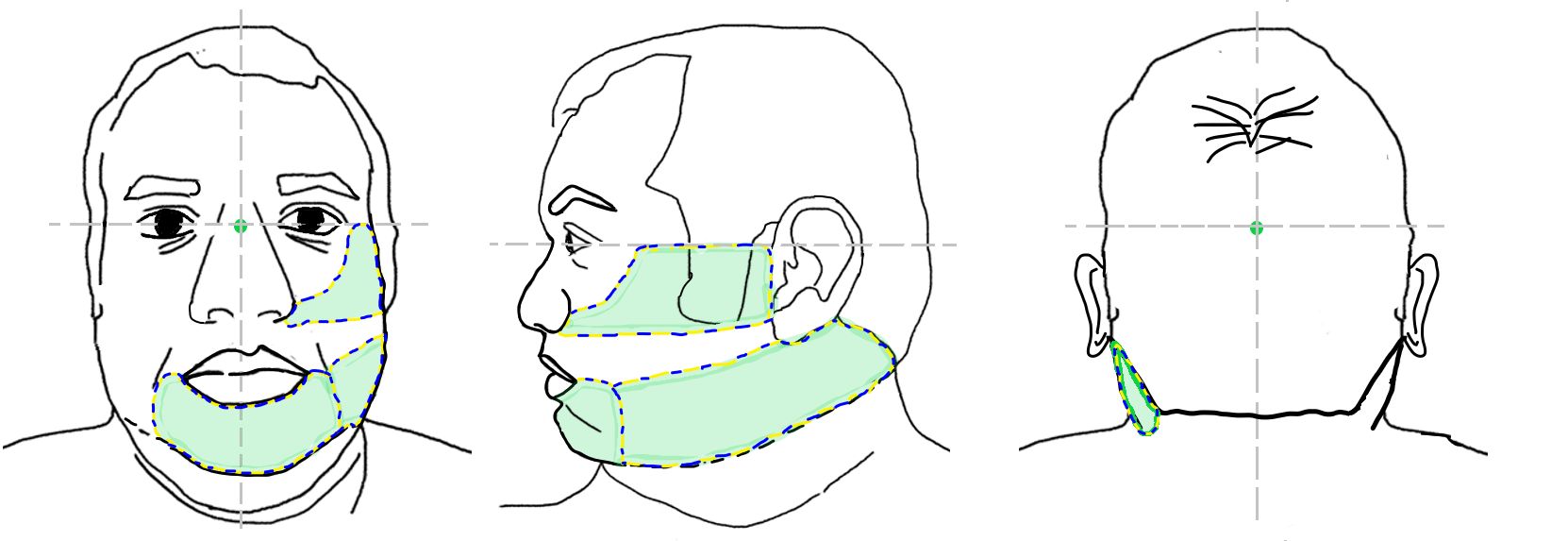}}   
        \vspace*{-0.2cm}
        \caption{3 on-skin regions 
        \\ 1) Cheekbone, 2) Chin, 3) Along the Jawline}
        \label{fig:-recommended-3-on-skin-regions}
    \end{subfigure}%
    \begin{subfigure}[b]{0.5\textwidth}
        \centering
        {\includegraphics[width=0.8\textwidth]{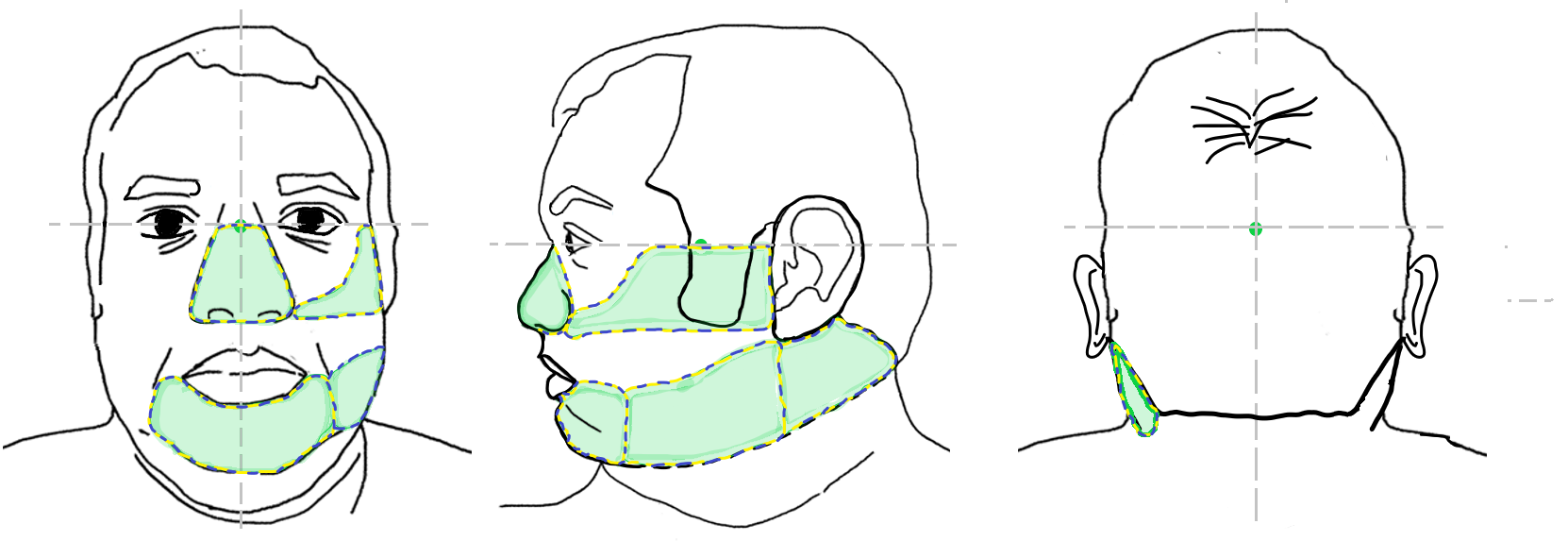}}
        \vspace*{-0.2cm}
        \caption{5 on-skin regions 
        \\ 1) Cheekbone, 2) Nose, 3) Chin 4) Jawline 5) Under ear}
        \label{fig:-recommended-5-on-skin-regions}
    \end{subfigure}%
    \vspace*{-0.3cm}
    \caption{Recommended areas~\footref{side-view-footnote} for re-using gestures in 3 and 5 on-skin regions. Dotted lines indicate region boundaries.}
    \label{fig:-highlighted face areas3}
\end{figure}

\begin{figure}[tb!]
    \centering 
    \captionsetup{justification=centering}
    \begin{subfigure}[b]{0.45\textwidth}
        \centering
        {\includegraphics[width=0.9\textwidth]{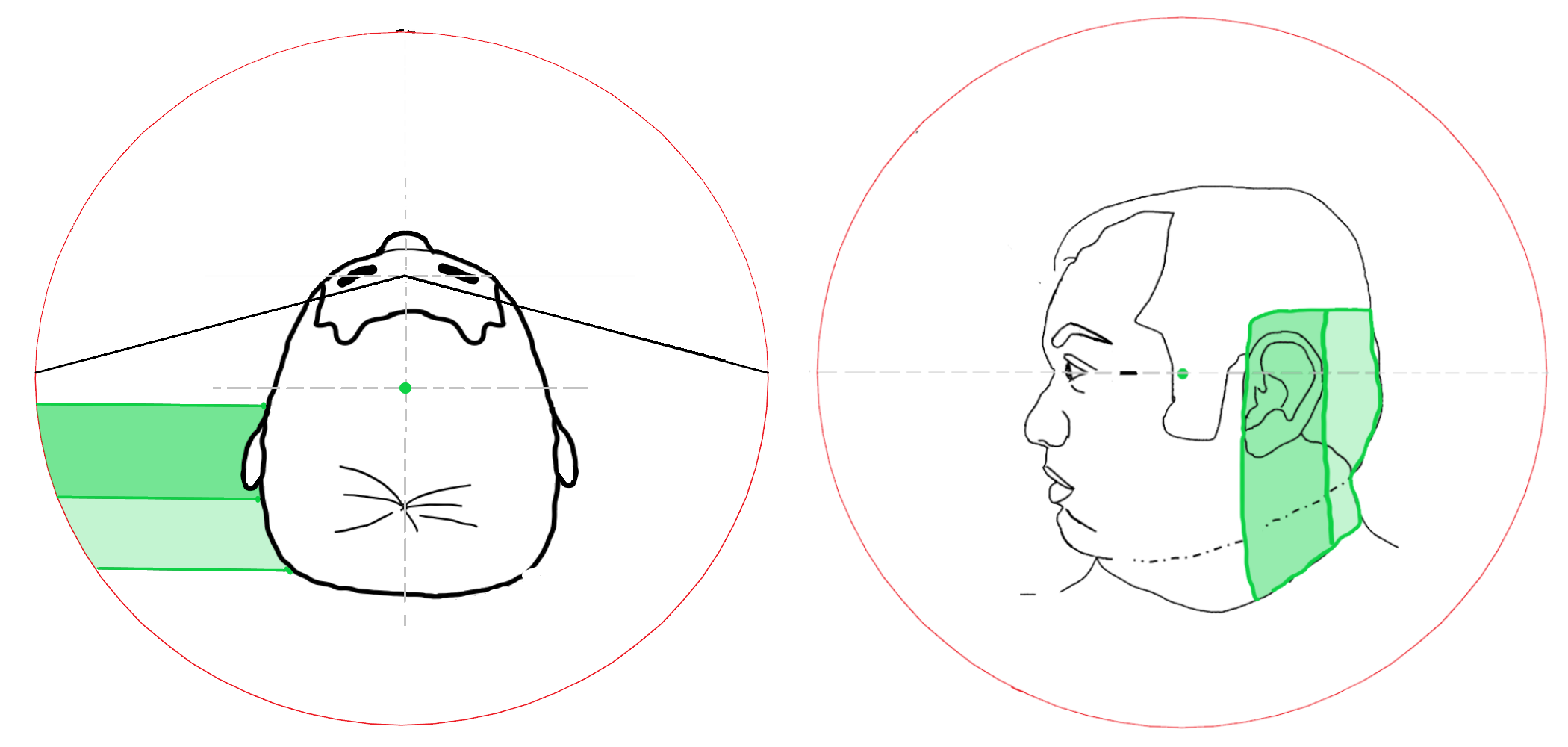}}   
        \vspace*{-0.2cm}
        \caption{Mid-air above ear}
        \label{fig:-recommended-3-mid-air-ear}
    \end{subfigure}%
    \begin{subfigure}[b]{0.45\textwidth}
        \centering
        {\includegraphics[width=0.9\textwidth]{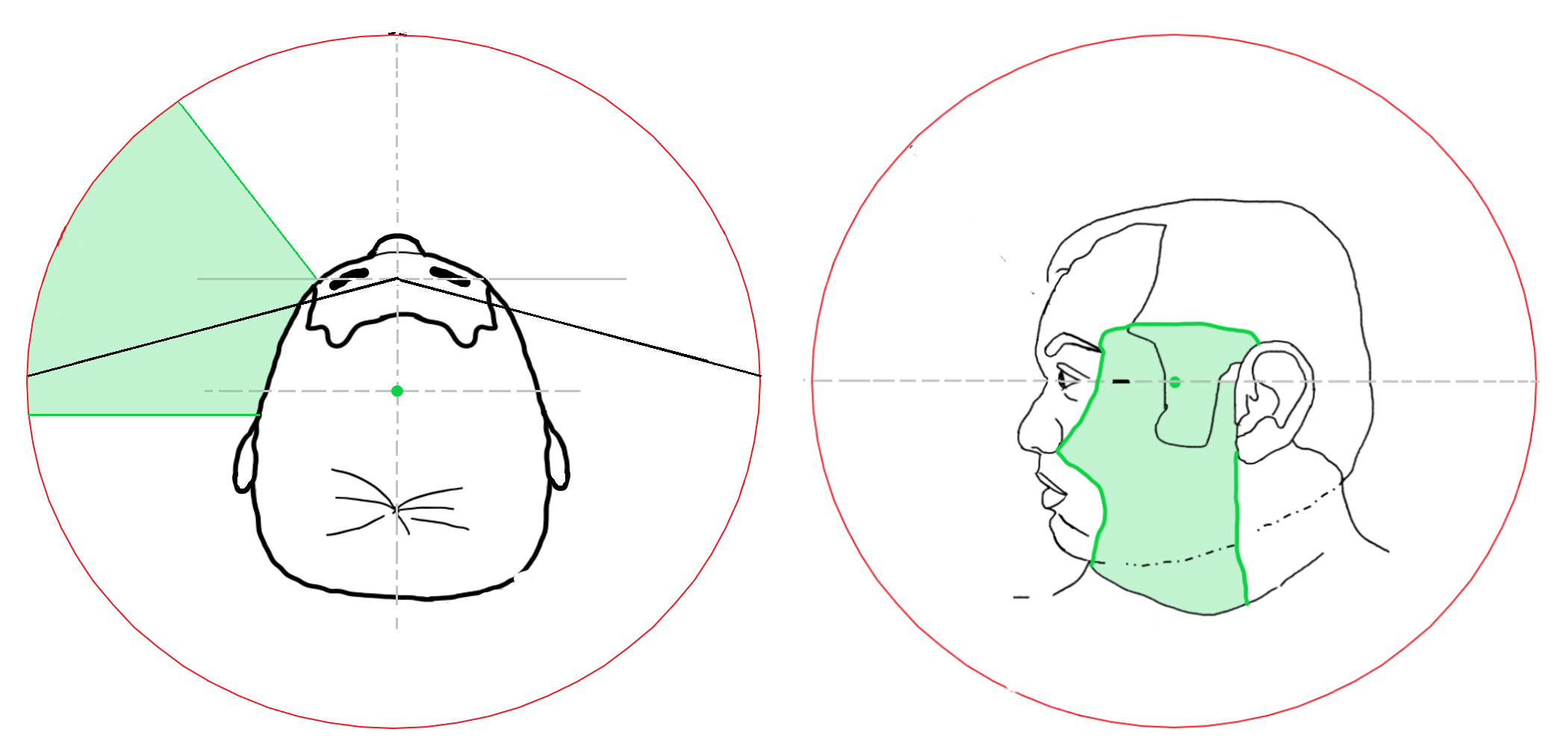}}
        \vspace*{-0.2cm}
        \caption{Mid-air above cheek}
        \label{fig:-recommended-5-mid-air-cheek}
    \end{subfigure}%
    \\
    \begin{subfigure}[b]{0.75\textwidth}
        \centering
        {\includegraphics[width=0.9\textwidth]{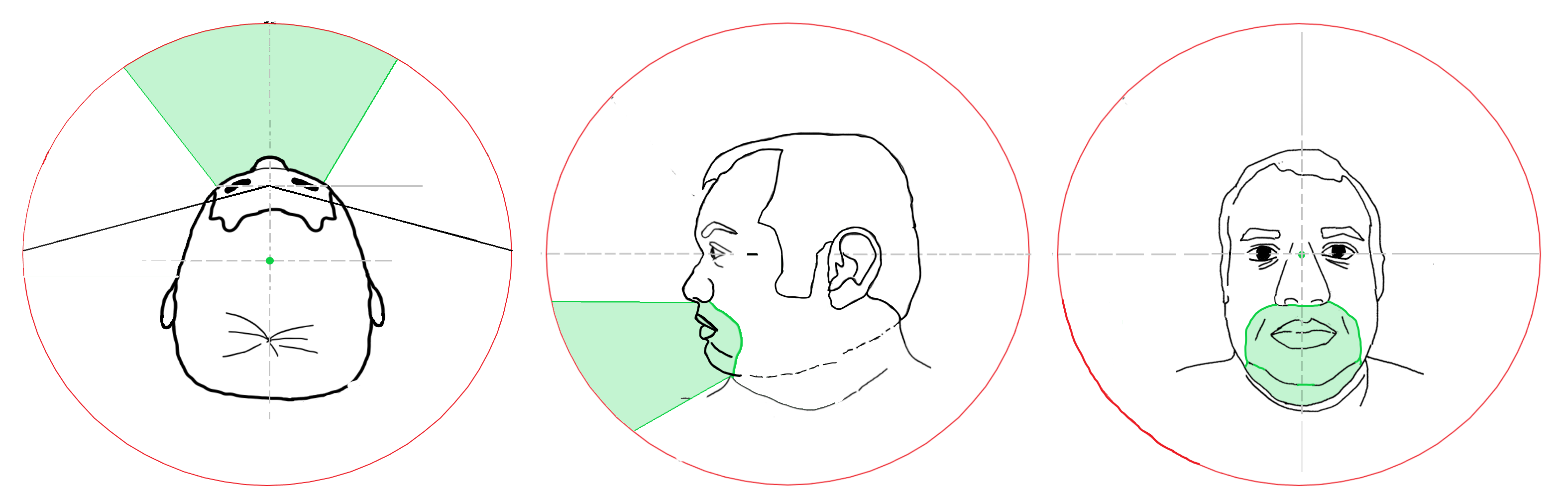}}
        \vspace*{-0.2cm}
        \caption{Mid-air above mouth and chin}
        \label{fig:-recommended-5-mid-air-mouth}
    \end{subfigure}%
    \vspace*{-0.3cm}
    \caption{Recommended areas~\footref{side-view-footnote} for re-using gestures in 3 mid-air regions. The red circle represents DV1 measurement boundary (30cm) from the center of the head discussed in Section \ref{subsection: Apparatus}.}
    \label{fig:-highlighted face areas4}
\end{figure}

Based on our preliminary insights into around-ear design space, we present the following design implications for implementing on-skin and mid-air, off-device gestures for earable devices:

\begin{itemize}
    \item The regions for on-skin and mid-air gesture reuse should be limited to 5 and 3, respectively, and \rev{mid-air gestures are preferred for 3 regions}.
    Figures \ref{fig:-highlighted face areas3} (on-skin) and \ref{fig:-highlighted face areas4} (mid-air) propose approximate gesture regions based on Figure \ref{fig: Most and Least Preferred Regions for On-Skin and MidAir space} and qualitative observations on region boundary patterns.
    \item For on-skin gestures, most frequent tasks should be mapped to the \textit{cheekbone} area due to its location within peripheral vision within the field of view, reduced obstacle and friction from lack of facial hair, and being outside of eyewear \rev{regions and} underlying bone support provides better tactile feedback.
    \item For mid-air gestures, the most frequent tasks should be mapped to the area from the \textit{temple} to the \textit{jaw} through the \textit{cheek} area (Figure \ref{fig:-recommended-5-mid-air-cheek}), as uni-manual gesture motion with this region mimics natural hand-to-face interactions, and the location is within the peripheral view.
    \item The \textit{temple} region (Figure \ref{fig:-avoidance-around-temple}) should be avoided for on-skin gestures to avoid potential eyewear frame displacement issues. 
    The highlighted region can be a buffer between on-skin gesture regions above and below the temple. 
    However, such restriction does not apply to mid-air gestures.
    \item Designing touch gestures below the \textit{eye} to the \textit{corner of the mouth} (Figure 6a) should be avoided because of potential eyewear issues.
    \item Gesture regions should be loosely packed as much as possible, and facial landmarks and contour lines should be leveraged while defining gesture regions.
\end{itemize}

\section{Limitations and Future Work}

We used motion tracking (Vicon camera array) to study around-ear gestures. 
However, exploring embedded earbud motion sensors for gesture recognition around the device could offer a more realistic approach. 
Our primary research focused on the exploration of reusing around-ear gestures by gesture space segmentation rather than the supporting technology itself. 
Notably, no commercially available Earable devices currently recognize both on-skin and mid-air unimanual gestures around the device simultaneously. 
Developing such a device for detecting gestures in these interaction spaces represents a compelling avenue for future research. 
In addition, We excluded interaction techniques on the ear itself, which was well-explored in previous research \cite{Earput2013Lissermann}. 
Adding on-ear interaction could complicate our study with numerous
combinations. 
Nevertheless, comparing around-ear and on-ear interactions could broaden insights into gesture-based Earable device interactions.

Due to the length of our study, we limited our gesture reuse exploration to only a sitting stance with the manual encumbrance of the dominant hand. 
However, it is necessary to understand participant preference for gesture space, number of regions, and region boundary definition in other real-world scenarios (\eg,~ standing, walking with the encumbered dominant and non-dominant hand, walking without manual encumbrance) for a more thorough understanding of around-ear unimanual gesture space for supporting off-device earable gestures in the real world.
In the future, we plan to explore around-ear, off-device earable gesture reuse with the aforementioned gesture classes across 3 to 5 gesture regions to build upon our initial \rev{work and} present a more comprehensive set of insights for earable interaction designers.  
We also plan to look into gesture-task mapping for proposed gesture reuse in various usage scenarios across \textit{smart home environments}, \textit{public display interactions}, and \textit{AR/VR interactions}, among others.
For example, an \textit{on-cheek} region can control ceiling lights in a \textit{smart home} environment. In contrast, a region on the \textit{chin} can manage the TV, and the region around the \textit{jawline} can interact with the smartphone. 
This can provide useful insight into the association of gesture region location with task importance in various scenarios and build upon our initial work for supporting proposed off-device earable interaction techniques in the real world.

The observations in Section \ref{subsection:- Gesture Region Area} set apart proposed around-ear earable interaction techniques with other gesture-based interfaces operating entirely within the field of view. 
Fitt’s law \cite{fitts1954information} assumes target visibility relates gesture region size directly with gesture performance. 
As evidenced by the example scenarios (Section \ref{subsection:- Gesture Region Area}), some target regions in our study fall outside the field of view, and gesture region size alone can not account for gesture performance for proposed around-ear interaction in earable devices. 
More recent complex models accounting for factors like location \cite{Cao_Peephole_Pointing, Rohs_Target} have been proposed to account for target regions outside the field of view. 
Ens \cite{Ens_HMD_Fitts} explores these extensions of Fitt’s Law in a head-worn display (HWD) and the head-motion context in a user study. However, their findings can not be directly translated into an earable context where the head motion to target mid-air and on-skin spaces outside of field-of-view using unimanual gestures is impossible.
More complex gesture performance models extending Fitt’s Law addressing region size, location, tactile feedback (on-skin gesture), or hand proprioception (mid-air gestures) need to be developed in around-ear, earable contexts. 
As a first attempt to explore uni-manual around-ear off-device gestures for earables, this work focuses on investigating users’ performances in different interaction spaces and how they generally vary for different numbers of regions. 
More in-depth study regarding the specific properties of the gesture region targets is left for future work.

\section{Conclusion}

In this paper, we experimented with gesture reuse for mid-air and on-skin-around-ear interaction space for Earable to increase the gestural input vocabulary.
Such an approach could counter the limited on-device interaction space and avoid the pitfalls with other potential interaction spaces such as ear-bending, facial expression, head movements, or voice-based interactions.
\rev{Our exploration found that tactile feedback switches user preference from mid-air to on-skin spaces for more gesture regions.}
For both spaces, we summarize that the interaction space should be segmented into at most 5 regions for gesture reuse, and for mid-air spaces, 3 regions are preferred. 
Besides our observations concerning gesture class and location preferences across different numbers of around-ear regions, our gesture motion analysis study revealed the overlapping area of preference for an increasing number of regions for different gestures, along with a heatmap of the most and least preferred regions around the ear - providing the ubicomp researcher and interaction designer communities a roadmap to map the most frequent tasks performed with Earable to certain gesture regions in mid-air and on-skin space. 
Although our work is by no means an exhaustive study of gesture reuse for off-device earable \rev{interaction, further} research in this direction could complement existing research into earable gesture recognition technology to make \rev{the} proposed off-device interaction market ready for \rev{commercialization and} popular among end users.

\bibliographystyle{ACM-Reference-Format}
\bibliography{main.bib}

\clearpage 

\appendix
\label{section: appendix}
\begin{appendices}

\section{GRDA Application}
\label{appendix:grda_application}

\begin{figure}[H]
     \centering
     \includegraphics[keepaspectratio, height=0.45\textheight]{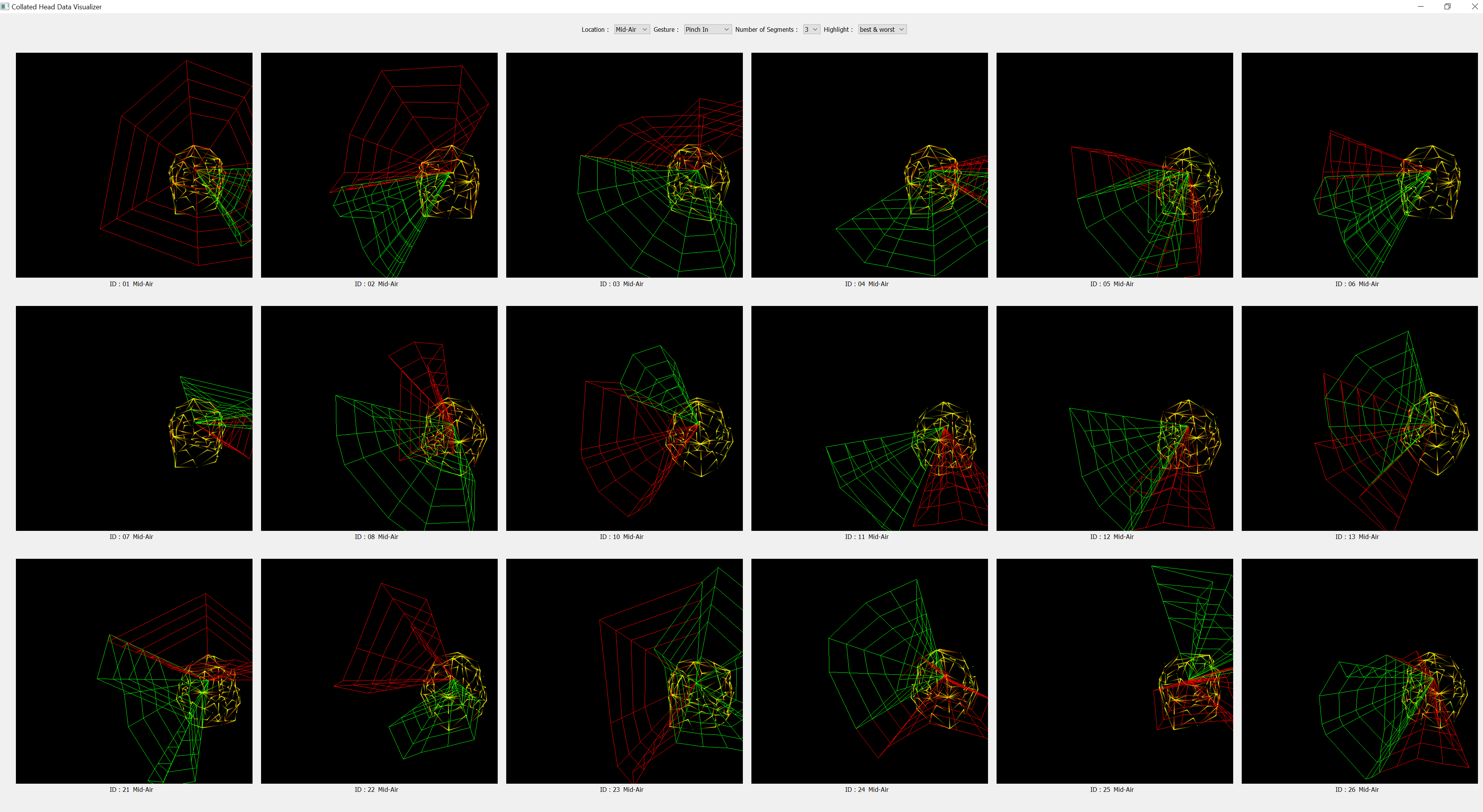}
     \captionsetup{justification=centering}
     \caption{GRDA application allowing observation of 3D head model and best (green)~/~worst (red) area comparison side by side}
     \label{fig:-3DViews-BestArea}
\end{figure}

\begin{figure}[H]
     \centering
     \includegraphics[keepaspectratio, width=0.6\textwidth]{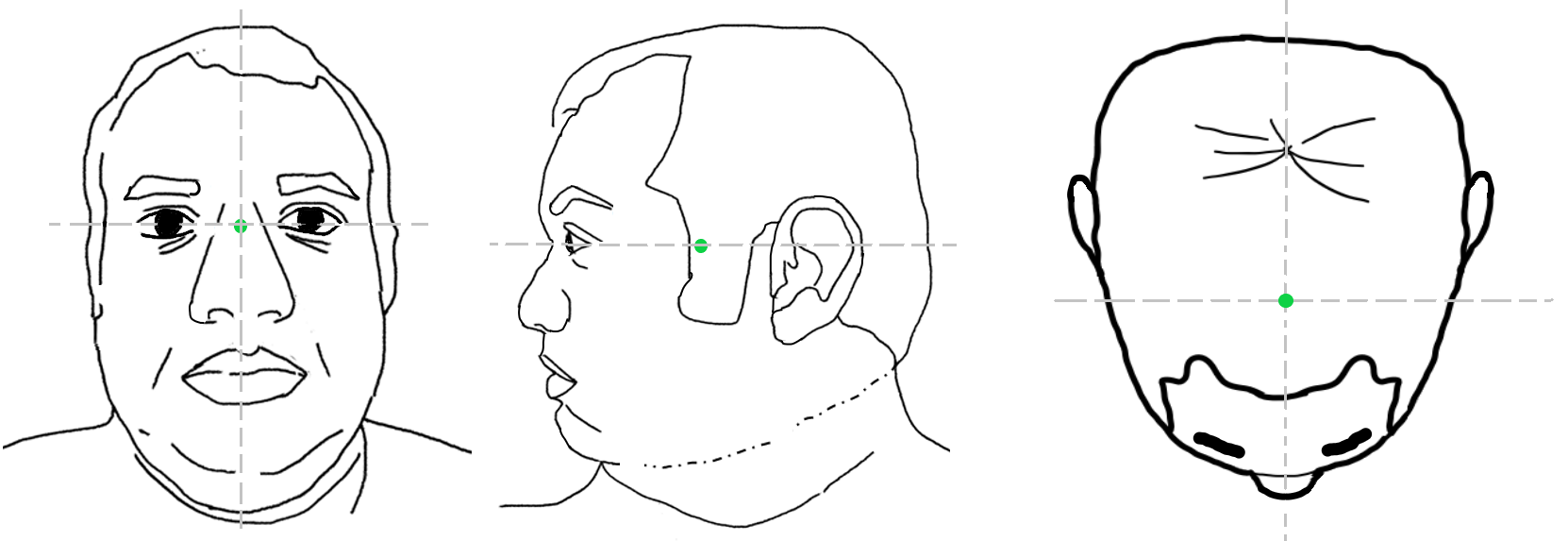}
     \captionsetup{justification=centering}
     \caption{Location of the center (marked in green) of the head with respect to front, side, and top view of the head for measuring distances in GRDA application. The dashed line in grey represents axis lines in 3 dimensions.}
     \label{fig:-head-center}
\end{figure}

\newpage

\section{Statistical Analysis}
\label{appendix:stat_analysis}

\subsection{Mean and Median Values}
\label{appendix:mean_and_media}

\begin{table}[H]
\centering
\tiny
\begin{tabular}{|c|cccccc|}
\hline
\multirow{3}{*}{Gesture} & \multicolumn{6}{c|}{\textbf{Gesture Time (DV1)}}                                                                                                                                                                                                                                                                                                                                                                                                    \\ \cline{2-7} 
                         & \multicolumn{3}{c|}{Mean}                                                                                                                                                                                                      & \multicolumn{3}{c|}{Median}                                                                                                                                                                               \\ \cline{2-7} 
                         & \multicolumn{1}{c|}{3 Regions} & \multicolumn{1}{c|}{5 Regions} & \multicolumn{1}{c|}{7 Regions} & \multicolumn{1}{c|}{3 Regions} & \multicolumn{1}{c|}{5 Regions} & \multicolumn{1}{c|}{7 Regions} \\ \hline
All gestures On-skin           & \multicolumn{1}{c|}{3712.687}                                            & \multicolumn{1}{c|}{3816.186}                                            & \multicolumn{1}{c|}{3953.525}                                            & \multicolumn{1}{c|}{3703.757}                                            & \multicolumn{1}{c|}{3819.503}                                            & 3922.156                                             \\ \hline
All gestures Mid-air           & \multicolumn{1}{c|}{3536.410}                                            & \multicolumn{1}{c|}{3672.517}                                            & \multicolumn{1}{c|}{3782.245}                                            & \multicolumn{1}{c|}{3495.404}                                            & \multicolumn{1}{c|}{3593.735}                                            & 3774.913                                            \\ \hline
Swipe up On-skin           & \multicolumn{1}{c|}{3493.432}                                            & \multicolumn{1}{c|}{3672.132}                                            & \multicolumn{1}{c|}{3893.627}                                            & \multicolumn{1}{c|}{3480.184}                                            & \multicolumn{1}{c|}{3691.117}                                            & 3892.43                                             \\ \hline
Swipe up Mid-air           & \multicolumn{1}{c|}{3243.691}                                            & \multicolumn{1}{c|}{3418.072}                                            & \multicolumn{1}{c|}{3460.292}                                            & \multicolumn{1}{c|}{3304.489}                                            & \multicolumn{1}{c|}{3429.727}                                            & 3431.065                                            \\ \hline
Swipe up On-skin           & \multicolumn{1}{c|}{3657.619}                                            & \multicolumn{1}{c|}{3750.83}                                             & \multicolumn{1}{c|}{3891.866}                                            & \multicolumn{1}{c|}{3660.206}                                            & \multicolumn{1}{c|}{3756.051}                                            & 3905.35                                             \\ \hline
Swipe up Mid-air           & \multicolumn{1}{c|}{3300.891}                                            & \multicolumn{1}{c|}{3471.041}                                            & \multicolumn{1}{c|}{3713.693}                                            & \multicolumn{1}{c|}{3293.885}                                            & \multicolumn{1}{c|}{3475.328}                                            & 3705.572                                            \\ \hline
Swipe front On-skin        & \multicolumn{1}{c|}{3459.888}                                            & \multicolumn{1}{c|}{3577.415}                                            & \multicolumn{1}{c|}{3719.333}                                            & \multicolumn{1}{c|}{3492.858}                                            & \multicolumn{1}{c|}{3561.259}                                            & 3713.978                                            \\ \hline
Swipe front Mid-air        & \multicolumn{1}{c|}{3366.385}                                            & \multicolumn{1}{c|}{3509.414}                                            & \multicolumn{1}{c|}{3864.546}                                            & \multicolumn{1}{c|}{3373.323}                                            & \multicolumn{1}{c|}{3514.852}                                            & 3688.63                                             \\ \hline
Swipe back On-skin         & \multicolumn{1}{c|}{3349.439}                                            & \multicolumn{1}{c|}{3523.131}                                            & \multicolumn{1}{c|}{3603.241}                                            & \multicolumn{1}{c|}{3343.3}                                              & \multicolumn{1}{c|}{3521.587}                                            & 3587.61                                             \\ \hline
Swipe back Mid-air         & \multicolumn{1}{c|}{3394.372}                                            & \multicolumn{1}{c|}{3494.341}                                            & \multicolumn{1}{c|}{3587.924}                                            & \multicolumn{1}{c|}{3381.961}                                            & \multicolumn{1}{c|}{3504.524}                                            & 3593.381                                            \\ \hline
Pinch in On-skin           & \multicolumn{1}{c|}{4195.593}                                            & \multicolumn{1}{c|}{4193.104}                                            & \multicolumn{1}{c|}{4281.091}                                            & \multicolumn{1}{c|}{4194.622}                                            & \multicolumn{1}{c|}{4171.085}                                            & 4273.56                                             \\ \hline
Pinch in Mid-air           & \multicolumn{1}{c|}{4052.149}                                            & \multicolumn{1}{c|}{4149.953}                                            & \multicolumn{1}{c|}{4215.33}                                             & \multicolumn{1}{c|}{4036.132}                                            & \multicolumn{1}{c|}{4118.786}                                            & 4207.033                                            \\ \hline
Pinch out On-skin          & \multicolumn{1}{c|}{4187.546}                                            & \multicolumn{1}{c|}{4265.589}                                            & \multicolumn{1}{c|}{4375.578}                                            & \multicolumn{1}{c|}{4177.971}                                            & \multicolumn{1}{c|}{4253.219}                                            & 4389.343                                            \\ \hline
Pinch out Mid-air          & \multicolumn{1}{c|}{3992.793}                                            & \multicolumn{1}{c|}{4102.66}                                             & \multicolumn{1}{c|}{4110.475}                                            & \multicolumn{1}{c|}{3992.889}                                            & \multicolumn{1}{c|}{4118.463}                                            & 4108.293                                            \\ \hline
Tap On-skin               & \multicolumn{1}{c|}{3650.159}                                            & \multicolumn{1}{c|}{3759.947}                                            & \multicolumn{1}{c|}{3900.269}                                            & \multicolumn{1}{c|}{3646.839}                                            & \multicolumn{1}{c|}{3798.125}                                            & 3898.539                                            \\ \hline
Tap Mid-air               & \multicolumn{1}{c|}{3424.899}                                            & \multicolumn{1}{c|}{3549.574}                                            & \multicolumn{1}{c|}{3676.62}                                             & \multicolumn{1}{c|}{3412.574}                                            & \multicolumn{1}{c|}{3544.269}                                            & 3682.643                                            \\ \hline
\end{tabular}%
\captionsetup{justification=centering}
\caption{Mean and median gesture time (DV1) in milliseconds (ms).}
\label{tab:DV1_mean_and_median}
\end{table}

\vspace{-0.5cm}
\begin{table}[H]
\centering
\tiny
\begin{tabular}{|c|llllll|}
\hline
\multirow{3}{*}{Gesture} & \multicolumn{6}{c|}{\textbf{Gesture Path Length (DV2)}}                                                                                                                                                                                                                                                                                                                                                                                                         \\ \cline{2-7} 
                         & \multicolumn{3}{c|}{Mean}                                                                                                                                                                                                      & \multicolumn{3}{c|}{Median}                                                                                                                                                                                                    \\ \cline{2-7} 
                         & \multicolumn{1}{c|}{3 Regions} & \multicolumn{1}{c|}{5 Regions} & \multicolumn{1}{c|}{7 Regions} & \multicolumn{1}{c|}{3 Regions} & \multicolumn{1}{c|}{5 Regions} & \multicolumn{1}{c|}{7 Regions} \\ \hline
All gestures On-skin       & \multicolumn{1}{c|}{1422.108}                                            & \multicolumn{1}{c|}{1391.224}                                            & \multicolumn{1}{c|}{1409.538}                                            & \multicolumn{1}{c|}{1348.711}                                            & \multicolumn{1}{c|}{1362.364}                                            & 1367.667                                                                 \\ \hline
All gestures Mid-air       & \multicolumn{1}{c|}{1517.834}                                             & \multicolumn{1}{c|}{1428.447}                                            & \multicolumn{1}{c|}{1445.943}                                            & \multicolumn{1}{c|}{1393.449}                                            & \multicolumn{1}{c|}{1400.107}                                           & 1390.451                                                                 \\ \hline
Swipe up On-skin           & \multicolumn{1}{c|}{1464.479}                                            & \multicolumn{1}{c|}{1440.299}                                            & \multicolumn{1}{c|}{1402.672}                                            & \multicolumn{1}{c|}{1327.588}                                            & \multicolumn{1}{c|}{1380.54}                                             & 1347.581                                                                 \\ \hline
Swipe up Mid-air           & \multicolumn{1}{c|}{1613.46}                                             & \multicolumn{1}{c|}{1509.911}                                            & \multicolumn{1}{c|}{1538.699}                                            & \multicolumn{1}{c|}{1609.951}                                            & \multicolumn{1}{c|}{1556.02}                                             & 1528.444                                                                 \\ \hline
Swipe up On-skin           & \multicolumn{1}{c|}{1517.034}                                            & \multicolumn{1}{c|}{1418.078}                                            & \multicolumn{1}{c|}{1486.041}                                            & \multicolumn{1}{c|}{1517.034}                                            & \multicolumn{1}{c|}{1418.078}                                            & 1486.041                                                                 \\ \hline
Swipe up Mid-air           & \multicolumn{1}{c|}{1703.035}                                            & \multicolumn{1}{c|}{1512.876}                                            & \multicolumn{1}{c|}{1551.334}                                            & \multicolumn{1}{c|}{1703.035}                                            & \multicolumn{1}{c|}{1512.876}                                            & 1551.334                                                                 \\ \hline
Swipe front On-skin        & \multicolumn{1}{c|}{1465.024}                                            & \multicolumn{1}{c|}{1544.209}                                            & \multicolumn{1}{c|}{1558.211}                                            & \multicolumn{1}{c|}{1476.635}                                            & \multicolumn{1}{c|}{1484.915}                                            & 1476.364                                                                 \\ \hline
Swipe front Mid-air        & \multicolumn{1}{c|}{1738.389}                                            & \multicolumn{1}{c|}{1675.337}                                            & \multicolumn{1}{c|}{1612.322}                                            & \multicolumn{1}{c|}{1739.876}                                            & \multicolumn{1}{c|}{1627.532}                                            & 1597.57                                                                  \\ \hline
Swipe back On-skin         & \multicolumn{1}{c|}{1511.529}                                            & \multicolumn{1}{c|}{1424.764}                                            & \multicolumn{1}{c|}{1480.937}                                            & \multicolumn{1}{c|}{1472.201}                                            & \multicolumn{1}{c|}{1408.686}                                            & 1436.66                                                                  \\ \hline
Swipe back Mid-air         & \multicolumn{1}{c|}{1590.3}                                              & \multicolumn{1}{c|}{1502.576}                                            & \multicolumn{1}{c|}{1550.981}                                            & \multicolumn{1}{c|}{1484.618}                                            & \multicolumn{1}{c|}{1507.982}                                            & 1476.923                                                                 \\ \hline
Pinch in On-skin           & \multicolumn{1}{c|}{1333.953}                                            & \multicolumn{1}{c|}{1236.612}                                            & \multicolumn{1}{c|}{1229.839}                                            & \multicolumn{1}{c|}{1243.066}                                            & \multicolumn{1}{c|}{1228.667}                                            & 1197.21                                                                  \\ \hline
Pinch in Mid-air           & \multicolumn{1}{c|}{1146.523}                                            & \multicolumn{1}{c|}{1160.207}                                            & \multicolumn{1}{c|}{1154.841}                                            & \multicolumn{1}{c|}{1117.967}                                            & \multicolumn{1}{c|}{1131.603}                                            & 1093.265                                                                 \\ \hline
Pinch out On-skin          & \multicolumn{1}{c|}{1269.022}                                            & \multicolumn{1}{c|}{1237.692}                                            & \multicolumn{1}{c|}{1260.57}                                             & \multicolumn{1}{c|}{1223.834}                                            & \multicolumn{1}{c|}{1239.05}                                             & 1242.855                                                                 \\ \hline
Pinch out Mid-air          & \multicolumn{1}{c|}{1277.779}                                            & \multicolumn{1}{c|}{1134.834}                                            & \multicolumn{1}{c|}{1247.689}                                            & \multicolumn{1}{c|}{1235.037}                                            & \multicolumn{1}{c|}{1108.712}                                            & 1167.867                                                                 \\ \hline
Tap On-skin               & \multicolumn{1}{c|}{1393.713}                                            & \multicolumn{1}{c|}{1436.913}                                            & \multicolumn{1}{c|}{1448.494}                                            & \multicolumn{1}{c|}{1415.095}                                            & \multicolumn{1}{c|}{1440.751}                                            & 1417.567                                                                 \\ \hline
Tap Mid-air               & \multicolumn{1}{c|}{1555.35}                                             & \multicolumn{1}{c|}{1503.385}                                            & \multicolumn{1}{c|}{1465.733}                                            & \multicolumn{1}{c|}{1413.159}                                            & \multicolumn{1}{c|}{1510.262}                                            & 1462.826                                                                 \\ \hline
\end{tabular}%
\captionsetup{justification=centering}
\caption{Mean and median gesture path length (DV2) in millimeters (mm)}
\label{tab:DV2_mean_and_median}
\end{table}

\vspace{-0.5cm}
\begin{table}[H]
\centering
\tiny
\begin{tabular}{|c|llllll|}
\hline
\multirow{3}{*}{Gesture} & \multicolumn{6}{c|}{\textbf{Gesture Accuracy (DV3)}}                                                                                                                                                                                                                                                                                                                                                                                                            \\ \cline{2-7} 
                         & \multicolumn{3}{c|}{Mean}                                                                                                                                                                                                      & \multicolumn{3}{c|}{Median}                                                                                                                                                                                                    \\ \cline{2-7} 
                         & \multicolumn{1}{c|}{3 Regions} & \multicolumn{1}{c|}{5 Regions} & \multicolumn{1}{c|}{7 Regions} & \multicolumn{1}{c|}{3 Regions} & \multicolumn{1}{c|}{5 Regions} & \multicolumn{1}{c|}{7 Regions} \\ \hline
All gestures On-skin       & \multicolumn{1}{c|}{95.767}                                              & \multicolumn{1}{c|}{91.746}                                              & \multicolumn{1}{c|}{81.406}                                               & \multicolumn{1}{c|}{100}                                                 & \multicolumn{1}{c|}{90.0}                                                 & \multicolumn{1}{c|}{78.571}                                                                   \\ \hline
All gestures Mid-air       & \multicolumn{1}{c|}{94.974}                                              & \multicolumn{1}{c|}{84.762}                                              & \multicolumn{1}{c|}{78.061}                                              & \multicolumn{1}{c|}{100}                                                 & \multicolumn{1}{c|}{90.0}                                                  & \multicolumn{1}{c|}{78.571}                                                                   \\ \hline
Swipe up On-skin           & \multicolumn{1}{c|}{99.074}                                              & \multicolumn{1}{c|}{95.556}                                              & \multicolumn{1}{c|}{82.54}                                               & \multicolumn{1}{c|}{100}                                                 & \multicolumn{1}{c|}{100}                                                 & \multicolumn{1}{c|}{85.714}                                                                   \\ \hline
Swipe up Mid-air           & \multicolumn{1}{c|}{93.518}                                              & \multicolumn{1}{c|}{81.667}                                              & \multicolumn{1}{c|}{78.968}                                              & \multicolumn{1}{c|}{100}                                                 & \multicolumn{1}{c|}{80}                                                  & \multicolumn{1}{c|}{78.571}                                                                   \\ \hline
Swipe up On-skin           & \multicolumn{1}{c|}{96.296}                                              & \multicolumn{1}{c|}{93.889}                                              & \multicolumn{1}{c|}{83.333}                                              & \multicolumn{1}{c|}{100}                                                 & \multicolumn{1}{c|}{90}                                                  & \multicolumn{1}{c|}{85.714}                                                                   \\ \hline
Swipe up Mid-air           & \multicolumn{1}{c|}{97.222}                                              & \multicolumn{1}{c|}{88.333}                                              & \multicolumn{1}{c|}{77.381}                                              & \multicolumn{1}{c|}{100}                                                 & \multicolumn{1}{c|}{90}                                                  & \multicolumn{1}{c|}{78.571}                                                                   \\ \hline
Swipe front On-skin        & \multicolumn{1}{c|}{97.222}                                              & \multicolumn{1}{c|}{94.444}                                              & \multicolumn{1}{c|}{83.73}                                               & \multicolumn{1}{c|}{100}                                                 & \multicolumn{1}{c|}{90}                                                  & \multicolumn{1}{c|}{85.714}                                                                   \\ \hline
Swipe front Mid-air        & \multicolumn{1}{c|}{97.222}                                              & \multicolumn{1}{c|}{86.111}                                              & \multicolumn{1}{c|}{75.793}                                              & \multicolumn{1}{c|}{100}                                                 & \multicolumn{1}{c|}{90}                                                  & \multicolumn{1}{c|}{71.429}                                                                  \\ \hline
Swipe back On-skin         & \multicolumn{1}{c|}{89.815}                                              & \multicolumn{1}{c|}{82.778}                                              & \multicolumn{1}{c|}{78.968}                                              & \multicolumn{1}{c|}{83.333}                                              & \multicolumn{1}{c|}{85}                                                  & \multicolumn{1}{c|}{75.0}                                                                      \\ \hline
Swipe back Mid-air         & \multicolumn{1}{c|}{96.296}                                              & \multicolumn{1}{c|}{87.778}                                              & \multicolumn{1}{c|}{86.905}                                              & \multicolumn{1}{c|}{100}                                                 & \multicolumn{1}{c|}{90}                                                  & \multicolumn{1}{c|}{85.714}                                                                   \\ \hline
Pinch in On-skin           & \multicolumn{1}{c|}{98.148}                                              & \multicolumn{1}{c|}{91.667}                                              & \multicolumn{1}{c|}{80.158}                                              & \multicolumn{1}{c|}{100}                                                 & \multicolumn{1}{c|}{90}                                                  & \multicolumn{1}{c|}{78.571}                                                                   \\ \hline
Pinch in Mid-air           & \multicolumn{1}{c|}{91.667}                                              & \multicolumn{1}{c|}{82.775}                                              & \multicolumn{1}{c|}{76.19}                                               & \multicolumn{1}{c|}{91.667}                                              & \multicolumn{1}{c|}{80}                                                  & \multicolumn{1}{c|}{71.429}                                                                   \\ \hline
Pinch out On-skin          & \multicolumn{1}{c|}{91.667}                                              & \multicolumn{1}{c|}{90.556}                                              & \multicolumn{1}{c|}{78.968}                                              & \multicolumn{1}{c|}{91.667}                                              & \multicolumn{1}{c|}{90}                                                  & \multicolumn{1}{c|}{78.871}                                                                   \\ \hline
Pinch out Mid-air          & \multicolumn{1}{c|}{92.593}                                              & \multicolumn{1}{c|}{80.556}                                              & \multicolumn{1}{c|}{78.147}                                              & \multicolumn{1}{c|}{100}                                                 & \multicolumn{1}{c|}{80}                                                  & \multicolumn{1}{c|}{78.571}                                                                   \\ \hline
Tap On-skin               & \multicolumn{1}{c|}{98.148}                                              & \multicolumn{1}{c|}{93.333}                                              & \multicolumn{1}{c|}{82.143}                                              & \multicolumn{1}{c|}{100}                                                 & \multicolumn{1}{c|}{90}                                                  & \multicolumn{1}{c|}{78.571}                                                                   \\ \hline
Tap Mid-air               & \multicolumn{1}{c|}{96.296}                                              & \multicolumn{1}{c|}{86.111}                                              & \multicolumn{1}{c|}{73.016}                                              & \multicolumn{1}{c|}{100}                                                 & \multicolumn{1}{c|}{90}                                                  & \multicolumn{1}{c|}{71.429}                                                                   \\ \hline
\end{tabular}%
\captionsetup{justification=centering}
\caption{Mean and median gesture accuracy (DV3) in percentage (\%).}
\label{tab:DV3_mean_and_median}
\end{table}

\newpage
\subsection{Gesture Time (DV1) Analysis}
\label{appendix:DV1}

\begin{table}[H]
\centering
\tiny
\begin{tabular}{|c|c|cccccccc|}
\hline
                             &                                                                                                             & \multicolumn{8}{c|}{\textbf{Gesture Time (DV1)}}                                                                                                                                                                                                                                                                                                                                                                                                                                     \\ \cline{3-10} 
                             &                                                                                                             & \multicolumn{2}{c|}{}                                                                                      & \multicolumn{2}{c|}{}                                                                                  & \multicolumn{4}{c|}{On-skin Vs Mid-air Gestures}                                                                                                                                                                                                                 \\ \cline{7-10} 
                             &                                                                                                             & \multicolumn{2}{c|}{\multirow{-2}{*}{\begin{tabular}[c]{@{}c@{}}Shapiro-Wilk\\ Test Results\end{tabular}}} & \multicolumn{2}{c|}{\multirow{-2}{*}{\begin{tabular}[c]{@{}c@{}}Bartlett\\ Test Results\end{tabular}}} & \multicolumn{2}{c|}{\begin{tabular}[c]{@{}c@{}}Unpaired T-test\\ Results for Normally \\ Distributed Data\end{tabular}} & \multicolumn{2}{c|}{\begin{tabular}[c]{@{}c@{}}Mann-Whitney U Test \\ Results for Data with \\ Non-Normal Distribution\end{tabular}} \\ \cline{3-10} 
\multirow{-4}{*}{Gesture}    & \multirow{-4}{*}{\begin{tabular}[c]{@{}c@{}}\# of Gesture\\ Regions in\\ Interaction \\ Space\end{tabular}} & \multicolumn{1}{c|}{$W$}                 & \multicolumn{1}{c|}{$p$}                                        & \multicolumn{1}{c|}{${\chi^2(1)}$}               & \multicolumn{1}{c|}{$p$}                            & \multicolumn{1}{c|}{$t$}                                 & \multicolumn{1}{c|}{$p$}                                     & \multicolumn{1}{c|}{$W$}                                                  & $p$                                                      \\ \hline
                             & 3                                                                                                           & \multicolumn{1}{c|}{0.994}               & \multicolumn{1}{c|}{$<0.001$}                                   & \multicolumn{1}{c|}{\cellcolor[HTML]{C0C0C0}-}   & \multicolumn{1}{c|}{\cellcolor[HTML]{C0C0C0}-}      & \multicolumn{1}{c|}{\cellcolor[HTML]{C0C0C0}-}           & \multicolumn{1}{c|}{\cellcolor[HTML]{C0C0C0}-}               & \multicolumn{1}{c|}{220569}                                               & $<0.001$                                                  \\ \cline{2-10} 
                             & 5                                                                                                           & \multicolumn{1}{c|}{0.989}               & \multicolumn{1}{c|}{$<0.001$}                                   & \multicolumn{1}{c|}{\cellcolor[HTML]{C0C0C0}-}   & \multicolumn{1}{c|}{\cellcolor[HTML]{C0C0C0}-}      & \multicolumn{1}{c|}{\cellcolor[HTML]{C0C0C0}-}           & \multicolumn{1}{c|}{\cellcolor[HTML]{C0C0C0}-}               & \multicolumn{1}{c|}{614211}                                               & $<0.001$                                                             \\ \cline{2-10} 
\multirow{-3}{*}{All gestures} & 7                                                                                                         & \multicolumn{1}{c|}{0.998}               & \multicolumn{1}{c|}{$<0.001$}                                   & \multicolumn{1}{c|}{\cellcolor[HTML]{C0C0C0}-}   & \multicolumn{1}{c|}{\cellcolor[HTML]{C0C0C0}-}      & \multicolumn{1}{c|}{\cellcolor[HTML]{C0C0C0}-}           & \multicolumn{1}{c|}{\cellcolor[HTML]{C0C0C0}-}               & \multicolumn{1}{c|}{1195028}                                              & $<0.001$                                  \\ \hline
                             & 3                                                                                                           & \multicolumn{1}{c|}{0.993}               & \multicolumn{1}{c|}{\cellcolor[HTML]{F8FF00}0.423}              & \multicolumn{1}{c|}{0.216}                       & \multicolumn{1}{c|}{\cellcolor[HTML]{F8FF00}0.642}  & \multicolumn{1}{c|}{$t(213.57) = 5.966$}                 & \multicolumn{1}{c|}{$<0.001$}                                & \multicolumn{1}{c|}{\cellcolor[HTML]{C0C0C0}-}                            & \cellcolor[HTML]{C0C0C0}-                                \\ \cline{2-10} 
                             & 5                                                                                                           & \multicolumn{1}{c|}{0.995}               & \multicolumn{1}{c|}{\cellcolor[HTML]{F8FF00}0.230}              & \multicolumn{1}{c|}{6.562}                       & \multicolumn{1}{c|}{0.010}                          & \multicolumn{1}{c|}{\cellcolor[HTML]{C0C0C0}-}           & \multicolumn{1}{c|}{\cellcolor[HTML]{C0C0C0}-}               & \multicolumn{1}{c|}{8509}                                                 & $<0.001$                                                 \\ \cline{2-10} 
\multirow{-3}{*}{Swipe up}    & 7                                                                                                           & \multicolumn{1}{c|}{0.998}               & \multicolumn{1}{c|}{\cellcolor[HTML]{F8FF00}0.738}              & \multicolumn{1}{c|}{0.022}                       & \multicolumn{1}{c|}{\cellcolor[HTML]{F8FF00}0.881}  & \multicolumn{1}{c|}{$t(501.96) = 13.514$}                & \multicolumn{1}{c|}{$<0.001$}                                & \multicolumn{1}{c|}{\cellcolor[HTML]{C0C0C0}-}                           & \cellcolor[HTML]{C0C0C0}-                                \\ \hline
                             & 3                                                                                                           & \multicolumn{1}{c|}{0.984}               & \multicolumn{1}{c|}{0.016}                                      & \multicolumn{1}{c|}{\cellcolor[HTML]{C0C0C0}-}   & \multicolumn{1}{c|}{\cellcolor[HTML]{C0C0C0}-}      & \multicolumn{1}{c|}{\cellcolor[HTML]{C0C0C0}-}           & \multicolumn{1}{c|}{\cellcolor[HTML]{C0C0C0}-}               & \multicolumn{1}{c|}{2611}                                                 & $<0.001$                                                 \\ \cline{2-10} 
                             & 5                                                                                                           & \multicolumn{1}{c|}{0.973}               & \multicolumn{1}{c|}{$<0.001$}                                   & \multicolumn{1}{c|}{\cellcolor[HTML]{C0C0C0}-}   & \multicolumn{1}{c|}{\cellcolor[HTML]{C0C0C0}-}      & \multicolumn{1}{c|}{\cellcolor[HTML]{C0C0C0}-}           & \multicolumn{1}{c|}{\cellcolor[HTML]{C0C0C0}-}               & \multicolumn{1}{c|}{8544}                                                 & $<0.001$                                                 \\ \cline{2-10} 
\multirow{-3}{*}{Swipe down}  & 7                                                                                                           & \multicolumn{1}{c|}{0.997}               & \multicolumn{1}{c|}{\cellcolor[HTML]{F8FF00}0.414}              & \multicolumn{1}{c|}{0.129}                       & \multicolumn{1}{c|}{\cellcolor[HTML]{F8FF00}0.719}  & \multicolumn{1}{c|}{$t(501.74) = 7.008$}                 & \multicolumn{1}{c|}{$<0.001$}                                & \multicolumn{1}{c|}{\cellcolor[HTML]{C0C0C0}-}                           & \cellcolor[HTML]{C0C0C0}-                                \\ \hline
                             & 3                                                                                                           & \multicolumn{1}{c|}{0.990}               & \multicolumn{1}{c|}{\cellcolor[HTML]{F8FF00}0.162}              & \multicolumn{1}{c|}{0.348}                       & \multicolumn{1}{c|}{\cellcolor[HTML]{F8FF00}0.555}  & \multicolumn{1}{c|}{$t(213.3) =  2.064$}                 & \multicolumn{1}{c|}{0.040}                                   & \multicolumn{1}{c|}{\cellcolor[HTML]{C0C0C0}-}                            & \cellcolor[HTML]{C0C0C0}-                                \\ \cline{2-10} 
                             & 5                                                                                                           & \multicolumn{1}{c|}{0.997}               & \multicolumn{1}{c|}{\cellcolor[HTML]{F8FF00}0.842}              & \multicolumn{1}{c|}{37.386}                      & \multicolumn{1}{c|}{$<0.001$}                       & \multicolumn{1}{c|}{\cellcolor[HTML]{C0C0C0}-}           & \multicolumn{1}{c|}{\cellcolor[HTML]{C0C0C0}-}               & \multicolumn{1}{c|}{14043}                                                & 0.03                                                     \\ \cline{2-10} 
\multirow{-3}{*}{Swipe front} & 7                                                                                                           & \multicolumn{1}{c|}{0.997}               & \multicolumn{1}{c|}{\cellcolor[HTML]{F8FF00}0.411}              & \multicolumn{1}{c|}{0.818}                       & \multicolumn{1}{c|}{\cellcolor[HTML]{F8FF00}0.366}  & \multicolumn{1}{c|}{$t(500.37) =  1.458$}                & \multicolumn{1}{c|}{\cellcolor[HTML]{F8FF00}0.146}           & \multicolumn{1}{c|}{\cellcolor[HTML]{C0C0C0}-}                           & \cellcolor[HTML]{C0C0C0}-                                \\ \hline
                             & 3                                                                                                           & \multicolumn{1}{c|}{0.995}               & \multicolumn{1}{c|}{\cellcolor[HTML]{F8FF00}0.641}              & \multicolumn{1}{c|}{7.973}                       & \multicolumn{1}{c|}{0.005}                          & \multicolumn{1}{c|}{\cellcolor[HTML]{C0C0C0}-}           & \multicolumn{1}{c|}{\cellcolor[HTML]{C0C0C0}-}               & \multicolumn{1}{c|}{6359}                                                 & \cellcolor[HTML]{F8FF00}0.252                           \\ \cline{2-10} 
                             & 5                                                                                                           & \multicolumn{1}{c|}{0.993}               & \multicolumn{1}{c|}{\cellcolor[HTML]{F8FF00}0.068}              & \multicolumn{1}{c|}{35.707}                      & \multicolumn{1}{c|}{$<0.001$}                       & \multicolumn{1}{c|}{\cellcolor[HTML]{C0C0C0}-}           & \multicolumn{1}{c|}{\cellcolor[HTML]{C0C0C0}-}               & \multicolumn{1}{c|}{14972}                                                & \cellcolor[HTML]{F8FF00}0.214                            \\ \cline{2-10} 
\multirow{-3}{*}{Swipe back}  & 7                                                                                                           & \multicolumn{1}{c|}{0.998}               & \multicolumn{1}{c|}{\cellcolor[HTML]{F8FF00}0.868}              & \multicolumn{1}{c|}{21.787}                      & \multicolumn{1}{c|}{$<0.001$}                       & \multicolumn{1}{c|}{\cellcolor[HTML]{C0C0C0}-}           & \multicolumn{1}{c|}{\cellcolor[HTML]{C0C0C0}-}               & \multicolumn{1}{c|}{31009}                                               & \cellcolor[HTML]{F8FF00}0.650                            \\ \hline
                             & 3                                                                                                           & \multicolumn{1}{c|}{0.994}               & \multicolumn{1}{c|}{\cellcolor[HTML]{F8FF00}0.548}              & \multicolumn{1}{c|}{2.589}                       & \multicolumn{1}{c|}{\cellcolor[HTML]{F8FF00}0.108}  & \multicolumn{1}{c|}{$t(208.98) =  3.703$}                & \multicolumn{1}{c|}{$<0.001$}                                & \multicolumn{1}{c|}{\cellcolor[HTML]{C0C0C0}-}                            & \cellcolor[HTML]{C0C0C0}-                                \\ \cline{2-10} 
                             & 5                                                                                                           & \multicolumn{1}{c|}{0.993}               & \multicolumn{1}{c|}{\cellcolor[HTML]{F8FF00}0.071}              & \multicolumn{1}{c|}{0.715}                       & \multicolumn{1}{c|}{\cellcolor[HTML]{F8FF00}0.398}  & \multicolumn{1}{c|}{$t(356.57) =  1.411$}                & \multicolumn{1}{c|}{\cellcolor[HTML]{F8FF00}0.159}           & \multicolumn{1}{c|}{\cellcolor[HTML]{C0C0C0}-}                            & \cellcolor[HTML]{C0C0C0}-                                \\ \cline{2-10} 
\multirow{-3}{*}{Pinch in}    & 7                                                                                                           & \multicolumn{1}{c|}{0.996}               & \multicolumn{1}{c|}{\cellcolor[HTML]{F8FF00}0.276}              & \multicolumn{1}{c|}{0.391}                       & \multicolumn{1}{c|}{\cellcolor[HTML]{F8FF00}0.532}  & \multicolumn{1}{c|}{$t(501.22) =  2.423$}                & \multicolumn{1}{c|}{0.016}                                   & \multicolumn{1}{c|}{\cellcolor[HTML]{C0C0C0}-}                           & \cellcolor[HTML]{C0C0C0}-                                \\ \hline
                             & 3                                                                                                           & \multicolumn{1}{c|}{0.989}               & \multicolumn{1}{c|}{\cellcolor[HTML]{F8FF00}0.091}              & \multicolumn{1}{c|}{1.792}                       & \multicolumn{1}{c|}{\cellcolor[HTML]{F8FF00}0.181}  & \multicolumn{1}{c|}{$t(210.49) =  4.496$}                & \multicolumn{1}{c|}{$<0.001$}                                & \multicolumn{1}{c|}{\cellcolor[HTML]{C0C0C0}-}                            & \cellcolor[HTML]{C0C0C0}-                                \\ \cline{2-10} 
                             & 5                                                                                                           & \multicolumn{1}{c|}{0.997}               & \multicolumn{1}{c|}{\cellcolor[HTML]{F8FF00}0.629}              & \multicolumn{1}{c|}{0.703}                       & \multicolumn{1}{c|}{\cellcolor[HTML]{F8FF00}0.402}  & \multicolumn{1}{c|}{$t(356.6) =  5.021$}                 & \multicolumn{1}{c|}{$<0.001$}                                & \multicolumn{1}{c|}{\cellcolor[HTML]{C0C0C0}-}                            & \cellcolor[HTML]{C0C0C0}-                                \\ \cline{2-10} 
\multirow{-3}{*}{Pinch out}   & 7                                                                                                           & \multicolumn{1}{c|}{0.997}               & \multicolumn{1}{c|}{\cellcolor[HTML]{F8FF00}0.601}              & \multicolumn{1}{c|}{0.337}                       & \multicolumn{1}{c|}{\cellcolor[HTML]{F8FF00}0.562}  & \multicolumn{1}{c|}{$t(501.33) =  10.105$}               & \multicolumn{1}{c|}{$<0.001$}                                & \multicolumn{1}{c|}{\cellcolor[HTML]{C0C0C0}-}                           & \cellcolor[HTML]{C0C0C0}-                                \\ \hline
                             & 3                                                                                                           & \multicolumn{1}{c|}{0.995}               & \multicolumn{1}{c|}{\cellcolor[HTML]{F8FF00}0.756}              & \multicolumn{1}{c|}{3.408}                       & \multicolumn{1}{c|}{\cellcolor[HTML]{F8FF00}0.065}  & \multicolumn{1}{c|}{$t(207.47) =  4.969$}                & \multicolumn{1}{c|}{$<0.001$}                                & \multicolumn{1}{c|}{\cellcolor[HTML]{C0C0C0}-}                            & \cellcolor[HTML]{C0C0C0}-                                \\ \cline{2-10} 
                             & 5                                                                                                           & \multicolumn{1}{c|}{0.994}               & \multicolumn{1}{c|}{\cellcolor[HTML]{F8FF00}0.179}              & \multicolumn{1}{c|}{36.576}                      & \multicolumn{1}{c|}{$<0.001$}                       & \multicolumn{1}{c|}{\cellcolor[HTML]{C0C0C0}-}           & \multicolumn{1}{c|}{\cellcolor[HTML]{C0C0C0}-}               & \multicolumn{1}{c|}{8214}                                                 & $<0.001$                                                 \\ \cline{2-10} 
\multirow{-3}{*}{Tap}        & 7                                                                                                           & \multicolumn{1}{c|}{0.994}               & \multicolumn{1}{c|}{0.022}                                      & \multicolumn{1}{c|}{31.425}                      & \multicolumn{1}{c|}{$<0.001$}                       & \multicolumn{1}{c|}{\cellcolor[HTML]{C0C0C0}-}           & \multicolumn{1}{c|}{\cellcolor[HTML]{C0C0C0}-}               & \multicolumn{1}{c|}{17018}                                                & $<0.001$                                                 \\ \hline
\end{tabular}%
\captionsetup{justification=centering}
\caption{Statistical Test Report for analyzing the effect of interaction space choice on gesture time (DV1) \rev{for} different number of gesture (Swipe/Pinch/Tap) regions around the ear. \\Yellow boxes represent statistical insignificance.}
\label{tab:DV1_vs_RQ1}
\end{table}

\vspace*{-1 cm}
\begin{table}[H]
\centering
\tiny
\begin{tabular}{|c|ccccccccccc|}
\hline
                          & \multicolumn{11}{c|}{\textbf{Gesture Time (DV1)}}                                                                                                                                                                                                                                                                                                                                                                                                                                                                                                                                                                                                                                                                        \\ \cline{2-12} 
                          & \multicolumn{2}{c|}{\begin{tabular}[c]{@{}c@{}}Shapiro- Wilk\\ Test Results\end{tabular}} & \multicolumn{2}{c|}{\begin{tabular}[c]{@{}c@{}}Bartlett- \\ Test Results\end{tabular}}              & \multicolumn{2}{c|}{\begin{tabular}[c]{@{}c@{}}One-Way ANOVA \\ Results for Normally \\ Distributed Data\end{tabular}} & \multicolumn{2}{c|}{\begin{tabular}[c]{@{}c@{}}Kruskal- Wallis Test \\ Results for Data with \\ Non-Normal Distribution\end{tabular}} & \multicolumn{3}{c|}{\begin{tabular}[c]{@{}c@{}}Post-hoc \\ Test Results\end{tabular}}                                                                                                                                                                 \\ \cline{2-12} 
\multirow{-3}{*}{Gesture} & \multicolumn{1}{c|}{$W$}        & \multicolumn{1}{c|}{$p$}                                & \multicolumn{1}{c|}{${\chi^2(2)}$}             & \multicolumn{1}{c|}{$p$}                           & \multicolumn{1}{c|}{$F(2, 502)$}                           & \multicolumn{1}{c|}{$p$}                                  & \multicolumn{1}{c|}{${\chi^2(2)}$}                                & \multicolumn{1}{c|}{$p$}                                          & \multicolumn{1}{c|}{\begin{tabular}[c]{@{}c@{}}3 Region \\ vs \\ 5 Region\end{tabular}} & \multicolumn{1}{c|}{\begin{tabular}[c]{@{}c@{}}5 Region \\ vs\\ 7 Region\end{tabular}} & \begin{tabular}[c]{@{}c@{}}3 Region \\ vs \\ 7 Region\end{tabular} \\ \hline
All gestures On-skin        & \multicolumn{1}{c|}{0.999}      & \multicolumn{1}{c|}{\cellcolor[HTML]{F8FF00}0.097}      & \multicolumn{1}{c|}{54.765}                     & \multicolumn{1}{c|}{$<0.001$}                         & \multicolumn{1}{c|}{\cellcolor[HTML]{C0C0C0}-}             & \multicolumn{1}{c|}{\cellcolor[HTML]{C0C0C0}-}            & \multicolumn{1}{c|}{157.96}                                       & \multicolumn{1}{c|}{\textless{}0.001}                             & \multicolumn{1}{c|}{\textless{}0.001}                                                   & \multicolumn{1}{c|}{\textless{}0.001}                                                  & \textless{}0.001                                                   \\ \hline
All gestures Mid-air        & \multicolumn{1}{c|}{0.994}      & \multicolumn{1}{c|}{$<0.001$}                           & \multicolumn{1}{c|}{\cellcolor[HTML]{C0C0C0}-} & \multicolumn{1}{c|}{\cellcolor[HTML]{C0C0C0}-}     & \multicolumn{1}{c|}{\cellcolor[HTML]{C0C0C0}-}             & \multicolumn{1}{c|}{\cellcolor[HTML]{C0C0C0}-}            & \multicolumn{1}{c|}{210.23}                                       & \multicolumn{1}{c|}{\textless{}0.001}                             & \multicolumn{1}{c|}{\textless{}0.001}                                                   & \multicolumn{1}{c|}{\textless{}0.001}                                                  & \textless{}0.001                                                   \\ \hline
Swipe up On-skin            & \multicolumn{1}{c|}{0.992}      & \multicolumn{1}{c|}{0.004}                              & \multicolumn{1}{c|}{\cellcolor[HTML]{C0C0C0}-} & \multicolumn{1}{c|}{\cellcolor[HTML]{C0C0C0}-}     & \multicolumn{1}{c|}{\cellcolor[HTML]{C0C0C0}-}             & \multicolumn{1}{c|}{\cellcolor[HTML]{C0C0C0}-}            & \multicolumn{1}{c|}{102.32}                                       & \multicolumn{1}{c|}{\textless{}0.001}                             & \multicolumn{1}{c|}{\textless{}0.001}                                                   & \multicolumn{1}{c|}{\textless{}0.001}                                                  & \textless{}0.001                                                   \\ \hline
Swipe up Mid-air            & \multicolumn{1}{c|}{0.997}      & \multicolumn{1}{c|}{\cellcolor[HTML]{F8FF00}0.570}      & \multicolumn{1}{c|}{9.657}                     & \multicolumn{1}{c|}{0.008}                         & \multicolumn{1}{c|}{\cellcolor[HTML]{C0C0C0}-}             & \multicolumn{1}{c|}{\cellcolor[HTML]{C0C0C0}-}            & \multicolumn{1}{c|}{27.286}                                       & \multicolumn{1}{c|}{\textless{}0.001}                             & \multicolumn{1}{c|}{\textless{}0.001}                                                   & \multicolumn{1}{c|}{\cellcolor[HTML]{F8FF00}0.72}                                      & \textless{}0.001                                                   \\ \hline
Swipe down On-skin          & \multicolumn{1}{c|}{0.987}      & \multicolumn{1}{c|}{\textless{}0.001}                   & \multicolumn{1}{c|}{\cellcolor[HTML]{C0C0C0}-} & \multicolumn{1}{c|}{\cellcolor[HTML]{C0C0C0}-}     & \multicolumn{1}{c|}{\cellcolor[HTML]{C0C0C0}-}             & \multicolumn{1}{c|}{\cellcolor[HTML]{C0C0C0}-}            & \multicolumn{1}{c|}{27.458}                                       & \multicolumn{1}{c|}{\textless{}0.001}                             & \multicolumn{1}{c|}{\cellcolor[HTML]{F8FF00}0.252}                                      & \multicolumn{1}{c|}{\textless{}0.001}                                                  & {$<0.001$}                                                              \\ \hline
Swipe down Mid-air          & \multicolumn{1}{c|}{0.995}      & \multicolumn{1}{c|}{\cellcolor[HTML]{F8FF00}0.051}      & \multicolumn{1}{c|}{48.396}                    & \multicolumn{1}{c|}{\textless{}0.001}              & \multicolumn{1}{c|}{\cellcolor[HTML]{C0C0C0}-}             & \multicolumn{1}{c|}{\cellcolor[HTML]{C0C0C0}-}            & \multicolumn{1}{c|}{165.79}                                       & \multicolumn{1}{c|}{\textless{}0.001}                             & \multicolumn{1}{c|}{\textless{}0.001}                                                   & \multicolumn{1}{c|}{\textless{}0.001}                                                  & \textless{}0.001                                                   \\ \hline
Swipe front On-skin         & \multicolumn{1}{c|}{0.992}      & \multicolumn{1}{c|}{0.007}                              & \multicolumn{1}{c|}{\cellcolor[HTML]{C0C0C0}-} & \multicolumn{1}{c|}{\cellcolor[HTML]{C0C0C0}-}     & \multicolumn{1}{c|}{\cellcolor[HTML]{C0C0C0}-}             & \multicolumn{1}{c|}{\cellcolor[HTML]{C0C0C0}-}            & \multicolumn{1}{c|}{48.043}                                       & \multicolumn{1}{c|}{\textless{}0.001}                             & \multicolumn{1}{c|}{0.041}                                                              & \multicolumn{1}{c|}{\textless{}0.001}                                                  & \textless{}0.001                                                   \\ \hline
Swipe front Mid-air         & \multicolumn{1}{c|}{0.993}      & \multicolumn{1}{c|}{0.009}                              & \multicolumn{1}{c|}{\cellcolor[HTML]{C0C0C0}-} & \multicolumn{1}{c|}{\cellcolor[HTML]{C0C0C0}-}     & \multicolumn{1}{c|}{\cellcolor[HTML]{C0C0C0}-}             & \multicolumn{1}{c|}{\cellcolor[HTML]{C0C0C0}-}            & \multicolumn{1}{c|}{100.79}                                       & \multicolumn{1}{c|}{\textless{}0.001}                             & \multicolumn{1}{c|}{\textless{}0.001}                                                   & \multicolumn{1}{c|}{\textless{}0.001}                                                  & \textless{}0.001                                                   \\ \hline
Swipe back On-skin          & \multicolumn{1}{c|}{0.991}      & \multicolumn{1}{c|}{0.003}                              & \multicolumn{1}{c|}{\cellcolor[HTML]{C0C0C0}-} & \multicolumn{1}{c|}{\cellcolor[HTML]{C0C0C0}-}     & \multicolumn{1}{c|}{\cellcolor[HTML]{C0C0C0}-}             & \multicolumn{1}{c|}{\cellcolor[HTML]{C0C0C0}-}            & \multicolumn{1}{c|}{42.3}                                         & \multicolumn{1}{c|}{\textless{}0.001}                             & \multicolumn{1}{c|}{\textless{}0.001}                                                   & \multicolumn{1}{c|}{0.019}                                                             & \textless{}0.001                                                   \\ \hline
Swipe back Mid-air          & \multicolumn{1}{c|}{0.998}      & \multicolumn{1}{c|}{\cellcolor[HTML]{F8FF00}0.647}      & \multicolumn{1}{c|}{36.365}                    & \multicolumn{1}{c|}{\textless{}0.001}              & \multicolumn{1}{c|}{\cellcolor[HTML]{C0C0C0}-}             & \multicolumn{1}{c|}{\cellcolor[HTML]{C0C0C0}-}            & \multicolumn{1}{c|}{31.522}                                       & \multicolumn{1}{c|}{\textless{}0.001}                             & \multicolumn{1}{c|}{0.010}                                                              & \multicolumn{1}{c|}{0.002}                                                             & \textless{}0.001                                                   \\ \hline
Pinch in On-skin            & \multicolumn{1}{c|}{0.996}      & \multicolumn{1}{c|}{\cellcolor[HTML]{F8FF00}0.289}      & \multicolumn{1}{c|}{0.286}                     & \multicolumn{1}{c|}{\cellcolor[HTML]{F8FF00}0.867} & \multicolumn{1}{c|}{8.094}                                 & \multicolumn{1}{c|}{\textless{}0.001}                     & \multicolumn{1}{c|}{\cellcolor[HTML]{C0C0C0}-}                    & \multicolumn{1}{c|}{\cellcolor[HTML]{C0C0C0}-}                    & \multicolumn{1}{c|}{\cellcolor[HTML]{F8FF00}1.0}                                        & \multicolumn{1}{c|}{0.010}                                                             & 0.046                                                              \\ \hline
Pinch in Mid-air            & \multicolumn{1}{c|}{0.995}      & \multicolumn{1}{c|}{\cellcolor[HTML]{F8FF00}0.060}       & \multicolumn{1}{c|}{2.658}                     & \multicolumn{1}{c|}{\cellcolor[HTML]{F8FF00}0.265} & \multicolumn{1}{c|}{20.701}                                & \multicolumn{1}{c|}{\textless{}0.001}                     & \multicolumn{1}{c|}{\cellcolor[HTML]{C0C0C0}-}                    & \multicolumn{1}{c|}{\cellcolor[HTML]{C0C0C0}-}                    & \multicolumn{1}{c|}{0.015}                                                              & \multicolumn{1}{c|}{\textless{}0.001}                                                  & \cellcolor[HTML]{F8FF00}0.071                                      \\ \hline
Pinch out On-skin           & \multicolumn{1}{c|}{0.998}      & \multicolumn{1}{c|}{\cellcolor[HTML]{F8FF00}0.672}      & \multicolumn{1}{c|}{2.725}                     & \multicolumn{1}{c|}{\cellcolor[HTML]{F8FF00}0.256} & \multicolumn{1}{c|}{42.535}                                & \multicolumn{1}{c|}{\textless{}0.001}                     & \multicolumn{1}{c|}{\cellcolor[HTML]{C0C0C0}-}                    & \multicolumn{1}{c|}{\cellcolor[HTML]{C0C0C0}-}                    & \multicolumn{1}{c|}{\cellcolor[HTML]{F8FF00}0.112}                                      & \multicolumn{1}{c|}{\textless{}0.001}                                                  & \textless{}0.001                                                   \\ \hline
Pinch out Mid-air           & \multicolumn{1}{c|}{0.998}      & \multicolumn{1}{c|}{\cellcolor[HTML]{F8FF00}0.913}      & \multicolumn{1}{c|}{1.864}                     & \multicolumn{1}{c|}{\cellcolor[HTML]{F8FF00}0.394} & \multicolumn{1}{c|}{57.640}                                & \multicolumn{1}{c|}{\textless{}0.001}                     & \multicolumn{1}{c|}{\cellcolor[HTML]{C0C0C0}-}                    & \multicolumn{1}{c|}{\cellcolor[HTML]{C0C0C0}-}                    & \multicolumn{1}{c|}{0.008}                                                              & \multicolumn{1}{c|}{\cellcolor[HTML]{F8FF00}1.0}                                       & 0.002                                                              \\ \hline
Tap On-skin                & \multicolumn{1}{c|}{0.974}      & \multicolumn{1}{c|}{\textless{}0.001}                   & \multicolumn{1}{c|}{\cellcolor[HTML]{C0C0C0}-} & \multicolumn{1}{c|}{\cellcolor[HTML]{C0C0C0}-}     & \multicolumn{1}{c|}{\cellcolor[HTML]{C0C0C0}-}             & \multicolumn{1}{c|}{\cellcolor[HTML]{C0C0C0}-}            & \multicolumn{1}{c|}{50.330}                                       & \multicolumn{1}{c|}{\textless{}0.001}                             & \multicolumn{1}{c|}{0.025}                                                              & \multicolumn{1}{c|}{\textless{}0.001}                                                  & \textless{}0.001                                                   \\ \hline
Tap Mid-air                & \multicolumn{1}{c|}{0.996}      & \multicolumn{1}{c|}{\cellcolor[HTML]{F8FF00}0.117}      & \multicolumn{1}{c|}{47.823}                    & \multicolumn{1}{c|}{\textless{}0.001}              & \multicolumn{1}{c|}{\cellcolor[HTML]{C0C0C0}-}             & \multicolumn{1}{c|}{\cellcolor[HTML]{C0C0C0}-}            & \multicolumn{1}{c|}{50.277}                                       & \multicolumn{1}{c|}{\textless{}0.001}                             & \multicolumn{1}{c|}{\textless{}0.001}                                                   & \multicolumn{1}{c|}{\textless{}0.001}                                                  & \textless{}0.001                                                   \\ \hline
\end{tabular}%
\captionsetup{justification=centering}
\caption{Statistical Test Report for analyzing the effect of increasing the number of regions around the ear on gesture time (DV1) for off-device mid-air and on-skin gesture (Swipe/Pinch/Tap) reuse.\\Yellow boxes represent statistical insignificance.}
\label{tab:DV1_vs_RQ2}
\end{table}

\newpage
\subsection{Gesture Path Length (DV2) Analysis}
\label{appendix:DV2}

\begin{table}[H]
\centering
\tiny
\begin{tabular}{|c|c|cccccccc|}
\hline
                             &                                                                                                             & \multicolumn{8}{c|}{\textbf{Gesture Path Length (DV2)}}                                                                                                                                                                                                                                                                                                                                                                                                                              \\ \cline{3-10} 
                             &                                                                                                             & \multicolumn{2}{c|}{}                                                                                      & \multicolumn{2}{c|}{}                                                                                  & \multicolumn{4}{c|}{Onskin Vs Mid-air Gestures}                                                                                                                                                                                                                 \\ \cline{7-10} 
                             &                                                                                                             & \multicolumn{2}{c|}{\multirow{-2}{*}{\begin{tabular}[c]{@{}c@{}}Shapiro-Wilk\\ Test Results\end{tabular}}} & \multicolumn{2}{c|}{\multirow{-2}{*}{\begin{tabular}[c]{@{}c@{}}Bartlett\\ Test Results\end{tabular}}} & \multicolumn{2}{c|}{\begin{tabular}[c]{@{}c@{}}Unpaired T-test\\ Results for Normally \\ Distributed Data\end{tabular}} & \multicolumn{2}{c|}{\begin{tabular}[c]{@{}c@{}}Mann-Whitney U Test \\ Results for Data with \\ Non-Normal Distribution\end{tabular}} \\ \cline{3-10} 
\multirow{-4}{*}{Gesture}    & \multirow{-4}{*}{\begin{tabular}[c]{@{}c@{}}\# of Gesture\\ Regions in\\ Interaction \\ Space\end{tabular}} & \multicolumn{1}{c|}{$W$}                           & \multicolumn{1}{c|}{$p$}                              & \multicolumn{1}{c|}{${\chi^2(1)}$}                 & \multicolumn{1}{c|}{$p$}                          & \multicolumn{1}{c|}{$t$}                                   & \multicolumn{1}{c|}{$p$}                                   & \multicolumn{1}{c|}{$W$}                                                  & $p$                                                      \\ \hline
                             & 3                                                                                                           & \multicolumn{1}{c|}{0.952}                         & \multicolumn{1}{c|}{$<0.001$}                         & \multicolumn{1}{c|}{\cellcolor[HTML]{C0C0C0}-}     & \multicolumn{1}{c|}{\cellcolor[HTML]{C0C0C0}-}    & \multicolumn{1}{c|}{\cellcolor[HTML]{C0C0C0}-}             & \multicolumn{1}{c|}{\cellcolor[HTML]{C0C0C0}-}             & \multicolumn{1}{c|}{306747}                                               & 0.012                                                    \\ \cline{2-10} 
                             & 5                                                                                                           & \multicolumn{1}{c|}{0.930}                         & \multicolumn{1}{c|}{$<0.001$}                         & \multicolumn{1}{c|}{\cellcolor[HTML]{C0C0C0}-}     & \multicolumn{1}{c|}{\cellcolor[HTML]{C0C0C0}-}    & \multicolumn{1}{c|}{\cellcolor[HTML]{C0C0C0}-}             & \multicolumn{1}{c|}{\cellcolor[HTML]{C0C0C0}-}             & \multicolumn{1}{c|}{829754}                                                 & 0.048                                                   \\ \cline{2-10} 
\multirow{-3}{*}{All gestures} & 7                                                                                                           & \multicolumn{1}{c|}{0.850}                        & \multicolumn{1}{c|}{$<0.001$}                         & \multicolumn{1}{c|}{\cellcolor[HTML]{C0C0C0}-}     & \multicolumn{1}{c|}{\cellcolor[HTML]{C0C0C0}-}    & \multicolumn{1}{c|}{\cellcolor[HTML]{C0C0C0}-}             & \multicolumn{1}{c|}{\cellcolor[HTML]{C0C0C0}-}             & \multicolumn{1}{c|}{1626208}                                                 & 0.020                                                     \\ \hline
                             & 3                                                                                                           & \multicolumn{1}{c|}{0.893}                         & \multicolumn{1}{c|}{$<0.001$}                         & \multicolumn{1}{c|}{\cellcolor[HTML]{C0C0C0}-}     & \multicolumn{1}{c|}{\cellcolor[HTML]{C0C0C0}-}    & \multicolumn{1}{c|}{\cellcolor[HTML]{C0C0C0}-}             & \multicolumn{1}{c|}{\cellcolor[HTML]{C0C0C0}-}             & \multicolumn{1}{c|}{7150}                                                 & 0.004                                                    \\ \cline{2-10} 
                             & 5                                                                                                           & \multicolumn{1}{c|}{0.991}                         & \multicolumn{1}{c|}{0.03}                             & \multicolumn{1}{c|}{\cellcolor[HTML]{C0C0C0}-}     & \multicolumn{1}{c|}{\cellcolor[HTML]{C0C0C0}-}    & \multicolumn{1}{c|}{\cellcolor[HTML]{C0C0C0}-}             & \multicolumn{1}{c|}{\cellcolor[HTML]{C0C0C0}-}             & \multicolumn{1}{c|}{18247}                                                & 0.038                                                    \\ \cline{2-10} 
\multirow{-3}{*}{Swipe up}    & 7                                                                                                           & \multicolumn{1}{c|}{0.984}                         & \multicolumn{1}{c|}{$<0.001$}                         & \multicolumn{1}{c|}{\cellcolor[HTML]{C0C0C0}-}     & \multicolumn{1}{c|}{\cellcolor[HTML]{C0C0C0}-}    & \multicolumn{1}{c|}{\cellcolor[HTML]{C0C0C0}-}             & \multicolumn{1}{c|}{\cellcolor[HTML]{C0C0C0}-}             & \multicolumn{1}{c|}{37674}                                                & $<0.001$                                                 \\ \hline
                             & 3                                                                                                           & \multicolumn{1}{c|}{0.957}                         & \multicolumn{1}{c|}{$<0.001$}                         & \multicolumn{1}{c|}{\cellcolor[HTML]{C0C0C0}-}     & \multicolumn{1}{c|}{\cellcolor[HTML]{C0C0C0}-}    & \multicolumn{1}{c|}{\cellcolor[HTML]{C0C0C0}-}             & \multicolumn{1}{c|}{\cellcolor[HTML]{C0C0C0}-}             & \multicolumn{1}{c|}{6560}                                                 & \cellcolor[HTML]{F8FF00}0.113                            \\ \cline{2-10} 
                             & 5                                                                                                           & \multicolumn{1}{c|}{0.995}                         & \multicolumn{1}{c|}{0.267}                            & \multicolumn{1}{c|}{5.906}                         & \multicolumn{1}{c|}{0.015}                        & \multicolumn{1}{c|}{\cellcolor[HTML]{C0C0C0}-}             & \multicolumn{1}{c|}{\cellcolor[HTML]{C0C0C0}-}             & \multicolumn{1}{c|}{18055}                                                & \cellcolor[HTML]{F8FF00}0.060                            \\ \cline{2-10} 
\multirow{-3}{*}{Swipe down}  & 7                                                                                                           & \multicolumn{1}{c|}{0.983}                         & \multicolumn{1}{c|}{$<0.001$}                         & \multicolumn{1}{c|}{\cellcolor[HTML]{C0C0C0}-}     & \multicolumn{1}{c|}{\cellcolor[HTML]{C0C0C0}-}    & \multicolumn{1}{c|}{\cellcolor[HTML]{C0C0C0}-}             & \multicolumn{1}{c|}{\cellcolor[HTML]{C0C0C0}-}             & \multicolumn{1}{c|}{34748}                                                & \cellcolor[HTML]{F8FF00}{0.067}                                                    \\ \hline
                             & 3                                                                                                           & \multicolumn{1}{c|}{0.985}                         & \multicolumn{1}{c|}{0.024}                            & \multicolumn{1}{c|}{\cellcolor[HTML]{C0C0C0}-}     & \multicolumn{1}{c|}{\cellcolor[HTML]{C0C0C0}-}    & \multicolumn{1}{c|}{\cellcolor[HTML]{C0C0C0}-}             & \multicolumn{1}{c|}{\cellcolor[HTML]{C0C0C0}-}             & \multicolumn{1}{c|}{7293}                                                 & {$<0.001$}                                                    \\ \cline{2-10} 
                             & 5                                                                                                           & \multicolumn{1}{c|}{0.796}                         & \multicolumn{1}{c|}{$<0.001$}                         & \multicolumn{1}{c|}{\cellcolor[HTML]{C0C0C0}-}     & \multicolumn{1}{c|}{\cellcolor[HTML]{C0C0C0}-}    & \multicolumn{1}{c|}{\cellcolor[HTML]{C0C0C0}-}             & \multicolumn{1}{c|}{\cellcolor[HTML]{C0C0C0}-}             & \multicolumn{1}{c|}{18469}                                                & 0.021                                                    \\ \cline{2-10} 
\multirow{-3}{*}{Swipe front} & 7                                                                                                           & \multicolumn{1}{c|}{0.986}                         & \multicolumn{1}{c|}{$<0.001$}                         & \multicolumn{1}{c|}{\cellcolor[HTML]{C0C0C0}-}     & \multicolumn{1}{c|}{\cellcolor[HTML]{C0C0C0}-}    & \multicolumn{1}{c|}{\cellcolor[HTML]{C0C0C0}-}             & \multicolumn{1}{c|}{\cellcolor[HTML]{C0C0C0}-}             & \multicolumn{1}{c|}{33820}                                                & \cellcolor[HTML]{F8FF00}0.206                            \\ \hline
                             & 3                                                                                                           & \multicolumn{1}{c|}{0.961}                         & \multicolumn{1}{c|}{$<0.001$}                         & \multicolumn{1}{c|}{\cellcolor[HTML]{C0C0C0}-}     & \multicolumn{1}{c|}{\cellcolor[HTML]{C0C0C0}-}    & \multicolumn{1}{c|}{\cellcolor[HTML]{C0C0C0}-}             & \multicolumn{1}{c|}{\cellcolor[HTML]{C0C0C0}-}             & \multicolumn{1}{c|}{6108}                                                 & \cellcolor[HTML]{F8FF00}0.549                            \\ \cline{2-10} 
                             & 5                                                                                                           & \multicolumn{1}{c|}{0.991}                         & \multicolumn{1}{c|}{0.048}                            & \multicolumn{1}{c|}{\cellcolor[HTML]{C0C0C0}-}     & \multicolumn{1}{c|}{\cellcolor[HTML]{C0C0C0}-}    & \multicolumn{1}{c|}{\cellcolor[HTML]{C0C0C0}-}             & \multicolumn{1}{c|}{\cellcolor[HTML]{C0C0C0}-}             & \multicolumn{1}{c|}{18080}                                                & \cellcolor[HTML]{F8FF00}0.056                            \\ \cline{2-10} 
\multirow{-3}{*}{Swipe back}  & 7                                                                                                           & \multicolumn{1}{c|}{0.507}                         & \multicolumn{1}{c|}{$<0.001$}                         & \multicolumn{1}{c|}{\cellcolor[HTML]{C0C0C0}-}     & \multicolumn{1}{c|}{\cellcolor[HTML]{C0C0C0}-}    & \multicolumn{1}{c|}{\cellcolor[HTML]{C0C0C0}-}             & \multicolumn{1}{c|}{\cellcolor[HTML]{C0C0C0}-}             & \multicolumn{1}{c|}{33408}                                                & \cellcolor[HTML]{F8FF00}0.311                            \\ \hline
                             & 3                                                                                                           & \multicolumn{1}{c|}{0.900}                         & \multicolumn{1}{c|}{$<0.001$}                         & \multicolumn{1}{c|}{\cellcolor[HTML]{C0C0C0}-}     & \multicolumn{1}{c|}{\cellcolor[HTML]{C0C0C0}-}    & \multicolumn{1}{c|}{\cellcolor[HTML]{C0C0C0}-}             & \multicolumn{1}{c|}{\cellcolor[HTML]{C0C0C0}-}             & \multicolumn{1}{c|}{4262}                                                 & $<0.001$                                                 \\ \cline{2-10} 
                             & 5                                                                                                           & \multicolumn{1}{c|}{0.997}                         & \multicolumn{1}{c|}{0.785}                            & \multicolumn{1}{c|}{1.2158}                        & \multicolumn{1}{c|}{0.270}                        & \multicolumn{1}{c|}{$t(355.59)=2.081$}                     & \multicolumn{1}{c|}{0.038}                                 & \multicolumn{1}{c|}{\cellcolor[HTML]{C0C0C0}-}                            & \cellcolor[HTML]{C0C0C0}-                                \\ \cline{2-10} 
\multirow{-3}{*}{Pinch in}    & 7                                                                                                           & \multicolumn{1}{c|}{0.986}                         & \multicolumn{1}{c|}{$<0.001$}                         & \multicolumn{1}{c|}{\cellcolor[HTML]{C0C0C0}-}     & \multicolumn{1}{c|}{\cellcolor[HTML]{C0C0C0}-}    & \multicolumn{1}{c|}{\cellcolor[HTML]{C0C0C0}-}             & \multicolumn{1}{c|}{\cellcolor[HTML]{C0C0C0}-}             & \multicolumn{1}{c|}{27424}                                                & 0.008                                                    \\ \hline
                             & 3                                                                                                           & \multicolumn{1}{c|}{0.962}                         & \multicolumn{1}{c|}{$<0.001$}                         & \multicolumn{1}{c|}{\cellcolor[HTML]{C0C0C0}-}     & \multicolumn{1}{c|}{\cellcolor[HTML]{C0C0C0}-}    & \multicolumn{1}{c|}{\cellcolor[HTML]{C0C0C0}-}             & \multicolumn{1}{c|}{\cellcolor[HTML]{C0C0C0}-}             & \multicolumn{1}{c|}{5789}                                                 & \cellcolor[HTML]{F8FF00}0.926                            \\ \cline{2-10} 
                             & 5                                                                                                           & \multicolumn{1}{c|}{0.985}                         & \multicolumn{1}{c|}{$<0.001$}                         & \multicolumn{1}{c|}{\cellcolor[HTML]{C0C0C0}-}     & \multicolumn{1}{c|}{\cellcolor[HTML]{C0C0C0}-}    & \multicolumn{1}{c|}{\cellcolor[HTML]{C0C0C0}-}             & \multicolumn{1}{c|}{\cellcolor[HTML]{C0C0C0}-}             & \multicolumn{1}{c|}{12731}                                                & $<0.001$                                                 \\ \cline{2-10} 
\multirow{-3}{*}{Pinch out}   & 7                                                                                                           & \multicolumn{1}{c|}{0.975}                         & \multicolumn{1}{c|}{$<0.001$}                         & \multicolumn{1}{c|}{\cellcolor[HTML]{C0C0C0}-}     & \multicolumn{1}{c|}{\cellcolor[HTML]{C0C0C0}-}    & \multicolumn{1}{c|}{\cellcolor[HTML]{C0C0C0}-}             & \multicolumn{1}{c|}{\cellcolor[HTML]{C0C0C0}-}             & \multicolumn{1}{c|}{30187}                                                & \cellcolor[HTML]{F8FF00}0.339                            \\ \hline
                             & 3                                                                                                           & \multicolumn{1}{c|}{0.960}                         & \multicolumn{1}{c|}{$<0.001$}                         & \multicolumn{1}{c|}{\cellcolor[HTML]{C0C0C0}-}     & \multicolumn{1}{c|}{\cellcolor[HTML]{C0C0C0}-}    & \multicolumn{1}{c|}{\cellcolor[HTML]{C0C0C0}-}             & \multicolumn{1}{c|}{\cellcolor[HTML]{C0C0C0}-}             & \multicolumn{1}{c|}{6375}                                                 & \cellcolor[HTML]{F8FF00}0.238                            \\ \cline{2-10} 
                             & 5                                                                                                           & \multicolumn{1}{c|}{0.970}                         & \multicolumn{1}{c|}{$<0.001$}                         & \multicolumn{1}{c|}{\cellcolor[HTML]{C0C0C0}-}     & \multicolumn{1}{c|}{\cellcolor[HTML]{C0C0C0}-}    & \multicolumn{1}{c|}{\cellcolor[HTML]{C0C0C0}-}             & \multicolumn{1}{c|}{\cellcolor[HTML]{C0C0C0}-}             & \multicolumn{1}{c|}{17406}                                                & \cellcolor[HTML]{F8FF00}0.222                            \\ \cline{2-10} 
\multirow{-3}{*}{Tap}        & 7                                                                                                           & \multicolumn{1}{c|}{0.992}                         & \multicolumn{1}{c|}{0.018}                            & \multicolumn{1}{c|}{\cellcolor[HTML]{C0C0C0}-}     & \multicolumn{1}{c|}{\cellcolor[HTML]{C0C0C0}-}    & \multicolumn{1}{c|}{\cellcolor[HTML]{C0C0C0}-}             & \multicolumn{1}{c|}{\cellcolor[HTML]{C0C0C0}-}             & \multicolumn{1}{c|}{32853}                                                & \cellcolor[HTML]{F8FF00}0.501                            \\ \hline
\end{tabular}%
\captionsetup{justification=centering}
\caption{Statistical Test Report for analyzing the effect of interaction space choice on gesture path length (DV2) \rev{for} different number of gesture (Swipe/Pinch/Tap) regions around the ear.\\Yellow boxes represent statistical insignificance.}
\label{tab:DV2_vs_RQ1}
\end{table}

\vspace{-0.5cm}
\begin{table}[H]
\centering
\tiny
\begin{tabular}{|c|ccccccc|}
\hline
                          & \multicolumn{7}{c|}{\textbf{Gesture Path Length (DV2)}}                                                                                                                                                                                                                                                                                                                                                                                                                             \\ \cline{2-8} 
                          & \multicolumn{2}{c|}{\begin{tabular}[c]{@{}c@{}}Shapiro-Wilk\\ Test Results\end{tabular}} & \multicolumn{2}{c|}{\begin{tabular}[c]{@{}c@{}}Kruskal-Wallis\\ Results for Data with\\ Non-Normal Distribution\end{tabular}} & \multicolumn{3}{c|}{\begin{tabular}[c]{@{}c@{}}Post-hoc\\ Test Results\end{tabular}}                                                                                                                                                             \\ \cline{2-8} 
\multirow{-3}{*}{Gesture} & \multicolumn{1}{c|}{$W$}                      & \multicolumn{1}{c|}{$p$}                     & \multicolumn{1}{c|}{${\chi^2}(2)$}          & \multicolumn{1}{c|}{$p$}                                     & \multicolumn{1}{c|}{\begin{tabular}[c]{@{}c@{}}3 Region\\ Vs\\ 5 Region\end{tabular}} & \multicolumn{1}{c|}{\begin{tabular}[c]{@{}c@{}}5 Region\\ Vs\\ 7 Region\end{tabular}} & \begin{tabular}[c]{@{}c@{}}3 Region\\ Vs\\ 7 Region\end{tabular} \\ \hline
All gestures On-skin        & \multicolumn{1}{c|}{0.964}                  & \multicolumn{1}{c|}{$<0.001$}                 & \multicolumn{1}{c|}{1.419}                                      & \multicolumn{1}{c|}{\cellcolor[HTML]{F8FF00}0.492}          & \multicolumn{1}{c|}{\cellcolor[HTML]{C0C0C0}-}                                        & \multicolumn{1}{c|}{\cellcolor[HTML]{C0C0C0}-}                                        & \cellcolor[HTML]{C0C0C0}-                                        \\ \hline
All gestures Mid-air        & \multicolumn{1}{c|}{0.841}                  & \multicolumn{1}{c|}{$<0.001$}                 & \multicolumn{1}{c|}{6.889}                                      & \multicolumn{1}{c|}{0.032}                                 & \multicolumn{1}{c|}{0.025}                                                            & \multicolumn{1}{c|}{\cellcolor[HTML]{F8FF00}1.0}                                                            & \multicolumn{1}{c|}{\cellcolor[HTML]{F8FF00}{0.170}}                                        \\ \hline
Swipe up On-skin            & \multicolumn{1}{c|}{0.929}                  & \multicolumn{1}{c|}{$<0.001$}                 & \multicolumn{1}{c|}{1.231}                                      & \multicolumn{1}{c|}{\cellcolor[HTML]{F8FF00}0.54}          & \multicolumn{1}{c|}{\cellcolor[HTML]{C0C0C0}-}                                        & \multicolumn{1}{c|}{\cellcolor[HTML]{C0C0C0}-}                                        & \cellcolor[HTML]{C0C0C0}-                                        \\ \hline
Swipe up Mid-air            & \multicolumn{1}{c|}{0.991}                  & \multicolumn{1}{c|}{0.002}                 & \multicolumn{1}{c|}{2.12}                                       & \multicolumn{1}{c|}{\cellcolor[HTML]{F8FF00}0.347}         & \multicolumn{1}{c|}{\cellcolor[HTML]{C0C0C0}-}                                        & \multicolumn{1}{c|}{\cellcolor[HTML]{C0C0C0}-}                                        & \cellcolor[HTML]{C0C0C0}-                                        \\ \hline
Swipe down On-skin          & \multicolumn{1}{c|}{0.993}                  & \multicolumn{1}{c|}{0.011}                 & \multicolumn{1}{c|}{2.294}                                      & \multicolumn{1}{c|}{\cellcolor[HTML]{F8FF00}0.318}         & \multicolumn{1}{c|}{\cellcolor[HTML]{C0C0C0}-}                                        & \multicolumn{1}{c|}{\cellcolor[HTML]{C0C0C0}-}                                        & \cellcolor[HTML]{C0C0C0}-                                        \\ \hline
Swipe down Mid-air          & \multicolumn{1}{c|}{0.946}                  & \multicolumn{1}{c|}{$<0.001$}                 & \multicolumn{1}{c|}{4.117}                                      & \multicolumn{1}{c|}{\cellcolor[HTML]{F8FF00}0.128}         & \multicolumn{1}{c|}{\cellcolor[HTML]{C0C0C0}-}                                        & \multicolumn{1}{c|}{\cellcolor[HTML]{C0C0C0}-}                                        & \cellcolor[HTML]{C0C0C0}-                                        \\ \hline
Swipe front On-skin         & \multicolumn{1}{c|}{0.935}                  & \multicolumn{1}{c|}{$<0.001$}                 & \multicolumn{1}{c|}{0.686}                                      & \multicolumn{1}{c|}{\cellcolor[HTML]{F8FF00}0.71}          & \multicolumn{1}{c|}{\cellcolor[HTML]{C0C0C0}-}                                        & \multicolumn{1}{c|}{\cellcolor[HTML]{C0C0C0}-}                                        & \cellcolor[HTML]{C0C0C0}-                                        \\ \hline
Swipe front Mid-air         & \multicolumn{1}{c|}{0.891}                  & \multicolumn{1}{c|}{$<0.001$}                 & \multicolumn{1}{c|}{5.033}                                      & \multicolumn{1}{c|}{\cellcolor[HTML]{F8FF00}0.081}         & \multicolumn{1}{c|}{\cellcolor[HTML]{C0C0C0}-}                                        & \multicolumn{1}{c|}{\cellcolor[HTML]{C0C0C0}-}                                        & \cellcolor[HTML]{C0C0C0}-                                        \\ \hline
Swipe back On-skin          & \multicolumn{1}{c|}{0.99}                   & \multicolumn{1}{c|}{$<0.001$}                 & \multicolumn{1}{c|}{2.654}                                      & \multicolumn{1}{c|}{\cellcolor[HTML]{F8FF00}0.265}         & \multicolumn{1}{c|}{\cellcolor[HTML]{C0C0C0}-}                                        & \multicolumn{1}{c|}{\cellcolor[HTML]{C0C0C0}-}                                        & \cellcolor[HTML]{C0C0C0}-                                        \\ \hline
Swipe back Mid-air          & \multicolumn{1}{c|}{0.542}                  & \multicolumn{1}{c|}{$<0.001$}                 & \multicolumn{1}{c|}{0.29}                                       & \multicolumn{1}{c|}{\cellcolor[HTML]{F8FF00}0.865}         & \multicolumn{1}{c|}{\cellcolor[HTML]{C0C0C0}-}                                        & \multicolumn{1}{c|}{\cellcolor[HTML]{C0C0C0}-}                                        & \cellcolor[HTML]{C0C0C0}-                                        \\ \hline
Pinch in On-skin            & \multicolumn{1}{c|}{0.974}                  & \multicolumn{1}{c|}{$<0.001$}                 & \multicolumn{1}{c|}{2.025}                                      & \multicolumn{1}{c|}{\cellcolor[HTML]{F8FF00}0.363}         & \multicolumn{1}{c|}{\cellcolor[HTML]{C0C0C0}-}                                        & \multicolumn{1}{c|}{\cellcolor[HTML]{C0C0C0}-}                                        & \cellcolor[HTML]{C0C0C0}-                                        \\ \hline
Pinch in Mid-air            & \multicolumn{1}{c|}{0.975}                  & \multicolumn{1}{c|}{$<0.001$}                 & \multicolumn{1}{c|}{0.357}                                      & \multicolumn{1}{c|}{\cellcolor[HTML]{F8FF00}0.836}         & \multicolumn{1}{c|}{\cellcolor[HTML]{C0C0C0}-}                                        & \multicolumn{1}{c|}{\cellcolor[HTML]{C0C0C0}-}                                        & \cellcolor[HTML]{C0C0C0}-                                        \\ \hline
Pinch out On-skin           & \multicolumn{1}{c|}{0.984}                  & \multicolumn{1}{c|}{$<0.001$}                 & \multicolumn{1}{c|}{0.195}                                      & \multicolumn{1}{c|}{\cellcolor[HTML]{F8FF00}0.907}         & \multicolumn{1}{c|}{\cellcolor[HTML]{C0C0C0}-}                                        & \multicolumn{1}{c|}{\cellcolor[HTML]{C0C0C0}-}                                        & \cellcolor[HTML]{C0C0C0}-                                        \\ \hline
Pinch out Mid-air           & \multicolumn{1}{c|}{0.965}                  & \multicolumn{1}{c|}{$<0.001$}                 & \multicolumn{1}{c|}{12.02}                                      & \multicolumn{1}{c|}{0.002}                                 & \multicolumn{1}{c|}{0.004}                                                            & \multicolumn{1}{c|}{0.018}                                                            & \multicolumn{1}{c|}{\cellcolor[HTML]{F8FF00}{1.0}}                                        \\ \hline
Tap On-skin                & \multicolumn{1}{c|}{0.983}                  & \multicolumn{1}{c|}{$<0.001$}                 & \multicolumn{1}{c|}{0.172}                                      & \multicolumn{1}{c|}{\cellcolor[HTML]{F8FF00}0.918}         & \multicolumn{1}{c|}{\cellcolor[HTML]{C0C0C0}-}                                        & \multicolumn{1}{c|}{\cellcolor[HTML]{C0C0C0}-}                                        & \cellcolor[HTML]{C0C0C0}-                                        \\ \hline
Tap Mid-air                & \multicolumn{1}{c|}{0.971}                  & \multicolumn{1}{c|}{$<0.001$}                 & \multicolumn{1}{c|}{0.388}                                      & \multicolumn{1}{c|}{\cellcolor[HTML]{F8FF00}0.824}         & \multicolumn{1}{c|}{\cellcolor[HTML]{C0C0C0}-}                                        & \multicolumn{1}{c|}{\cellcolor[HTML]{C0C0C0}-}                                        & \cellcolor[HTML]{C0C0C0}-                                        \\ \hline
\end{tabular}%
\captionsetup{justification=centering}
\caption{Statistical Test Report for analyzing \rev{the} effect of increasing \rev{the} number of regions around the ear on gesture path length (DV2) for off-device mid-air and on-skin gesture (Swipe/Pinch/Tap) reuse. \\Yellow boxes represent statistical insignificance.}
\label{tab:DV2_vs_RQ2}
\end{table}

\newpage
\subsection{Gesture Accuracy (DV3) Analysis}
\label{appendix:DV3}

\begin{table}[H]
\centering
\tiny
\begin{tabular}{|c|c|ccccccccc|}
\hline
                             &                                                                                                             & \multicolumn{9}{c|}{\textbf{Gesture Accuracy (DV3)}}                                                                                                                                                                                                                                                                                                                                                                                                                         \\ \cline{3-11} 
                             &                                                                                                             & \multicolumn{2}{c|}{}                                                                                      & \multicolumn{2}{c|}{}                                                                                  & \multicolumn{5}{c|}{Onskin Vs Mid-air Gestures}                                                                                                                                                                                                         \\ \cline{7-11} 
                             &                                                                                                             & \multicolumn{2}{c|}{\multirow{-2}{*}{\begin{tabular}[c]{@{}c@{}}Shapiro-Wilk\\ Test Results\end{tabular}}} & \multicolumn{2}{c|}{\multirow{-2}{*}{\begin{tabular}[c]{@{}c@{}}Bartlett\\ Test Results\end{tabular}}} & \multicolumn{2}{c|}{\begin{tabular}[c]{@{}c@{}}Paired T-test\\ Results for Normally \\ Distributed Data\end{tabular}} & \multicolumn{3}{c|}{\begin{tabular}[c]{@{}c@{}}Wilcoxon Test \\ Results for Data with \\ Non-Normal Distribution\end{tabular}} \\ \cline{3-11} 
\multirow{-4}{*}{Gesture}    & \multirow{-4}{*}{\begin{tabular}[c]{@{}c@{}}\# of Gesture\\ Regions in\\ Interaction \\ Space\end{tabular}} & \multicolumn{1}{c|}{$W$}                           & \multicolumn{1}{c|}{$p$}                              & \multicolumn{1}{c|}{${\chi^2(1)}$}                 & \multicolumn{1}{c|}{$p$}                          & \multicolumn{1}{c|}{$t$}                                  & \multicolumn{1}{c|}{$p$}                                  & \multicolumn{1}{c|}{$W$}               & \multicolumn{1}{c|}{$Z$}                 & $p$                                        \\ \hline
                             & 3                                                                                                           & \multicolumn{1}{c|}{0.560}                         & \multicolumn{1}{c|}{$<0.001$}                         & \multicolumn{1}{c|}{\cellcolor[HTML]{C0C0C0}-}     & \multicolumn{1}{c|}{\cellcolor[HTML]{C0C0C0}-}    & \multicolumn{1}{c|}{\cellcolor[HTML]{C0C0C0}-}            & \multicolumn{1}{c|}{\cellcolor[HTML]{C0C0C0}-}            & \multicolumn{1}{c|}{470}               & \multicolumn{1}{c|}{0.884}               & \cellcolor[HTML]{F8FF00}0.461             \\ \cline{2-11} 
                             & 5                                                                                                           & \multicolumn{1}{c|}{0.857}                         & \multicolumn{1}{c|}{$<0.001$}                         & \multicolumn{1}{c|}{\cellcolor[HTML]{C0C0C0}-}     & \multicolumn{1}{c|}{\cellcolor[HTML]{C0C0C0}-}    & \multicolumn{1}{c|}{\cellcolor[HTML]{C0C0C0}-}            & \multicolumn{1}{c|}{\cellcolor[HTML]{C0C0C0}-}            & \multicolumn{1}{c|}{730.5}             & \multicolumn{1}{c|}{6.220}               & $<0.001$                                   \\ \cline{2-11} 
\multirow{-3}{*}{All gestures} & 7                                                                                                           & \multicolumn{1}{c|}{0.920}                       & \multicolumn{1}{c|}{$<0.001$}                         & \multicolumn{1}{c|}{\cellcolor[HTML]{C0C0C0}-}     & \multicolumn{1}{c|}{\cellcolor[HTML]{C0C0C0}-}    & \multicolumn{1}{c|}{\cellcolor[HTML]{C0C0C0}-}            & \multicolumn{1}{c|}{\cellcolor[HTML]{C0C0C0}-}            & \multicolumn{1}{c|}{1950.5}            & \multicolumn{1}{c|}{3.003}               & 0.002                                      \\ \hline
                             & 3                                                                                                           & \multicolumn{1}{c|}{0.514}                         & \multicolumn{1}{c|}{$<0.001$}                         & \multicolumn{1}{c|}{\cellcolor[HTML]{C0C0C0}-}     & \multicolumn{1}{c|}{\cellcolor[HTML]{C0C0C0}-}    & \multicolumn{1}{c|}{\cellcolor[HTML]{C0C0C0}-}            & \multicolumn{1}{c|}{\cellcolor[HTML]{C0C0C0}-}            & \multicolumn{1}{c|}{0}                 & \multicolumn{1}{c|}{2.450}               & 0.031                                      \\ \cline{2-11} 
                             & 5                                                                                                           & \multicolumn{1}{c|}{0.839}                         & \multicolumn{1}{c|}{$<0.001$}                         & \multicolumn{1}{c|}{\cellcolor[HTML]{C0C0C0}-}     & \multicolumn{1}{c|}{\cellcolor[HTML]{C0C0C0}-}    & \multicolumn{1}{c|}{\cellcolor[HTML]{C0C0C0}-}            & \multicolumn{1}{c|}{\cellcolor[HTML]{C0C0C0}-}            & \multicolumn{1}{c|}{0}                 & \multicolumn{1}{c|}{3.645}               & $<0.001$                                   \\ \cline{2-11} 
\multirow{-3}{*}{Swipe up}    & 7                                                                                                           & \multicolumn{1}{c|}{0.776}                         & \multicolumn{1}{c|}{$<0.001$}                         & \multicolumn{1}{c|}{\cellcolor[HTML]{C0C0C0}-}     & \multicolumn{1}{c|}{\cellcolor[HTML]{C0C0C0}-}    & \multicolumn{1}{c|}{\cellcolor[HTML]{C0C0C0}-}            & \multicolumn{1}{c|}{\cellcolor[HTML]{C0C0C0}-}            & \multicolumn{1}{c|}{12}               & \multicolumn{1}{c|}{2.335}               & 0.033                                      \\ \hline
                             & 3                                                                                                           & \multicolumn{1}{c|}{0.485}                         & \multicolumn{1}{c|}{$<0.001$}                         & \multicolumn{1}{c|}{\cellcolor[HTML]{C0C0C0}-}     & \multicolumn{1}{c|}{\cellcolor[HTML]{C0C0C0}-}    & \multicolumn{1}{c|}{\cellcolor[HTML]{C0C0C0}-}            & \multicolumn{1}{c|}{\cellcolor[HTML]{C0C0C0}-}            & \multicolumn{1}{c|}{4}                 & \multicolumn{1}{c|}{0.577}               & \cellcolor[HTML]{F8FF00}0.772              \\ \cline{2-11} 
                             & 5                                                                                                           & \multicolumn{1}{c|}{0.775}                         & \multicolumn{1}{c|}{$<0.001$}                         & \multicolumn{1}{c|}{\cellcolor[HTML]{C0C0C0}-}     & \multicolumn{1}{c|}{\cellcolor[HTML]{C0C0C0}-}    & \multicolumn{1}{c|}{\cellcolor[HTML]{C0C0C0}-}            & \multicolumn{1}{c|}{\cellcolor[HTML]{C0C0C0}-}            & \multicolumn{1}{c|}{13}                & \multicolumn{1}{c|}{2.516}               & 0.020                                      \\ \cline{2-11} 
\multirow{-3}{*}{Swipe down}  & 7                                                                                                           & \multicolumn{1}{c|}{0.871}                         & \multicolumn{1}{c|}{$<0.001$}                         & \multicolumn{1}{c|}{\cellcolor[HTML]{C0C0C0}-}     & \multicolumn{1}{c|}{\cellcolor[HTML]{C0C0C0}-}    & \multicolumn{1}{c|}{\cellcolor[HTML]{C0C0C0}-}            & \multicolumn{1}{c|}{\cellcolor[HTML]{C0C0C0}-}            & \multicolumn{1}{c|}{11}               & \multicolumn{1}{c|}{2.753}               & 0.007                                      \\ \hline
                             & 3                                                                                                           & \multicolumn{1}{c|}{0.451}                         & \multicolumn{1}{c|}{$<0.001$}                         & \multicolumn{1}{c|}{\cellcolor[HTML]{C0C0C0}-}     & \multicolumn{1}{c|}{\cellcolor[HTML]{C0C0C0}-}    & \multicolumn{1}{c|}{\cellcolor[HTML]{C0C0C0}-}            & \multicolumn{1}{c|}{\cellcolor[HTML]{C0C0C0}-}            & \multicolumn{1}{c|}{5}                 & \multicolumn{1}{c|}{0.0}                 & \cellcolor[HTML]{F8FF00}1.0                \\ \cline{2-11} 
                             & 5                                                                                                           & \multicolumn{1}{c|}{0.791}                         & \multicolumn{1}{c|}{$<0.001$}                         & \multicolumn{1}{c|}{\cellcolor[HTML]{C0C0C0}-}     & \multicolumn{1}{c|}{\cellcolor[HTML]{C0C0C0}-}    & \multicolumn{1}{c|}{\cellcolor[HTML]{C0C0C0}-}            & \multicolumn{1}{c|}{\cellcolor[HTML]{C0C0C0}-}            & \multicolumn{1}{c|}{0}                 & \multicolumn{1}{c|}{3.392}               & $<0.001$                                   \\ \cline{2-11} 
\multirow{-3}{*}{Swipe front} & 7                                                                                                           & \multicolumn{1}{c|}{0.861}                         & \multicolumn{1}{c|}{$<0.001$}                         & \multicolumn{1}{c|}{\cellcolor[HTML]{C0C0C0}-}     & \multicolumn{1}{c|}{\cellcolor[HTML]{C0C0C0}-}    & \multicolumn{1}{c|}{\cellcolor[HTML]{C0C0C0}-}            & \multicolumn{1}{c|}{\cellcolor[HTML]{C0C0C0}-}            & \multicolumn{1}{c|}{11}               & \multicolumn{1}{c|}{3.208}               & $<0.001$                                   \\ \hline
                             & 3                                                                                                           & \multicolumn{1}{c|}{0.628}                         & \multicolumn{1}{c|}{$<0.001$}                         & \multicolumn{1}{c|}{\cellcolor[HTML]{C0C0C0}-}     & \multicolumn{1}{c|}{\cellcolor[HTML]{C0C0C0}-}    & \multicolumn{1}{c|}{\cellcolor[HTML]{C0C0C0}-}            & \multicolumn{1}{c|}{\cellcolor[HTML]{C0C0C0}-}            & \multicolumn{1}{c|}{40}                & \multicolumn{1}{c|}{2.333}               & 0.039                                      \\ \cline{2-11} 
                             & 5                                                                                                           & \multicolumn{1}{c|}{0.827}                         & \multicolumn{1}{c|}{$<0.001$}                         & \multicolumn{1}{c|}{\cellcolor[HTML]{C0C0C0}-}     & \multicolumn{1}{c|}{\cellcolor[HTML]{C0C0C0}-}    & \multicolumn{1}{c|}{\cellcolor[HTML]{C0C0C0}-}            & \multicolumn{1}{c|}{\cellcolor[HTML]{C0C0C0}-}            & \multicolumn{1}{c|}{100}               & \multicolumn{1}{c|}{1.616}               & \cellcolor[HTML]{F8FF00}0.152              \\ \cline{2-11} 
\multirow{-3}{*}{Swipe back}  & 7                                                                                                           & \multicolumn{1}{c|}{0.830}                         & \multicolumn{1}{c|}{$<0.001$}                         & \multicolumn{1}{c|}{\cellcolor[HTML]{C0C0C0}-}     & \multicolumn{1}{c|}{\cellcolor[HTML]{C0C0C0}-}    & \multicolumn{1}{c|}{\cellcolor[HTML]{C0C0C0}-}            & \multicolumn{1}{c|}{\cellcolor[HTML]{C0C0C0}-}            & \multicolumn{1}{c|}{135}              & \multicolumn{1}{c|}{2.784}               & 0.003                                      \\ \hline
                             & 3                                                                                                           & \multicolumn{1}{c|}{0.580}                         & \multicolumn{1}{c|}{$<0.001$}                         & \multicolumn{1}{c|}{\cellcolor[HTML]{C0C0C0}-}     & \multicolumn{1}{c|}{\cellcolor[HTML]{C0C0C0}-}    & \multicolumn{1}{c|}{\cellcolor[HTML]{C0C0C0}-}            & \multicolumn{1}{c|}{\cellcolor[HTML]{C0C0C0}-}            & \multicolumn{1}{c|}{5}                 & \multicolumn{1}{c|}{2.333}               & 0.039                                      \\ \cline{2-11} 
                             & 5                                                                                                           & \multicolumn{1}{c|}{0.870}                         & \multicolumn{1}{c|}{$<0.001$}                         & \multicolumn{1}{c|}{\cellcolor[HTML]{C0C0C0}-}     & \multicolumn{1}{c|}{\cellcolor[HTML]{C0C0C0}-}    & \multicolumn{1}{c|}{\cellcolor[HTML]{C0C0C0}-}            & \multicolumn{1}{c|}{\cellcolor[HTML]{C0C0C0}-}            & \multicolumn{1}{c|}{8}                 & \multicolumn{1}{c|}{2.673}               & 0.008                                      \\ \cline{2-11} 
\multirow{-3}{*}{Pinch in}    & 7                                                                                                           & \multicolumn{1}{c|}{0.917}                         & \multicolumn{1}{c|}{$<0.001$}                         & \multicolumn{1}{c|}{\cellcolor[HTML]{C0C0C0}-}     & \multicolumn{1}{c|}{\cellcolor[HTML]{C0C0C0}-}    & \multicolumn{1}{c|}{\cellcolor[HTML]{C0C0C0}-}            & \multicolumn{1}{c|}{\cellcolor[HTML]{C0C0C0}-}            & \multicolumn{1}{c|}{40}               & \multicolumn{1}{c|}{0.729}               & \cellcolor[HTML]{F8FF00}0.481              \\ \hline
                             & 3                                                                                                           & \multicolumn{1}{c|}{0.636}                         & \multicolumn{1}{c|}{$<0.001$}                         & \multicolumn{1}{c|}{\cellcolor[HTML]{C0C0C0}-}     & \multicolumn{1}{c|}{\cellcolor[HTML]{C0C0C0}-}    & \multicolumn{1}{c|}{\cellcolor[HTML]{C0C0C0}-}            & \multicolumn{1}{c|}{\cellcolor[HTML]{C0C0C0}-}            & \multicolumn{1}{c|}{25}                & \multicolumn{1}{c|}{0.333}               & \cellcolor[HTML]{F8FF00}0.789              \\ \cline{2-11} 
                             & 5                                                                                                           & \multicolumn{1}{c|}{0.879}                         & \multicolumn{1}{c|}{$<0.001$}                         & \multicolumn{1}{c|}{\cellcolor[HTML]{C0C0C0}-}     & \multicolumn{1}{c|}{\cellcolor[HTML]{C0C0C0}-}    & \multicolumn{1}{c|}{\cellcolor[HTML]{C0C0C0}-}            & \multicolumn{1}{c|}{\cellcolor[HTML]{C0C0C0}-}            & \multicolumn{1}{c|}{11}                & \multicolumn{1}{c|}{2.922}               & 0.004                                      \\ \cline{2-11} 
\multirow{-3}{*}{Pinch out}   & 7                                                                                                           & \multicolumn{1}{c|}{0.909}                         & \multicolumn{1}{c|}{0.006}                            & \multicolumn{1}{c|}{\cellcolor[HTML]{C0C0C0}-}     & \multicolumn{1}{c|}{\cellcolor[HTML]{C0C0C0}-}    & \multicolumn{1}{c|}{\cellcolor[HTML]{C0C0C0}-}            & \multicolumn{1}{c|}{\cellcolor[HTML]{C0C0C0}-}            & \multicolumn{1}{c|}{69}               & \multicolumn{1}{c|}{0.044}               & \cellcolor[HTML]{F8FF00}0.961              \\ \hline
                             & 3                                                                                                           & \multicolumn{1}{c|}{0.451}                         & \multicolumn{1}{c|}{$<0.001$}                         & \multicolumn{1}{c|}{\cellcolor[HTML]{C0C0C0}-}     & \multicolumn{1}{c|}{\cellcolor[HTML]{C0C0C0}-}    & \multicolumn{1}{c|}{\cellcolor[HTML]{C0C0C0}-}            & \multicolumn{1}{c|}{\cellcolor[HTML]{C0C0C0}-}            & \multicolumn{1}{c|}{7}                 & \multicolumn{1}{c|}{0.817}               & \cellcolor[HTML]{F8FF00}0.688              \\ \cline{2-11} 
                             & 5                                                                                                           & \multicolumn{1}{c|}{0.806}                         & \multicolumn{1}{c|}{$<0.001$}                         & \multicolumn{1}{c|}{\cellcolor[HTML]{C0C0C0}-}     & \multicolumn{1}{c|}{\cellcolor[HTML]{C0C0C0}-}    & \multicolumn{1}{c|}{\cellcolor[HTML]{C0C0C0}-}            & \multicolumn{1}{c|}{\cellcolor[HTML]{C0C0C0}-}            & \multicolumn{1}{c|}{0}                 & \multicolumn{1}{c|}{3.118}               & 0.002                                      \\ \cline{2-11} 
\multirow{-3}{*}{Tap}        & 7                                                                                                           & \multicolumn{1}{c|}{0.915}                         & \multicolumn{1}{c|}{0.009}                            & \multicolumn{1}{c|}{\cellcolor[HTML]{C0C0C0}-}     & \multicolumn{1}{c|}{\cellcolor[HTML]{C0C0C0}-}    & \multicolumn{1}{c|}{\cellcolor[HTML]{C0C0C0}-}            & \multicolumn{1}{c|}{\cellcolor[HTML]{C0C0C0}-}            & \multicolumn{1}{c|}{10}                & \multicolumn{1}{c|}{3.057}               & 0.002                                      \\ \hline
\end{tabular}%
\captionsetup{justification=centering}
\caption{Statistical Test Report for analyzing \rev{the} effect of interaction space choice on gesture accuracy (DV3) for different numbers of gesture (Swipe/Pinch/Tap) regions around the ear.\\Yellow boxes represent statistical insignificance.}
\label{tab:DV3_vs_RQ1}
\end{table}

\vspace{-0.5cm}
\begin{table}[H]
\centering
\tiny
\begin{tabular}{|c|ccccccc|}
\hline
                          & \multicolumn{7}{c|}{Gesture Accuracy (DV3)}                                                                                                                                                                                                                                                                                                                                                                                                                             \\ \cline{2-8} 
                          & \multicolumn{2}{c|}{\begin{tabular}[c]{@{}c@{}}Shapiro-Wilk\\ Test Results\end{tabular}} & \multicolumn{2}{c|}{\begin{tabular}[c]{@{}c@{}}Friedman Test\\ Results for Data with\\ Non-Normal Distribution\end{tabular}} & \multicolumn{3}{c|}{\begin{tabular}[c]{@{}c@{}}Post-hoc\\ Test Results\end{tabular}}                                                                                                                                                             \\ \cline{2-8} 
\multirow{-3}{*}{Gesture} & \multicolumn{1}{c|}{$W$}                    & \multicolumn{1}{c|}{$p$}                   & \multicolumn{1}{c|}{${\chi^2}(2)$}                    & \multicolumn{1}{c|}{$p$}                                              & \multicolumn{1}{c|}{\begin{tabular}[c]{@{}c@{}}3 Region\\ Vs\\ 5 Region\end{tabular}} & \multicolumn{1}{c|}{\begin{tabular}[c]{@{}c@{}}5 Region\\ Vs\\ 7 Region\end{tabular}} & \begin{tabular}[c]{@{}c@{}}3 Region\\ Vs\\ 7 Region\end{tabular} \\ \hline
Swipe up On-skin            & \multicolumn{1}{c|}{0.874}                  & \multicolumn{1}{c|}{$<0.001$}              & \multicolumn{1}{c|}{117.05}                          & \multicolumn{1}{c|}{$<0.001$}                                         & \multicolumn{1}{c|}{$<0.001$}                                                         & \multicolumn{1}{c|}{$<0.001$}                                                         & \multicolumn{1}{c|}{$<0.001$}                                    \\ \hline
Swipe up Mid-air            & \multicolumn{1}{c|}{0.920}                  & \multicolumn{1}{c|}{$<0.001$}              & \multicolumn{1}{c|}{129.27}                          & \multicolumn{1}{c|}{$<0.001$}                                         & \multicolumn{1}{c|}{$<0.001$}                                                         & \multicolumn{1}{c|}{$<0.001$}                                                         & \multicolumn{1}{c|}{$<0.001$}                                    \\ \hline
Swipe up On-skin            & \multicolumn{1}{c|}{0.777}                  & \multicolumn{1}{c|}{$<0.001$}                 & \multicolumn{1}{c|}{23.311}                          & \multicolumn{1}{c|}{$<0.001$}                                            & \multicolumn{1}{c|}{\cellcolor[HTML]{F8FF00}0.4281}                                   & \multicolumn{1}{c|}{0.003}                                                            & {$<0.001$}                                                            \\ \hline
Swipe up Mid-air            & \multicolumn{1}{c|}{0.846}                  & \multicolumn{1}{c|}{$<0.001$}                 & \multicolumn{1}{c|}{20.333}                          & \multicolumn{1}{c|}{$<0.001$}                                            & \multicolumn{1}{c|}{0.002}                                                            & \multicolumn{1}{c|}{\cellcolor[HTML]{F8FF00}0.115}                                    & {$<0.001$}                                                            \\ \hline
Swipe down On-skin          & \multicolumn{1}{c|}{0.867}                  & \multicolumn{1}{c|}{$<0.001$}                 & \multicolumn{1}{c|}{23.311}                          & \multicolumn{1}{c|}{$<0.001$}                                            & \multicolumn{1}{c|}{\cellcolor[HTML]{F8FF00}0.4281}                                   & \multicolumn{1}{c|}{0.003}                                                            & {$<0.001$}                                                            \\ \hline
Swipe down Mid-air          & \multicolumn{1}{c|}{0.877}                  & \multicolumn{1}{c|}{$<0.001$}                 & \multicolumn{1}{c|}{20.333}                          & \multicolumn{1}{c|}{$<0.001$}                                            & \multicolumn{1}{c|}{0.003}                                                            & \multicolumn{1}{c|}{\cellcolor[HTML]{F8FF00}0.115}                                    & {$<0.001$}                                                            \\ \hline
Swipe front On-skin         & \multicolumn{1}{c|}{0.827}                  & \multicolumn{1}{c|}{$<0.001$}                 & \multicolumn{1}{c|}{24.125}                          & \multicolumn{1}{c|}{$<0.001$}                                            & \multicolumn{1}{c|}{\cellcolor[HTML]{F8FF00}0.077}                                    & \multicolumn{1}{c|}{$<0.001$}                                                            & {$<0.001$}                                                            \\ \hline
Swipe front Mid-air         & \multicolumn{1}{c|}{0.876}                  & \multicolumn{1}{c|}{$<0.001$}                 & \multicolumn{1}{c|}{32.444}                          & \multicolumn{1}{c|}{$<0.001$}                                            & \multicolumn{1}{c|}{$<0.001$}                                                            & \multicolumn{1}{c|}{0.002}                                                            & {$<0.001$}                                                            \\ \hline
Swipe back On-skin          & \multicolumn{1}{c|}{0.921}                  & \multicolumn{1}{c|}{$<0.001$}                 & \multicolumn{1}{c|}{5.778}                           & \multicolumn{1}{c|}{0.045}                    & \multicolumn{1}{c|}{0.033}                                        & \multicolumn{1}{c|}{\cellcolor[HTML]{F8FF00}0.180}                                        & {0.008}                                       \\ \hline
Swipe back Mid-air          & \multicolumn{1}{c|}{0.871}                  & \multicolumn{1}{c|}{$<0.001$}                 & \multicolumn{1}{c|}{10.111}                          & \multicolumn{1}{c|}{0.006}                                            & \multicolumn{1}{c|}{0.027}                                                            & \multicolumn{1}{c|}{\cellcolor[HTML]{F8FF00}1}                                        & 0.002                                                            \\ \hline
Pinch in On-skin            & \multicolumn{1}{c|}{0.847}                  & \multicolumn{1}{c|}{$<0.001$}                 & \multicolumn{1}{c|}{24.121}                          & \multicolumn{1}{c|}{$<0.001$}                                            & \multicolumn{1}{c|}{0.032}                                                            & \multicolumn{1}{c|}{0.007}                                                            & {$<0.001$}                                                            \\ \hline
Pinch in Mid-air            & \multicolumn{1}{c|}{0.923}                  & \multicolumn{1}{c|}{0.002}                 & \multicolumn{1}{c|}{12.333}                          & \multicolumn{1}{c|}{0.002}                                            & \multicolumn{1}{c|}{\cellcolor[HTML]{F8FF00}0.133}                                    & \multicolumn{1}{c|}{0.025}                                                            & 0.003                                                            \\ \hline
Pinch out On-skin           & \multicolumn{1}{c|}{0.901}                  & \multicolumn{1}{c|}{$<0.001$}                 & \multicolumn{1}{c|}{6.971}                           & \multicolumn{1}{c|}{0.031}                                            & \multicolumn{1}{c|}{\cellcolor[HTML]{F8FF00}1}                                        & \multicolumn{1}{c|}{0.003}                                                            & 0.009                                                            \\ \hline
Pinch out Mid-air           & \multicolumn{1}{c|}{0.928}                  & \multicolumn{1}{c|}{0.003}                 & \multicolumn{1}{c|}{10.333}                          & \multicolumn{1}{c|}{0.006}                                            & \multicolumn{1}{c|}{0.017}                                                            & \multicolumn{1}{c|}{\cellcolor[HTML]{F8FF00}1}                                        & 0.004                                                            \\ \hline
Tap On-skin                & \multicolumn{1}{c|}{0.831}                  & \multicolumn{1}{c|}{$<0.001$}                 & \multicolumn{1}{c|}{19.5}                            & \multicolumn{1}{c|}{$<0.001$}                                            & \multicolumn{1}{c|}{0.036}                                                            & \multicolumn{1}{c|}{0.002}                                                            & {$<0.001$}                                                            \\ \hline
Tap Mid-air                & \multicolumn{1}{c|}{0.905}                  & \multicolumn{1}{c|}{$<0.001$}                 & \multicolumn{1}{c|}{29.778}                          & \multicolumn{1}{c|}{$<0.001$}                                            & \multicolumn{1}{c|}{0.003}                                                            & \multicolumn{1}{c|}{$<0.001$}                                                            & {$<0.001$}                                                            \\ \hline
\end{tabular}%
\captionsetup{justification=centering}
\caption{Statistical Test Report for analyzing \rev{the} effect of increasing \rev{the} number of regions around the ear on gesture accuracy (DV3) for off-device mid-air and on-skin gestures (Swipe/Pinch/Tap).\\Yellow boxes represent statistical insignificance.}
\label{tab:DV3_vs_RQ2}
\end{table}

\end{appendices}

\end{document}